\documentclass[fleqn,usenatbib]{mnras}

\usepackage{newtxtext,newtxmath}

\usepackage{url}
\usepackage{color}
\usepackage{tabulary}
\usepackage{multirow}
\usepackage{amsfonts}
\usepackage{array}
\usepackage{ulem}
\usepackage[T1]{fontenc}
\usepackage{ae,aecompl}

\usepackage{graphicx}	
\usepackage{amsmath}	
\usepackage{amssymb}	
\usepackage{cases}

\graphicspath{ {pictures/} }
\newcommand{\beq}{\begin{equation}}
\newcommand{\eeq}{\end{equation}}
\newcommand{\simlt}{\mathrel{\hbox{\rlap{\hbox{\lower4pt\hbox{$\sim$}}}\hbox{$<$}}}}
\newcommand{\simgt}{\mathrel{\hbox{\rlap{\hbox{\lower4pt\hbox{$\sim$}}}\hbox{$>$}}}}

\newcommand{\erg}{\;\mathrm{erg}}
\newcommand{\s}{\;\mathrm{s}}

\newcommand{\Msol}{\;\mathrm{M}_{\odot}}
\newcommand{\Rsol}{\;\mathrm{R}_{\odot}}
\newcommand{\gram}{\;\mathrm{g}}
\newcommand{\cm}{\;\mathrm{cm}}
\newcommand{\cmcube}{\;\mathrm{cm}^{-3}}
\newcommand{\km}{\;\mathrm{km}}
\newcommand{\AU}{\;\mathrm{AU}}
\newcommand{\pc}{\;\mathrm{pc}}
\newcommand{\yr}{\;\mathrm{yr}}
\newcommand{\Myr}{\;\mathrm{Myr}}

\def\apjl{ApJL}
\def\apj{ApJ}
\def\mnras{M.N.R.A.S.}
\def\aap{A\&A}

\def\araa{Ann. Rev. A\&A}

\def\apjs{ApJ Supp.}

\defcitealias{Gualandris+2004}{GPZE04}
\defcitealias{Ryu+2016}{RTP}

\title[runaway stars]{Formation of runaway stars in a star-cluster potential}

\author[T. Ryu et al.]{
Taeho Ryu$^{1}$\thanks{email: taeho.ryu@stonybrook.edu}, Nathan W. C. Leigh$^{2}$, Rosalba Perna$^{1}$
\\
$^{1}$Department of Physics and Astronomy, Stony Brook University, Stony Brook, NY 11794-3800, USA\\
$^{2}$Department of Astrophysics, American Museum of Natural History, Central Park West and 79th Street, New York, NY 10024}

\date{Accepted XXX. Received YYY; in original form ZZZ}

\pubyear{2016}

\begin{document}
\label{firstpage}
\pagerange{\pageref{firstpage}--\pageref{lastpage}}
\maketitle

\begin{abstract}

We study the formation of runaway stars due to binary-binary (2+2)
interactions in young star-forming clusters and/or associations.  This
is done using a combination of analytic methods and numerical
simulations of 2+2 scattering interactions, both in isolation and in a
homogeneous background potential.  We focus on interactions that
produce two single stars and a binary, and study the outcomes as a
function of the depth of the background potential, within a range
typical of cluster cores. As reference parameters for the
observational properties, we use those observed for the system of
runaway stars AE Aur and $\mu$~Col and binary $\iota$~Ori.  We find
that the outcome fractions have no appreciable dependence on the depth
of the potential, and neither do the velocities of the ejected single
stars. However, as the potential gets deeper and a larger fraction of
binaries remain trapped, two binary populations emerge, with the
escaped component having higher speeds and shorter semi-major axes
than the trapped one.  Additionally, we find that the relative angles
between the ejected products are generally large. In particular, the
angle between the ejected fastest star and the escaped binary is typically
$\gtrsim 120-135^{\circ}$, with a peak at around $160^{\circ}$. However,
as the potential gets deeper, the angle distribution becomes broader.
Finally, we discuss the implications of our results for the
interpretation of the properties of the runaway stars AE Aur and
$\mu$~Col.
\end{abstract}




\section{introduction}

\label{sec: introduction}

Runaway stars are a population of fast-moving stars, characterized by
speeds $\gtrsim 30$~km~s$^{-1}$.  They are generally of the O and B spectral
type, and are often found at some distance from star-forming regions
\citep{Blaauw1961,Stone1979}.  However, velocity measurements and reconstruction of
trajectories for a number of these stars have shown that they likely
originated in stellar clusters, and hence they were ejected from them 
at high speeds (i.e. \citealt{Hoogerwerf+2000}).

The origin of the high velocities of these runaway stars has remained
elusive, with two competing mechanisms for their formation being
mostly discussed in the literature: (a) Ejection of a star in a binary
system when its companion goes off as a supernova explosion
\citep{Zwicky1957,Blaauw1961}; (b) Dynamical formation as a result of
scattering in a close encounter between stars in a star
  cluster \citep{Poveda+1967,Gies+1986} or resulting from infalling star
  clusters interacting with massive black holes in galactic centres
  \citep{Dolcetta+2015,Fragione+2016,Fragione+2017}.

Here we focus on the latter mechanism, especially encounters between two binary stars. 
We investigate
the production of runaway stars from 2+2 scatterings, from which the
2+1+1 outcome is the most probable \citep[e.g.][]{Leigh+2016}. The single stars 
in the tail of the high-$v$ distribution are candidates
for runaway stars. 

Work by \citet{Ryu+2017} has shown that the outcome products of 2+2
scatterings, and their properties, are affected by the presence of a
background potential, representing the gravitational influence of the region in which the
scatterings occur. Motivated by this, we focus here on an 
investigation of the products of the 2+1+1 outcome in the presence of a background
potential, which we identify as the parent star cluster of the
runaway stars.

Understanding the mechanism by which runaway stars form relies on a
proper reconstruction of the velocities immediately following the
scattering, as a function of the observable properties of the ejected
stars, such as their ejection velocities, the angle between them, the
timescale since the scattering event.  All of these quantities
can be affected by the presence of a background potential, as shown in
previous work.

Here we quantify the effect of the background potential
on the 2+1+1 outcome products of 2+2 scatterings by focusing on the
specific case of the runway stars in the Trapezium cluster, which have
been monitored extensively \citep{Blaauw+1954}. This system is comprised by a set of two
single runaway stars, AE Aur and $\mu$ Col, and the spectroscopic
binary $\iota$ Ori. Their trajectories in the Galactic potential have
been traced back in time and found to intersect at the position of the
Trapezium cluster about 2.5~yr ago. \citet{Gualandris+2004}  
supported the idea that the system was created via the encounter of two
low-eccentricity binaries with comparable binding energy, as a result
of which the wider binary was ionized, and another star was swapped
into the original $\iota$ Ori binary. These simulations, while
modeling the $N$-body interactions among the stars, did not however
consider the effects that the Trapezium background potential would
have on the outcome products. Here we perform such an investigation:
in particular, we perform numerical 2+2 scattering experiments in a
background potential and examine in detail the properties of the 2+1+1
products.

The results presented in this paper have important implications for large samples of runaway 
stars, as are expected to emerge from the GAIA database \citep[e.g.][]{Kenyon+2014}.
No scattering experiments conducted to date (known to the authors) have quantified 
the effects of the host star cluster environment in determining the properties 
of runaway stars formed from dynamical scatterings.  Here, we address this issue 
directly. In addition to considering the potential of the Trapezium cluster, we also
generalize our study by considering deeper potentials (than observed in Trapezium).
For completeness, we note that \citet{Oh+2015} studied
the dependency of dynamical ejections of O Stars on the masses of very
young star clusters, while \citet{Perets+2012}
investigated the ejection of runaway stars from star clusters using
extensive $N$-body simulations.

The paper describes the numerical method and the simulation set up in
Sec.2; results are presented in Sec.3, while Sec.4 is devoted to a discussion
and conclusions.

\section{Method}
\label{sec:method}
In this section we first describe the numerical scattering simulations
of binary-binary encounters in a background potential. Next we detail
our choices of the initial conditions, physically motivated by
various observations.

\subsection{Overview of the scattering experiments}
\label{sec:overview}
We investigate the effects of a continuous background potential on the
formation of runaway O/B-type stars by performing numerical scattering
experiments of two (non-identical) binaries.  In order to cleanly
explore the effects of the background potential, we conservatively fix
the initial parameters of the two binaries, including the stellar
masses and the orbital parameters. This allows us to separate the
influence of the background potential from that of the initial binary
parameters.  Additionally, using the same initial conditions, we also
run simulations without any background potential as representative of
the case of purely (i.e., isolated) stellar dynamics. This allows us to identify the
outcome products which are mostly dependent on the background
potential, and the ones which are not.

In this study, we use two different codes: For the simulations with
the background potential, we use the $N$-body code developed by
\citealt[][hereafter RTP]{Ryu+2016}, and subsequently used for
scattering experiments by \citet{Ryu+2017}.  For the cases without
a background potential, and which include the study of finite size
effects, we use the \texttt{FEWBODY} code. The reason for using
different codes was simply motivated by running times and allocated
computing resources.  \texttt{FEWBODY} is generally faster when
compared to the \citetalias{Ryu+2016} code without a background potential, and hence it
allowed us to run a large number of simulations, as well as to study
the effect of the finite radii of the stars, while running in parallel
the \citetalias{Ryu+2016} code with a background potential on separate computing
resources. We remark that the two codes give statistically consistent
results when the \citetalias{Ryu+2016} code is run without a background potential.
In the following, we briefly summarize the main features of both codes.
For more details, we refer the reader to the respective original papers.

In the \citetalias{Ryu+2016} code,  the equations of motion of the
four stars are solved with the 4th - order \& 5 - stage
Runge-Kutta-Fehlberg method \citep{Fehlberg} using adaptive time steps
with error control tolerance\footnote{In the Runge-Kutta-Fehlberg
  method, the error can be controlled by using the higher-order
  embedded method. The error, defined as the difference between the two
  solutions from the 4th - order and 5th - order calculations, is
  estimated at each time step.  The following time step is
  automatically determined to yield an error less than a given error
  control tolerance.} of $10^{-11}$.  The numerical method uses a very
precise and stable integration among the large class of 
Runge-Kutta schemes, particularly by adopting the Butcher tableau for
Fehlberg's 4(5) method.  However, a small error control tolerance can
lead to a small time step size even for trivial
calculations. Therefore, in order to boost the computation speed up to
an acceptable level throughout the simulations, we additionally set a
minimum value for the time step, $10^{-7}\times \tau_{\rm dyn,
  \,min}$, where $\tau_{\rm dyn, \,min}$ is the smallest value of the
dynamical time between any two stars in the simulation at a given time
step.  This may cause the actual numerical errors to be higher than
what they would have been if set by the error control tolerance.  Hence, we manually 
monitor the errors to yield an acceptable computation speed while still maintaining
small numerical errors in the total energy.  Given the initial total
energy of the system $E(t=0)$, the numerical error in the total energy
of the system ($|[E(t)-E(t=0)]/E(t=0)|$ where $E(t)$ is defined in the
equation \ref{eq:totalenergy}) never exceeds $10^{-4}-10^{-6}$ in all
simulations.  
We use this code to perform $10^{3}$ scattering experiments of
two binaries in a background potential of varying depth.

For simulations without a background potential, we use the
\texttt{FEWBODY} numerical scattering code\footnote{For the source
  code, see http://fewbody.sourceforge.net}.  The code integrates the
usual $N$-body equations in configuration- (i.e., position-) space in
order to advance the system forward in time, using the eighth-order
Runge-Kutta Prince-Dormand integration method with ninth-order error
estimate and adaptive time-step.  For more details about the
\texttt{FEWBODY} code, we refer the reader to \citet{Fregeau+2004}.

We use different criteria in the \texttt{FEWBODY} code and 
the \citetalias{Ryu+2016} code
to decide when a given encounter is complete. 
In \texttt{FEWBODY} code,
we use the same criteria as \citet{Fregeau+2004}. 
To first order, this is defined as the point
at which the separately bound hierarchies that make up the system are
no longer interacting with each other or evolving internally.  More
specifically, the integration is terminated when the top-level
hierarchies have positive relative velocity and the corresponding
top-level $N$-body system has positive total energy.  Each hierarchy
must also be dynamically stable and experience a tidal perturbation
from other nodes within the same hierarchy that is less than the
critical value adopted by \texttt{FEWBODY}, called the tidal tolerance
parameter.  As in \citet{Leigh+2016} and \citet{GellerLeigh2015}, we adopt a
stricter value for the tidal tolerance parameter $\delta =$ 10$^{-7}$
than is provided by the default in \texttt{FEWBODY}, for all
simulations without the background potential analyzed in this paper. 
This is chosen to ensure the
correct outcome classifications at very low relative velocities, as
well as reasonable computer integration times.  We note that this
could lead to a slight over-estimate of the total encounter lifetimes
(see \citealt{GellerLeigh2015} for more details). However, 
each simulation with a 
background potential is run for $t=4\Myr$ using \citetalias{Ryu+2016} code.

\subsection{Background gravitational potential}
\label{sec:backgroundpotential}

\begin{table}
	\centering
	\setlength\extrarowheight{4pt}
	\begin{tabulary}{0.7\linewidth}{ c| c }
		\hline
		Mass (binary 1, binary 2)& $[39\Msol+16\Msol]$, $[19\Msol+16\Msol]$\\
		Initial separations $r_{12}$ & $1000\AU$\\
		Relative velocity $v_{\rm rel}$ & $2\km/\s$\\
	 Eccentricity  $e$ & $f(e)\sim e$\\
	  $e_{\rm min},e_{\rm max}$ & $0,~0.99$\\
	 Impact parameter $b$ & $f(b)\sim b^{2}$\\
	 	  $b_{\rm min},b_{\rm max}$ & $2,~220-230\AU$\\
	 Semimajor axis ($a_{1}$, $a_{2}$)& $\sim3.2\AU$, $\sim1.6\AU$ \\
	 Code termination time & $t=4\Myr$ (for runs with $V_{\rm bg}$)\\
	\multirow{2}{*}{Stellar radii (binary 1, binary2)} & $[16\Rsol+6\Rsol]$, $[11\Rsol+8\Rsol]$ \\
	 & (only for runs without $V_{\rm bg}$)\\
	 Density $\rho$ & $1\bar{\rho}$, $10^{3}\bar{\rho}$\\
	 Total background mass $M_{\rm bg}$ & $3\times10^{3}\Msol$, $10^{4}\Msol$\\
		\hline
	\end{tabulary}
	\caption{The initial conditions of our scattering
          experiments. The masses and radii of the four stars
          are chosen assuming them as proxies for $\iota$ Ori, AE Aur
          and $\mu$ Col.  These runaways are believed to have formed during encounters
          in the Trapezium cluster. Note that in simulations with a 
          background potential we do not take into account the stellar
          radii and physical collisions. In those without the
          background potential we explore the frequency of the
          physical collisions between stars given the stellar radii as
          shown above. $\bar{\rho}=1.66\times10^{-10}\gram\cmcube=2450\Msol\pc^{-3}$. }
	\label{tab:modelparameter}
\end{table}

\begin{table*}
	\centering
	\setlength\extrarowheight{4pt}
	\begin{tabulary}{0.7\linewidth}{ c| c c c c c c c ccc}
		\hline
		Model& $\rho$ & $M_{\rm bg}$ & $r_{\rm bg}$& $\alpha=E_{\rm b, 1}/E_{\rm b, 2}$ & $[b_{\rm min},~ b_{\rm max}]$  & $R_{\rm \star}$ & collisions? & $r_{\rm 12}$ & $v_{\rm rel}$& $v_{\rm esc} (r=r_{\rm bg})$\\
		\hline\hline
		Model 0 & 0 & 0 & 0 & 1 & [0, $230\AU$]     & 0 & no & $3000-6000\AU$ & $2\km\s^{-1}$& - \\
		Model 0-1 & 0 & 0& 0  & 1  &[0, $230\AU$]     & non-zero& yes& $3000-6000\AU$ & $2\km\s^{-1}$& -\\
		Model 1& 1 & $3\times10^{3}\Msol$& $0.66\pc$ & 1 &[0, $230\AU$]& 0& no& $1000\AU$ & $2\km\s^{-1}$& $6.2\km\s^{-1}$\\
		Model 1-1& 1 & $3\times10^{3}\Msol$& $0.66\pc$& 2&[0, $230\AU$]& 0& no& $1000\AU$ & $2\km\s^{-1}$ & $6.2\km\s^{-1}$\\
		Model 1-2& 1 & $3\times10^{3}\Msol$& $0.66\pc$& 0.5&[0, $230\AU$]& 0& no& $1000\AU$ & $2\km\s^{-1}$ & $6.2\km\s^{-1}$\\
		Model 2& 1 & $10^{4}\Msol$& $1.0\pc$& 1 &[0, $230\AU$]& 0& no& $1000\AU$ & $2\km\s^{-1}$&$9.3\km\s^{-1}$\\					
		Model 3& 1000& $3\times10^{3}\Msol$& $0.066\pc$& 1 &[0, $230\AU$]& 0& no& $1000\AU$ & $2\km\s^{-1}$ & $20\km\s^{-1}$\\
		\hline
	\end{tabulary}
	\caption{Model parameters used in our study. (From the first column to the last) 
the model name, the density $\rho$ (in units of 
$\bar\rho$), the total background mass $M_{\rm bg}$, 
the width of the potential ($r_{\rm bg}$), the ratio of the binding energies of the 
two initial binaries ($\alpha$), the range of the impact parameter $b$ drawn from 
the distribution of $f(b)\sim b^{2}$, the stellar radius ($R_{\star}$, \citetalias{Gualandris+2004}),
 the inclusion of 
the physical collisions, the initial separation $r_{\rm 12}$, 
the initial relative velocity $v_{\rm rel}$  and the escape velocity $v_{\rm esc}$ 
at $r=r_{\rm bg}$.}
	\label{tab:modelparameter2}
\end{table*}
We adopt the same background potential model of \citet{Ryu+2017} in
this study. We will briefly describe the essential ingredients of the
model here, but see \citet{Ryu+2017} for more details.

We consider a spherically symmetry potential with a constant density $\rho$ and an 
outer boundary $r_{\rm bg}$. Then, for a given total background mass $M_{\rm bg}$, 
the outer boundary is automatically determined. The background mass enclosed in a 
spherical volume of radius $r$ can be written as
\begin{align}
M_{\rm en,bg}(r)&=
\begin{cases}
\frac{4\uppi}{3}\rho r^{3} \hfill \hspace{0.8in} r \leq r_{\rm bg}\,;\\
\frac{4\uppi}{3}\rho r_{\rm bg}^{3}=M_{\rm bg} \hfill r>r_{\rm bg}\,.
\end{cases}
\label{eq:engasmass}
\end{align}
In the following, $\rho$ will be expressed in units of 
$\bar\rho=1.66\times10^{-19}\gram \cmcube=2450\Msol/\pc^{3}$, which corresponds to a density
of $n=10^{5}\cm^{-3}$ assuming a mean molecular weight of unity.

The gravitational force imparted by a background mass on a given star particle 
at $r$ follows the analytic formula: 
\begin{align}
\label{eq:backforce}
\textbf{\textit{f}}_{\rm bg}(r)&=-\frac{Gm M_{\rm en}(r)}{r^{3}}\textbf{\textit{r}}\nonumber\\ 
&=
\begin{cases}
-\frac{4}{3}\uppi G m\rho \;\textbf{\textit{r}} \hfill \hspace{0.3in} r \leq r_{\rm bg}\,;\\
-\frac{4}{3}\uppi G m\rho \Big(\frac{r_{\rm bg}}{r}\Big)^{3}\textbf{\textit{r}} \hfill \hspace{0.3in} r > r_{\rm bg}\,,\\
\end{cases}
\end{align}
where $m$ is the mass of the star and $\textbf{\textit{r}}$ is the vector pointing from 
the system CM to the star. Accordingly, the background potential has the following form:
\begin{align}
\label{eq:backpotential}
V_{\rm bg}(r)&=
\begin{cases}
\frac{2}{3}\uppi G m\rho (r^{2}-3r_{\rm bg}^{2})\hfill \hspace{0.3in} r \leq r_{\rm bg}\,;\\
-\frac{GmM_{\rm bg}}{r}=-\frac{4}{3}\uppi G m\rho \frac{r_{\rm bg}^{3}}{r} \hfill \hspace{0.3in} r > r_{\rm bg}\,.\\
\end{cases}
\end{align}

The total energy $E(t)$ of the four stars in the system at time $t$,
including the contribution from $V_{\rm bg}(r)$, can be written as
\begin{equation}
\label{eq:totalenergy}
E(t)=\sum_{i=1}^{4}\frac{1}{2}m_{i}v_{i}^{2}-\sum_{\substack{i,j=1\\(i>j)}}^{4}
\frac{Gm_{i}m_{j}}{|\textbf{\textit{r}}_{i}-\textbf{\textit{r}}_{j}|}+\sum_{i=1}^{4} V_{\rm bg}(r_{i})\,,
\end{equation}
where $m_{i}$ is the mass of each star and $v_{i}$ is the velocity of each star with 
respect to the system CM.

As in \citet{Ryu+2017}, we consider only the gravitational
force imparted by the static background potential. The dissipative
effects from the background mass such as dynamical friction or
decelerations due to mass accretion are not taken into account.

\subsection{Initial conditions and simulation setup}
\label{Initialconditions}

We consider direct encounters of two binaries with different masses but 
the same binding energy in the presence/absence of a background
potential. We define the ratio of binding energies of the 
two initial binaries as $\alpha=E_{\rm b,1}/E_{\rm b,2}$, 
 where $E_{\rm b,1} (E_{\rm b,2})$ is the binding energy of the more (less) massive initial binary. This choice is motivated by \citet[][hereafter GPZE04]{Gualandris+2004}, who 
 find that the distributions of velocities and semimajor axes do not change 
 significantly for $\alpha\leq3$. They also find that 
 the formation of the 
 $\iota$ Ori binary considered in this study is favored by such low ratios of $\alpha$.
We consider two values for the density of the background potential ($\rho=1$ and
$10^{3}$ in units of $\bar{\rho}$, \citealt{Pfalzner2009})
and two values for the total mass of the background matter ($M_{\rm
  bg}=3\times10^{3}\Msol$ and $10^{4}\Msol$).  These chosen values 
are typical for young star cluster cores (e.g
\citealt[]{Zwart+2010} and references therein), such as e.g. 30
Doradus cluster \citep[e.g][]{Hunter+1995,Andersen+2009,Selman2013}.

In this study, we examine the hypothesis of the dynamical ejection scenario. 
We choose the parameters of the initial binaries to match the observed properties of 
$\iota$ Ori binary and two runaway stars (AE Aur and $\mu$ Col), which 
are believed to have formed during a binary-binary encounter in the Trapezium cluster. 
Furthermore, we focus on one particular case where the end products of the encounters 
are a binary consisting of $\iota$ Ori A and $\iota$ Ori B and two single stars. We  
refer to this as the 2+1+1 outcome.
We assume one of the binaries (binary 1) consists of two stars with masses $m_{11}=39\Msol$ 
(a proxy for $\iota$ Ori A) and $m_{12}=16\Msol$ (AE Aur), 
while the other binary (binary 2) is composed 
of two stars with masses $m_{21}=19\Msol$ ($\iota$ Ori B) 
and $m_{22}=16\Msol$ ($\mu$ Col).  
Hereafter, for brevity we denote the four stars by $S_{11}$, $S_{12}$, $S_{21}$ and $S_{22}$. 
These specific couplings and the masses of the initial binary components are motivated 
by \citetalias{Gualandris+2004} (see their Table 2). 
Accordingly, the total masses of the 
two binaries are $55\Msol$ and $35\Msol$, respectively, and 
the combined mass of the two binaries is $90\Msol$. 

We consider two different choices for the stellar radii, zero-size (point particles,
or stellar radii $R_{\star}=0$) and finite-sized spherical
particles. In the simulations with a background potential we only
consider point particles, whereas we consider both cases in the
simulations without a background potential. The stellar radii are
taken from Table 2 in \citetalias{Gualandris+2004}, i.e.,
$R_{\star,11}=16\Rsol$, $R_{\star,12}=6\Rsol$, $R_{\star,21}=11\Rsol$
and $R_{\star,22}=8\Rsol$. \citetalias{Gualandris+2004} found that
their results have a weak dependence on the choice of the stellar
radii. In the finite-size case, we take into account physical
collisions between stars. We assume that a physical collision happens
when the radii of the stars overlap.  Collisions are done in the
``sticky star'' approximation.  Here, stars are treated as rigid
spheres with radii equal to their stellar radii.  When the radii of
two stars overlap, they are merged together with no mass loss and
assuming conservation of linear momentum.  
After the collision, we assume that the radius of the product is equal to the 
sum of the colliding stars' radii.

We conservatively take the initial separation of the two binaries to be
$r_{12}=1000\AU$ and a relative initial velocity of $v_{\rm
  rel}=2\km/\s$, which is the mean observed dispersion velocity in the
Trapezium cluster \citep[e.g][]{Herbig86}. The impact parameter $b$ is
randomly drawn from a distribution $f(b)\sim b^{2}$ within the range 
$[b_{\rm min},b_{\rm max}]=[0,220-230\AU]$\footnote{$b_{\rm max}$ is
  smaller than the maximum impact parameter in
  \citetalias{Gualandris+2004} by a factor of 2-3 if we assume
  $\sigma=\uppi b_{\rm max}^{2}$.}, while the
eccentricities of the binaries are randomly generated
from a thermal distribution $f(e)\sim 2e$. The
mutual inclinations between the binary orbital planes, as well as
their initial phases, are randomly chosen.

Given the masses above, the total energy of the whole system, $E_{\rm
  present}$, is estimated from the observed velocities and the relative
positions of the four stars ($\iota$ Ori binary, AE Aur and $\iota$
Ori B, \citealt{Turon+1992}),  $E_{\rm present}\simeq
-(2-4)\times10^{48}\erg$\footnote{The absolute value of the energy
  ($|E_{\rm present}|$) may be slightly smaller (i.e., less bound) than what
  is estimated from the best fit values in
  \citealt{Turon+1992}. However, it is still within the observed range when the
measurement errors are taken into account.}.  In calculating the total energy,
we include the contribution of the background potential $V_{\rm bg}$ (Equation
\ref{eq:backpotential}) 
with total mass $M_{\rm bg}$ and with its origin coinciding with the CM of
the four stars. Finally, by
energy conservation, the total initial energies in the center of
mass (CM) frame of the four stars can be written as follows,

\begin{align}
\label{eq:totalinitalenergy}
E_{\rm present}&=\frac{1}{2}\mu v_{\rm rel}^{2} - \frac{G m_{1}m_{2}}{r_{12}}
+E_{\rm b,1}+E_{\rm b,2}\nonumber\\
&+\frac{2}{3}\uppi G \rho [m_{1} (r_{1}^{2}-3r_{\rm bg}^{2})+m_{2} (r_{2}^{2}-3r_{\rm bg}^{2})]\,,
\end{align}
where $\mu$ is the reduced mass of the two binaries, 
$\mu=55\times35/(55+35)\Msol\simeq21.3\Msol$ and $E_{\rm b,1} (E_{\rm b,2})$ 
is the binding energy of binary 1 (binary 2). Also, $r_{1}$ ($r_{2}$) in the last term 
(the background gravitational potential for the two binaries) is the distance from 
the origin to binary 1 (binary 2). In the CM frame, $r_{1}=m_{2}/(m_{1}+m_{2})r_{12}$ 
and  $r_{1}+r_{2}=r_{12}$\footnote{Note that for the gravitational potentials 
	(between the stars and the background), we consider each binary as a point particle
with a mass equal to the sum of the masses of the
two component stars of each binary. The errors arising 
from this treatment are negligible, $|\Delta V/V|\simeq (a/r_{12})^{2}\leq10^{-5}$.}. 
Given the estimated binding energy reservoir, the semimajor axes of the two binaries 
are chosen such that their binding energies are equal, $a_{1}\simeq3.2\AU$ and $a_{2}\simeq1.6\AU$. 
Most of the contribution to the total energy comes from the total binding energy 
$E_{\rm b}~(=E_{\rm b,1}+E_{\rm b,2}$) so that the semimajor axes for the two 
binaries have weak dependences on $\rho$ and $r_{12}$\footnote{$|V_{\rm bg}/E_{\rm b}|\leq10^{-3}$}. Each simulation is run for $4\Myr$.
This time is chosen to be comparable to and within the estimated uncertainties of the age of a young cluster
(e.g. the age of the Trapezium cluster is $\sim 2.5-3$~Myr.)

In the simulations without the background potential, we use the same
initial conditions described above except that $r_{12}$ is set to be
larger by a factor of a few
\footnote{We expect that the increase in $r_{12}$ by a
    factor of a few does not affect our results significantly. In our
    experiments, for such small initial relative velocities $v_{\rm
      rel}$, and $r_{12}\simeq 1000\AU$, the two initial binaries
    approach each other on an almost radial orbit, and experience a subsequent 
    head-on collision at the moment of impact. In other words, the
    instantaneous impact parameter at impact is smaller than the sum of the binary semimajor axes 
    by a factor of a few. This means that the total angular momentum has only 
    a weak dependence on the initial impact parameter $b$ and the initial separation 
    $r_{12}$. Therefore, for the same total energy, our results are
    robust to changes in $b$ and $r_{12}$ by factors of a few.}.  We refer to this set of simulations as Model 0. We consider three
sets of simulations with a background potential given different
choices of $M_{\rm bg}$ and $\rho$ (or, simply, the escape velocity
$v_{\rm esc}$). We refer to these as Model 1, Model 2 and Model 3. The
specific parameters for each of these models are summarized in Table
\ref{tab:modelparameter2}. In addition to those four main models, we
run simulations with different stellar sizes (Model 0-1) and binding
energy ratio (Model 1-1 and 1-2) for investigative purposes. However, in order
to avoid adding complexity in interpreting our results, we concentrate
on the four main models (without ``-1'' or ``-2'') when it comes to the orbital
parameters of the interaction.

\section{Results}
In this section, we present the results of our scattering experiments
between two binaries in the presence/absence of a background potential
including the outcome probabilities for our various models. In
particular, focusing on the formation of the most massive binaries
(which will be denoted by [$S_{11}+S_{21}$] binary) and two ejected
single stars ($S_{12}$ and $S_{22}$), we describe the statistical
properties of the simulated final binaries and single stars. 

\subsection{Overview }
\subsubsection{Outcome probability}
After an encounter between two binaries, four possible end products can emerge,

\begin{enumerate}
	\item \hspace{0.03in} a binary and two single stars (2+1+1)
	
	\item \hspace{0.01in} two binaries\hspace{0.79in} (2+2)
	
	\item a triple and a single star \hspace{0.23in}(3+1)
	
	\item four single stars\hspace{0.63in} (1+1+1+1)
\end{enumerate}

\begin{figure}
	\centering
	\includegraphics[width=8.7cm]{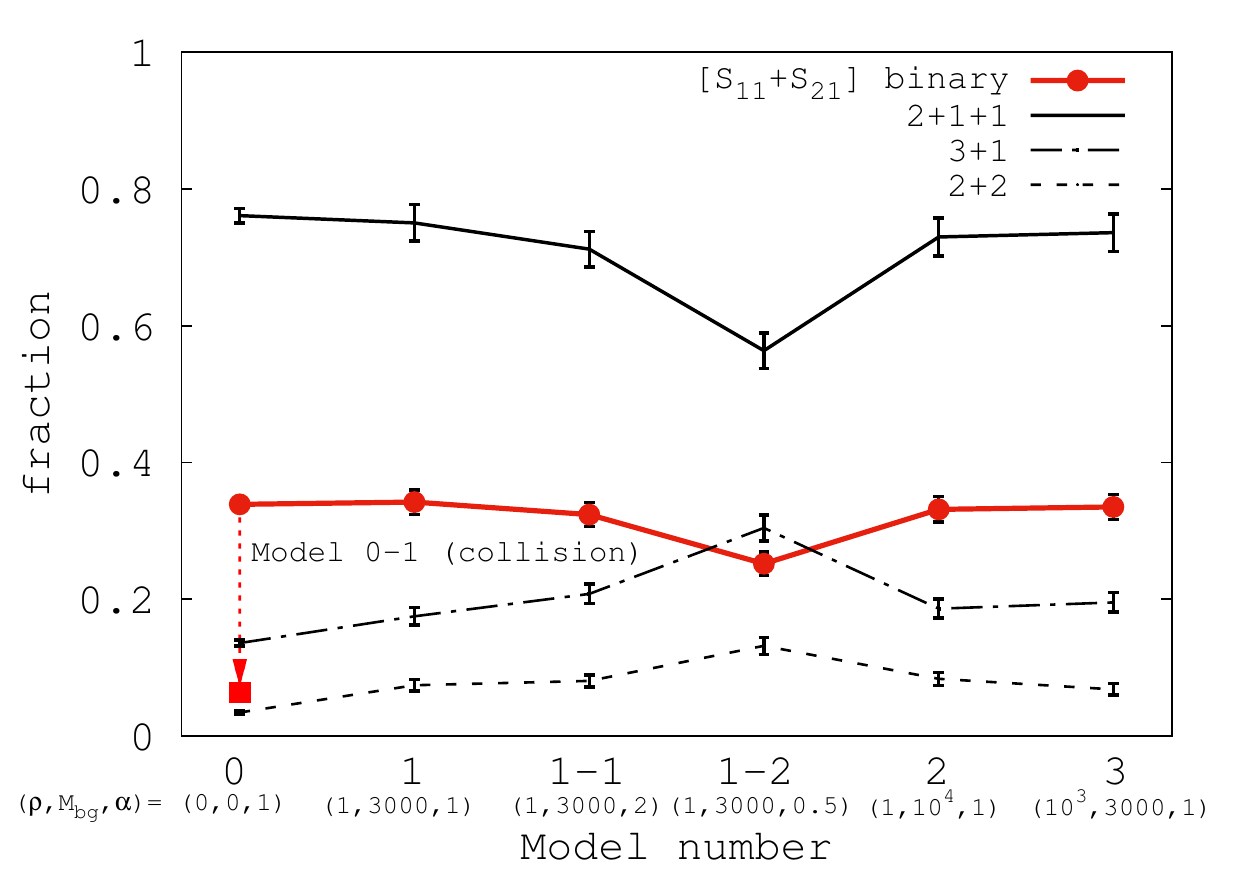}
	\caption{Fraction of each outcome from the scattering
         experiments for our models.  We remind the reader of the values of the parameters
         below each model name by indicating the density $\rho$, the total
          background mass $M_{\rm bg}$, and the ratio of the binding
          energies of the two initial binaries $\alpha$ (see Table
          \ref{tab:modelparameter2}). Different line types are adopted
          to distinguish each outcome: the 2+1+1 outcome (thin black
          solid line), the 2+2 outcome (dot-dashed line) and the 3+1
          outcome (dashed line). Among the possible outcomes, it is
          most common for the 2+1+1 outcome to emerge, especially when
          the two most massive stars form a binary (solid red line
          with circular dots).  The error bars indicate the Poisson
          uncertainties for each simulation set.  }
	\label{fig:outcomeprobability}
\end{figure}
Hereafter, we will exchangeably use what is given in the accompanying
parentheses to denote each outcome.

Figure \ref{fig:outcomeprobability} shows the outcome probabilities
for all the models.  The error bars indicate the Poisson uncertainties
for each simulation set.  A short description of the relevant model
parameters (density $\rho$, total background mass $M_{\rm bg}$ and
ratio of the binding energies of the two initial binaries $\alpha$;
see Table \ref{tab:modelparameter2}) is provided below each model
name. We use different line types for each outcome: the 2+1+1 outcome
(thin black solid line), the 3+1 outcome (dot-dashed line) and the 2+2
outcome (dashed line). In addition, we indicate the formation
probabilities of the most massive [$S_{\rm 11}+S_{\rm 21}$] binaries
by the thick solid red line with circular dots. Note that the
fractions of the 2+1+1 outcome (black solid line) include those for
the [$S_{\rm 11}+S_{\rm 21}$] binary as well as all other possible 2+1+1
outcomes.

As shown in the figure, the 2+1+1 outcome is the most probable ($\sim
0.77-0.78$) across all the models (Model 0 to Model 3). 
Additionally, within this outcome,
the [$S_{\rm 11}+S_{\rm 21}$] binary is the most likely to form
($\sim0.38$), followed by the formation of the second most massive
binaries, which are not shown in the figure. Note that since we
consider two stars of equal mass ($m_{12}=m_{22}=16\Msol$), two
different binaries of the same mass, ([$S_{11}+S_{12}$] or
[$S_{11}+S_{22}$] with $m_{\rm b}=55\Msol$), could form. Their outcome
probabilities are comparable ($\simeq 0.2$ for each binary).  The
outcome fractions of the least massive binaries ([$S_{21}+S_{12}$] or
[$S_{21}+S_{22}$] with a mass $m_{\rm b}=35\Msol$) are the lowest ($\sim
0.01-0.02$). However, when we take into account physical collisions
assuming finite sizes for the stars, the fraction of outcomes producing a [$S_{\rm
    11}+S_{\rm 21}$] binary (red square connected with the downward
dotted arrow from the fraction for Model 0) is reduced to $\sim0.064$
(decrease by a factor of 5-6). In reality, direct physical collisions
during encounters could occur \citep{LeighGeller2012,LeighGeller2015}
and their chances may increase in the presence of a background potential due to
more gravitationally-focused cross sections for the stars and longer-lasting
interactions accompanied by an increased probability of interruptions by other stars
\citep{GellerLeigh2015}.  In all models, the 1+1+1+1 outcome has not
emerged. Given the negative total energy in our simulations this is
expected because the 1+1+1+1 outcome only forms when the total energy
is positive.

Comparing Model 1 ($\alpha=1$) with Models 1-1 and 1-2 ($\alpha\neq1$), we find that the fraction of 2+1+1 outcomes (including that for the [$S_{11}+S_{21}$] binary) decreases whereas the fraction of 3+1 outcomes increases. These results are consistent with other studies \citep[e.g][]{Mikkola1984,Gualandris+2004,Leigh+2016}. As the binding energy ratio becomes different from unity, the closer binary acts more like a single star causing the interaction to resemble more a three-body exchange encounter, leading to the preferential formation of triples.  Given the different masses of the two initial binaries, the ratio between their semimajor axes $a_{1}/a_{2}$ becomes larger as $\alpha$ decreases, or $a_{2}\simeq(\alpha/2)a_{1}$. In other words, the semimajor axis of the less massive initial binary ($a_{2}$) is smallest for $\alpha=0.5$. This seems to explain the larger increase in the fraction of 3+1 outcomes for $\alpha=0.5$, since the less massive 
initial binary effectively acts more like a single star in this case. We emphasize that a more thorough investigation focusing on the effects of adopting different mass ratios and binding energy ratios is required to confirm these conclusions.

We do not observe significant differences in the outcome probabilities
between the various models (with/without the background potential and
for different depths of the background potential). This is because in
most of the simulations the stellar interactions tend to occur inside
a region where the stellar dynamics are governed by the stellar
gravitational potential, not by the background potential. Furthermore,
the ejection velocities of the single stars tend to be sufficiently high
that they are capable of escaping the background potential to
infinity without being trapped in the background potential, which we 
show in Figure \ref{fig:velocity_singlestar}.  This is consistent
with the results of \citet{Ryu+2017}. They performed numerical
scattering experiments between two identical binaries in a homogeneous
background potential, and investigated the effects of the
potential as a function of its density and the semimajor axes of the two colliding
binaries (see equation 17
in \citealt{Ryu+2017}). For tightly bound binaries ($a\simeq
1-3\AU$) and the densities considered in this study, they showed that
the outcomes are determined by stellar interactions taking place within a
region where the stellar masses are larger than the enclosed background
mass, corresponding to \textit{Region 1} in \citet{Ryu+2017}.
\begin{table}
	\centering
	\setlength\extrarowheight{2pt}
	\begin{tabulary}{0.3\linewidth}{ c c c c c p{0.7cm}p{0.7cm}}
		\hline
		Model & (E,2E) & (NE,2E) & (E,1E) & (NE,1E) & (E,NE) & (NE,NE)\\
		(unit)  &     $\times10^{-1}$     &    $\times10^{-1}$           &     $\times10^{-2}$      &    $\times10^{-3}$          &    $\times10^{-3}$         &  $\times10^{-3}$            \\
		\hline
		\hline
		1    & 9.26    & 0.455    & 1.99     & $8.52$  & 0    & 0 \\
		1-1& 9.55    & 0.327    & 1.19     & 0           &  0   & 0 \\
	    1-2& 8.73    & 0.900    & 3.77     & 0           &  0   & 0 \\
		2    & 8.24  &   1.31  &    4.15    & 3.19   & 0    & 0 \\
		3    & 4.46   & 4.15 &     4.62   &  83.1  & 0    & 9.23\\
		\hline
	\end{tabulary}
	\caption{The first/second slot in the parenthesis indicate
		whether binaries/single stars have escaped from the
		potential at $t=4\Myr$. For binaries, "E" and "NE" refer to
		the case where binaries have escaped and have not escaped,
		respectively. For single stars, we consider three
		possible cases, 1) 2E: two single stars have escaped, 2) 1E:
		only one star has escaped and 3) NE: neither
		of the two single stars have escaped.  We provide the plots showing
		the probabilities in Figure \ref{fig:escapeprobability}.}
	\label{tab:escapeprobability}
\end{table}
\begin{figure}
	\centering
	\includegraphics[width=8.4cm]{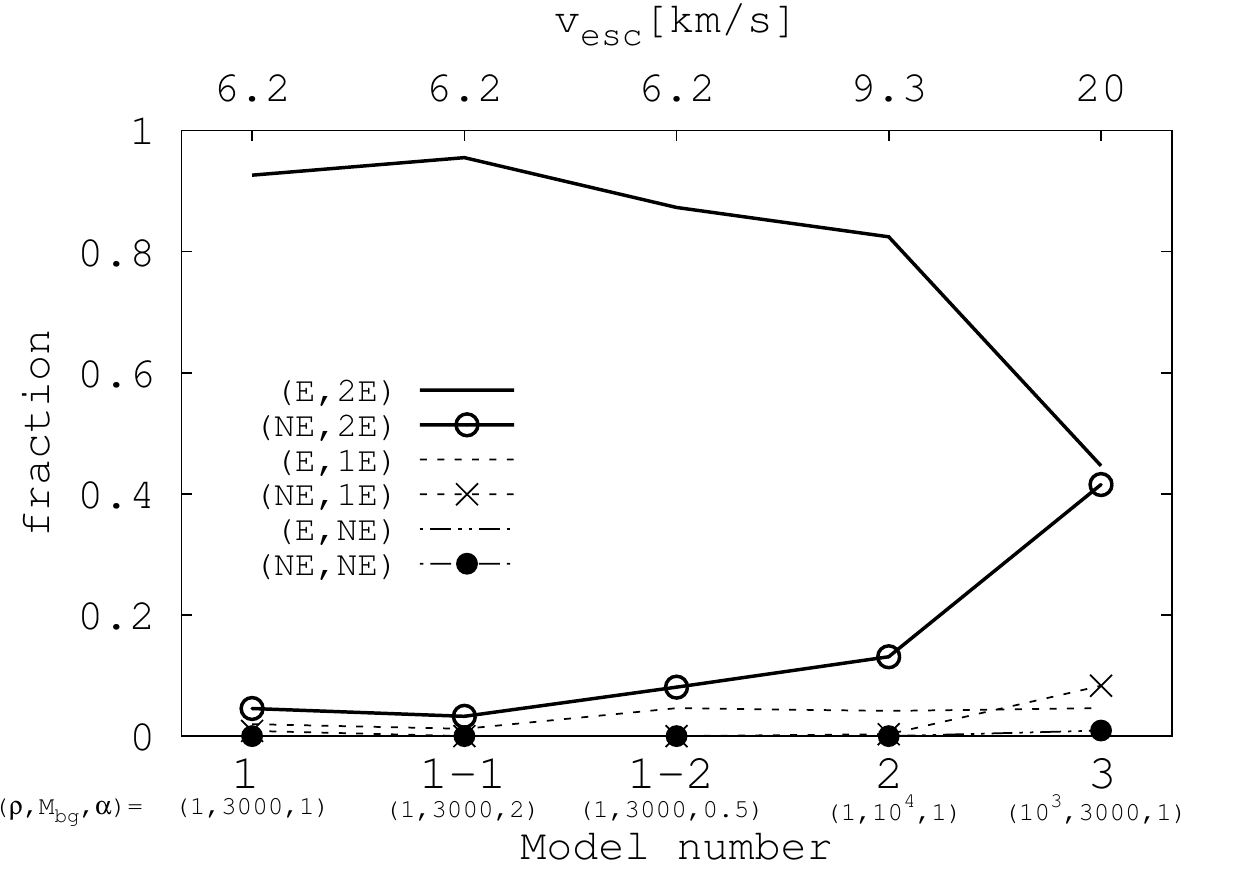}
	\caption{Escape probability for each outcome product (binary/single star) for models 
		with a background potential. The first/second slot in the parentheses indicates 
		whether binaries/single stars have escaped from the potential at $t=4\Myr$. 
		See the caption of Table \ref{tab:escapeprobability} or the text for the detailed  
		description of each outcome product. }
	\label{fig:escapeprobability}
\end{figure}

\begin{figure*}
	\centering
	\includegraphics[width=8.4cm]{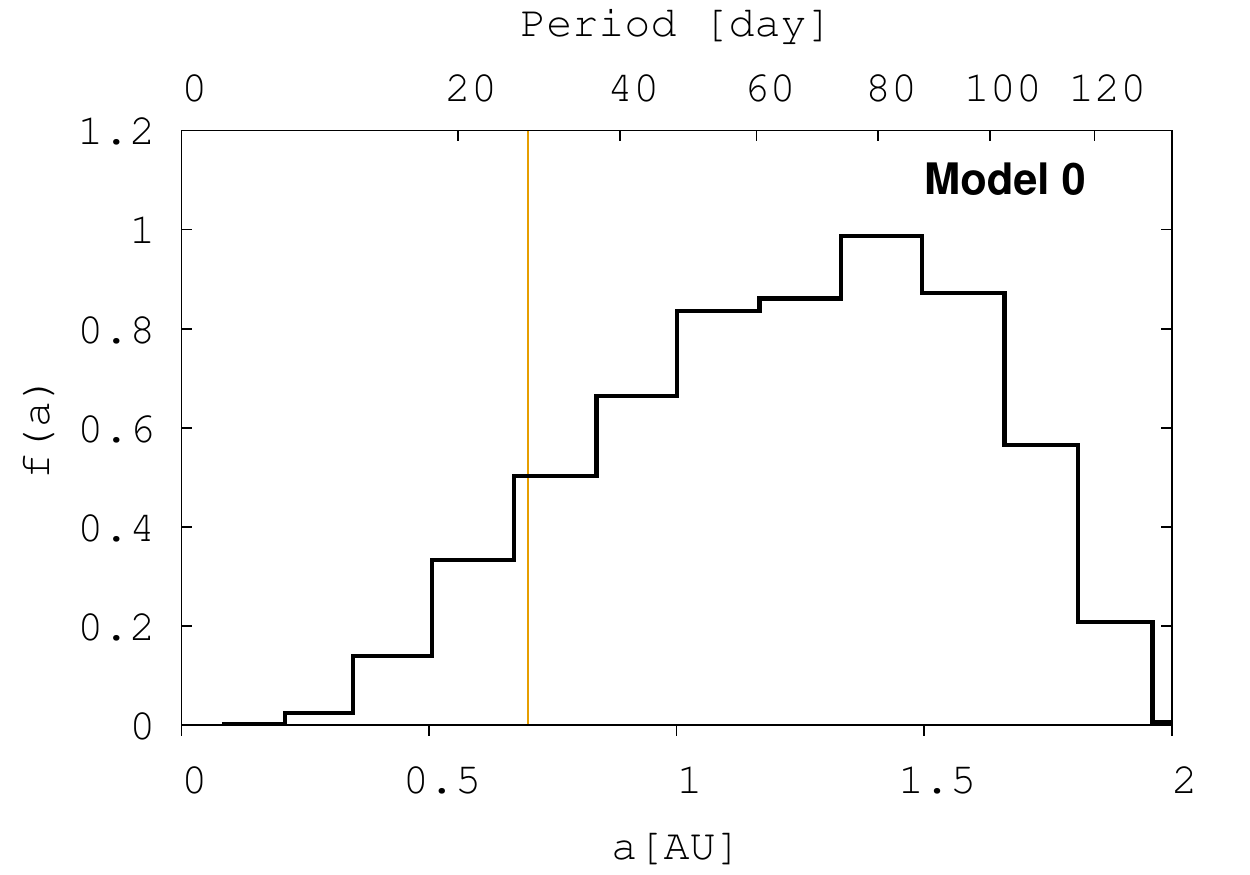}
	\includegraphics[width=8.4cm]{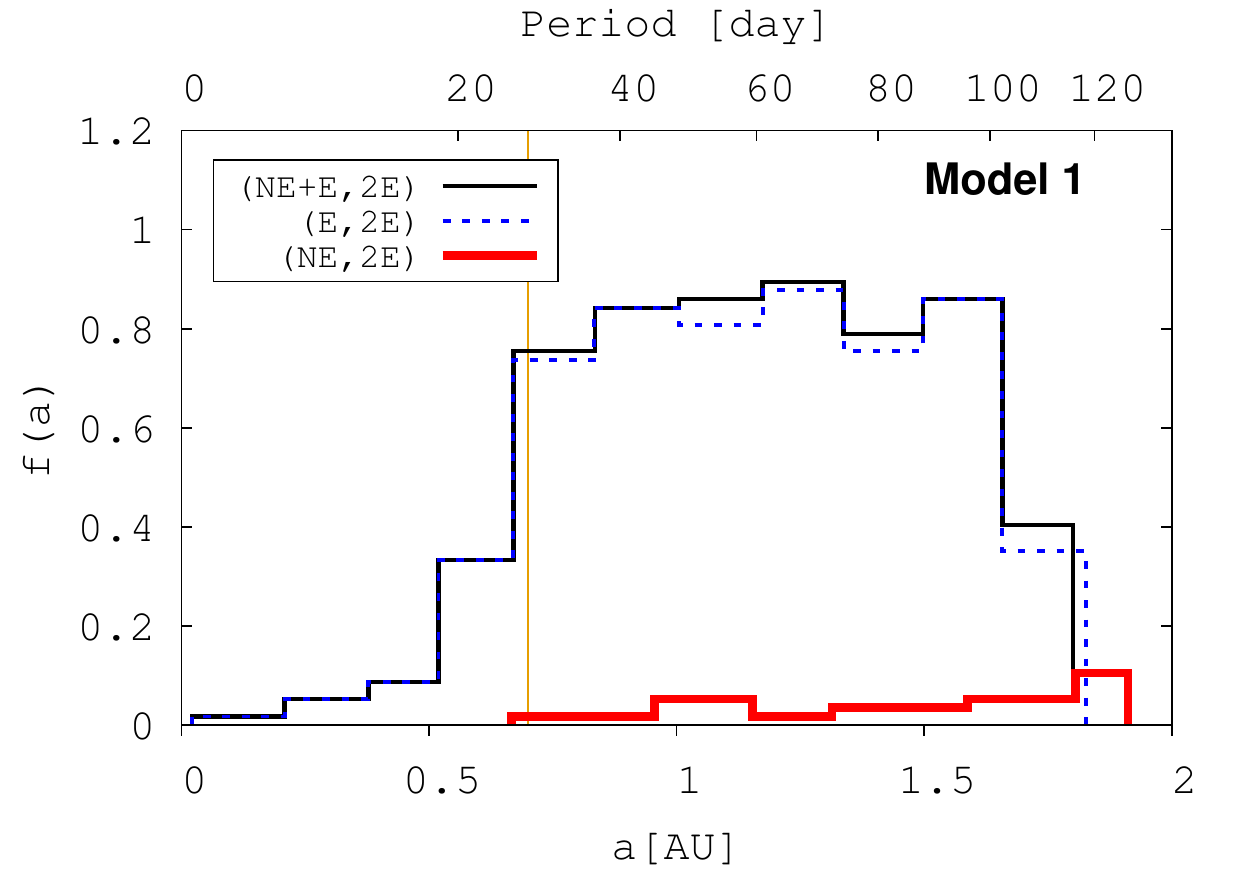}
	\includegraphics[width=8.4cm]{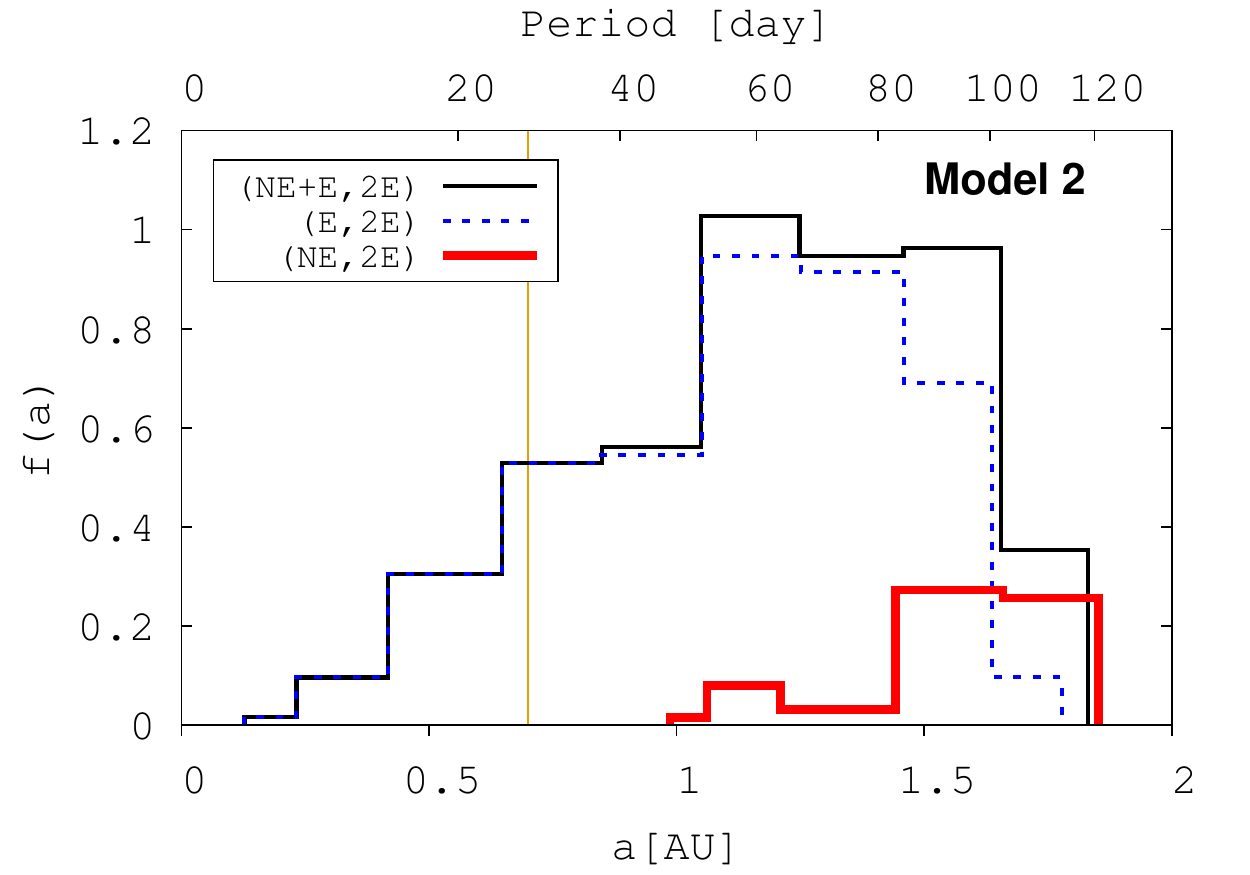}
	\includegraphics[width=8.4cm]{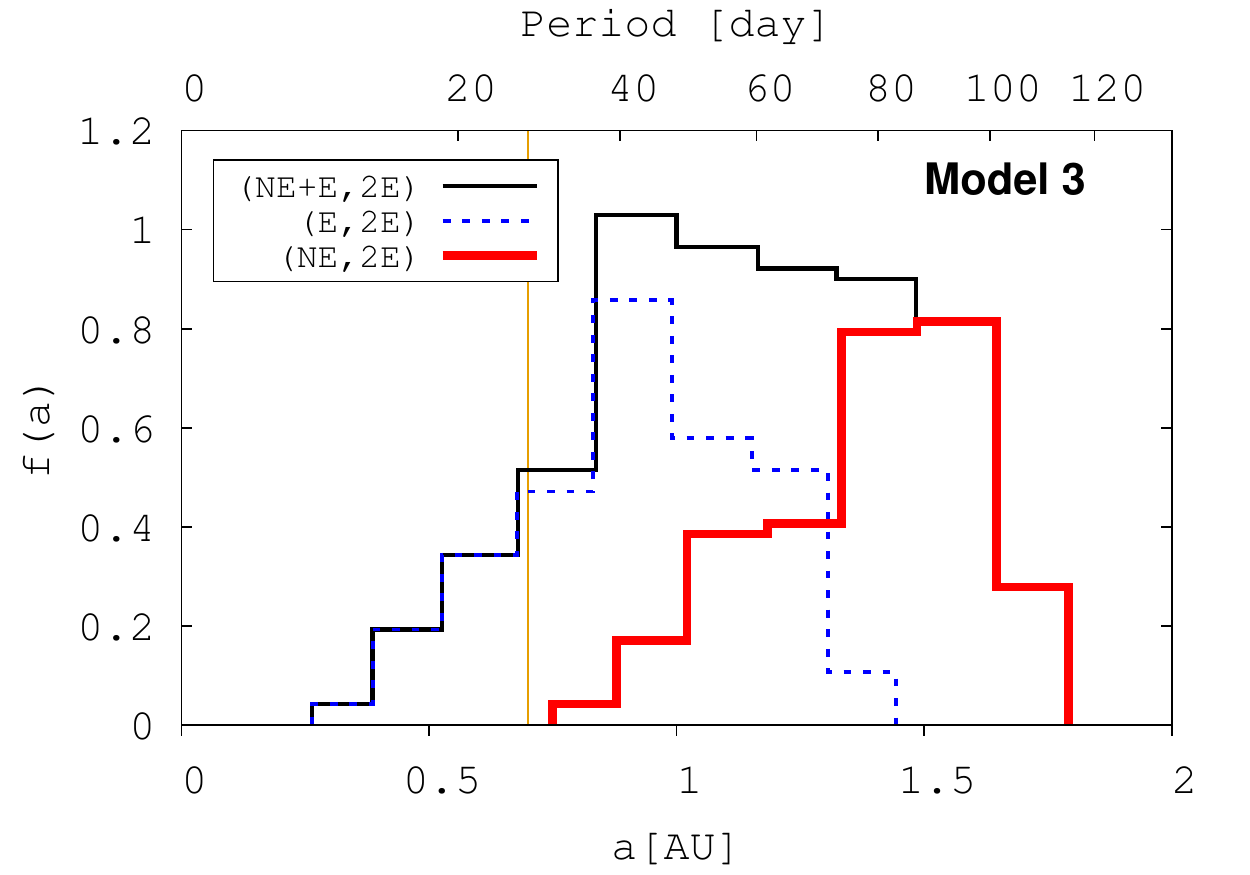}
	\caption{The distributions of the semimajor axes $a$ for the
		[$S_{11}+S_{21}$] binaries. The vertical gold lines indicate
		an observed value of $a\simeq 0.7\AU$ or a period of 29
		days.  In the plots for 
		Model 1 to Model 3, we show three different distributions
		with different line types: those for E-binaries (dotted
		blue line) and NE-binaries (thick red solid line) and the overall
		distributions (thin black solid line) as the sum of the two
		distributions.}
	\label{fig:semimajoraxis}
\end{figure*}

A comparison with the branching ratios of \citetalias{Gualandris+2004}
(see their Figure 3) shows that we find predominantly consistent outcome
fractions. On the one hand, they find an increase in the fraction of 3+1 outcomes as 
$\alpha$ becomes different from unity, as shown in Figure \ref{fig:outcomeprobability}. 
On the other hand, 
in our study, the 2+1+1 outcome for $\alpha=0.5-2$ is the
most probable, whereas in their study the most probable case is
"Flybys" (if the two original binaries remain bound) followed by 
the 2+1+1 outcome ("Ionizations+$\iota$ Orionis" in their Figure 3), which is 
a little bit lower ($0.3-0.4$ for $\alpha=0.5-2$) than in our simulations. 
However, since we are likely 
using different initial conditions such as impact parameters
\footnote{We miss many ``flyby" outcomes in our simulations due to having adopted smaller collisional cross sections at impact.} 
and initial separations, a direct comparison with their results 
is not straightforward.

\begin{figure*}
	\centering
	\includegraphics[width=8.5cm]{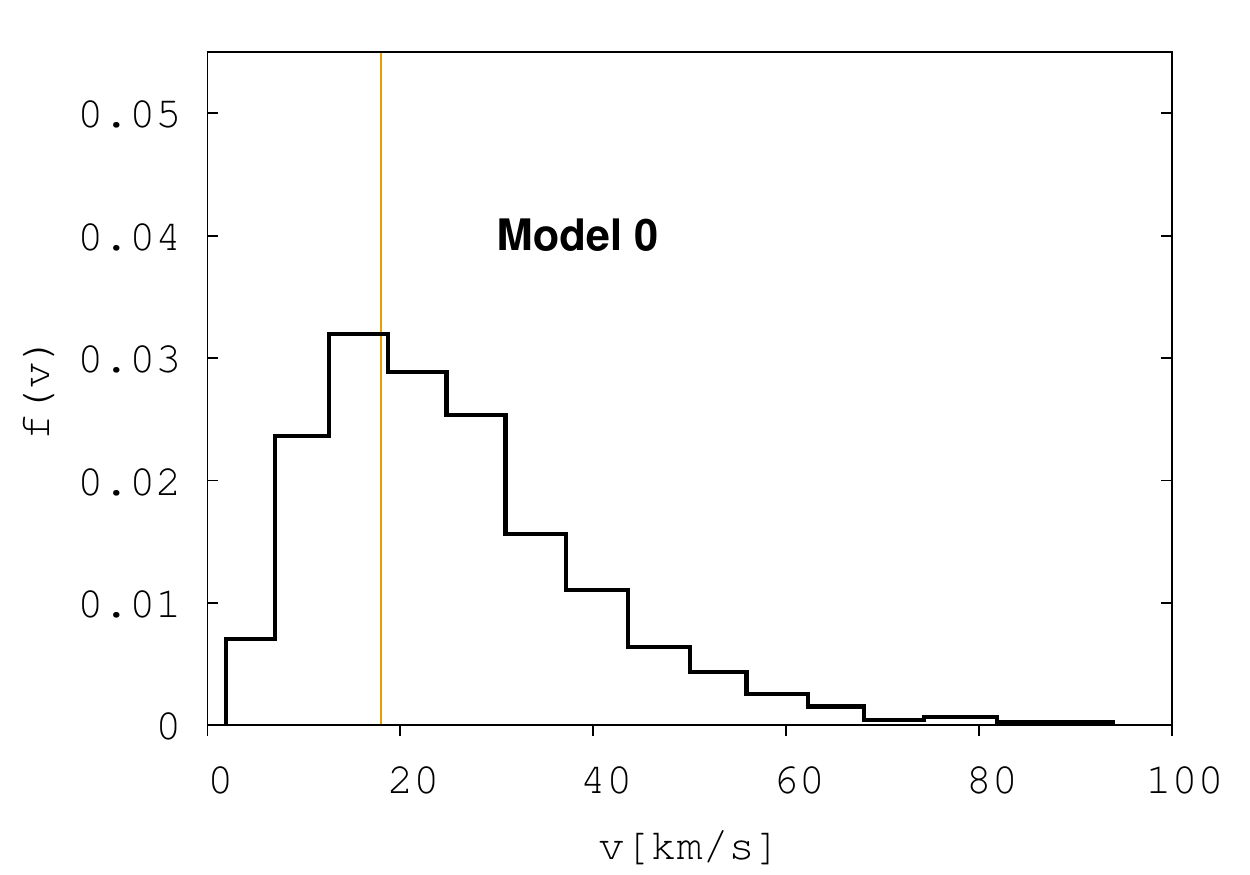}
	\includegraphics[width=8.5cm]{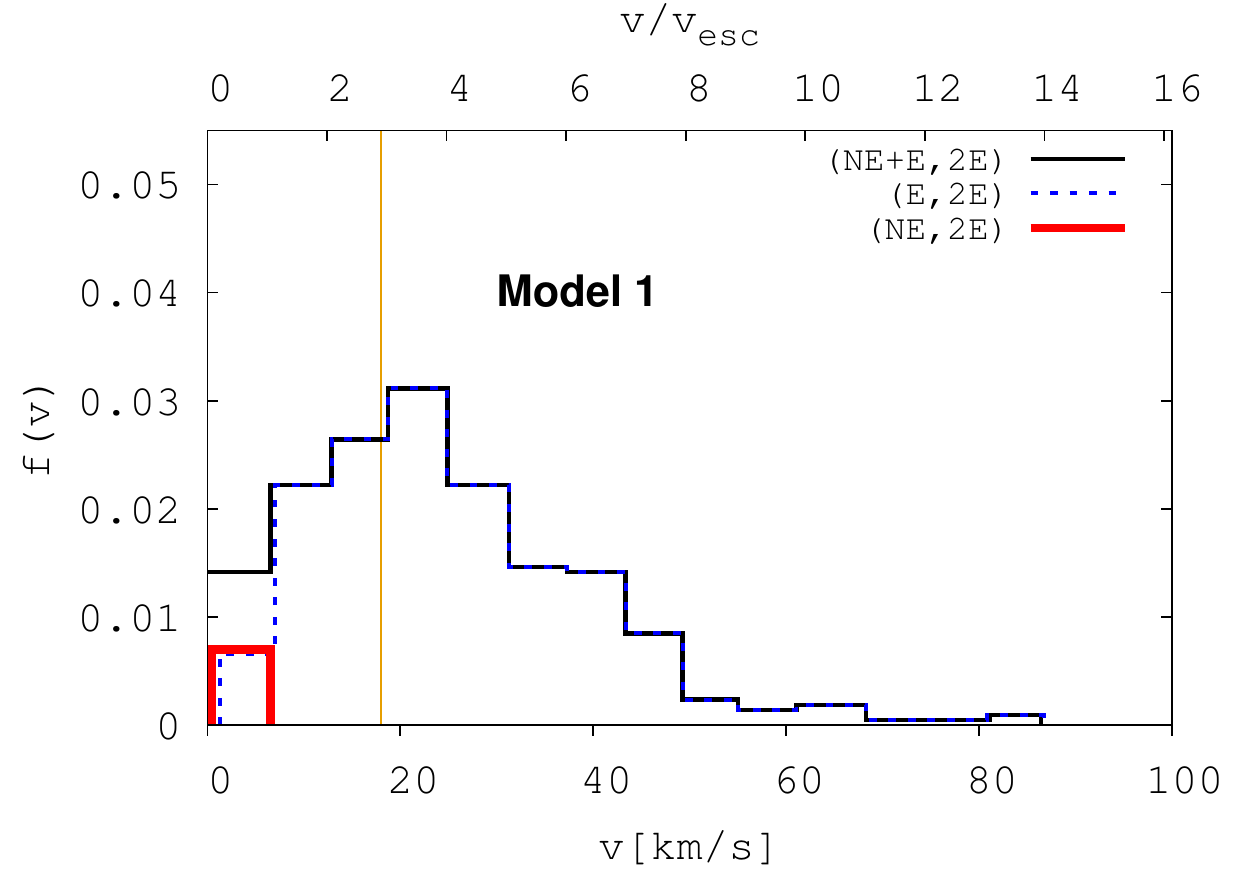}
	\includegraphics[width=8.5cm]{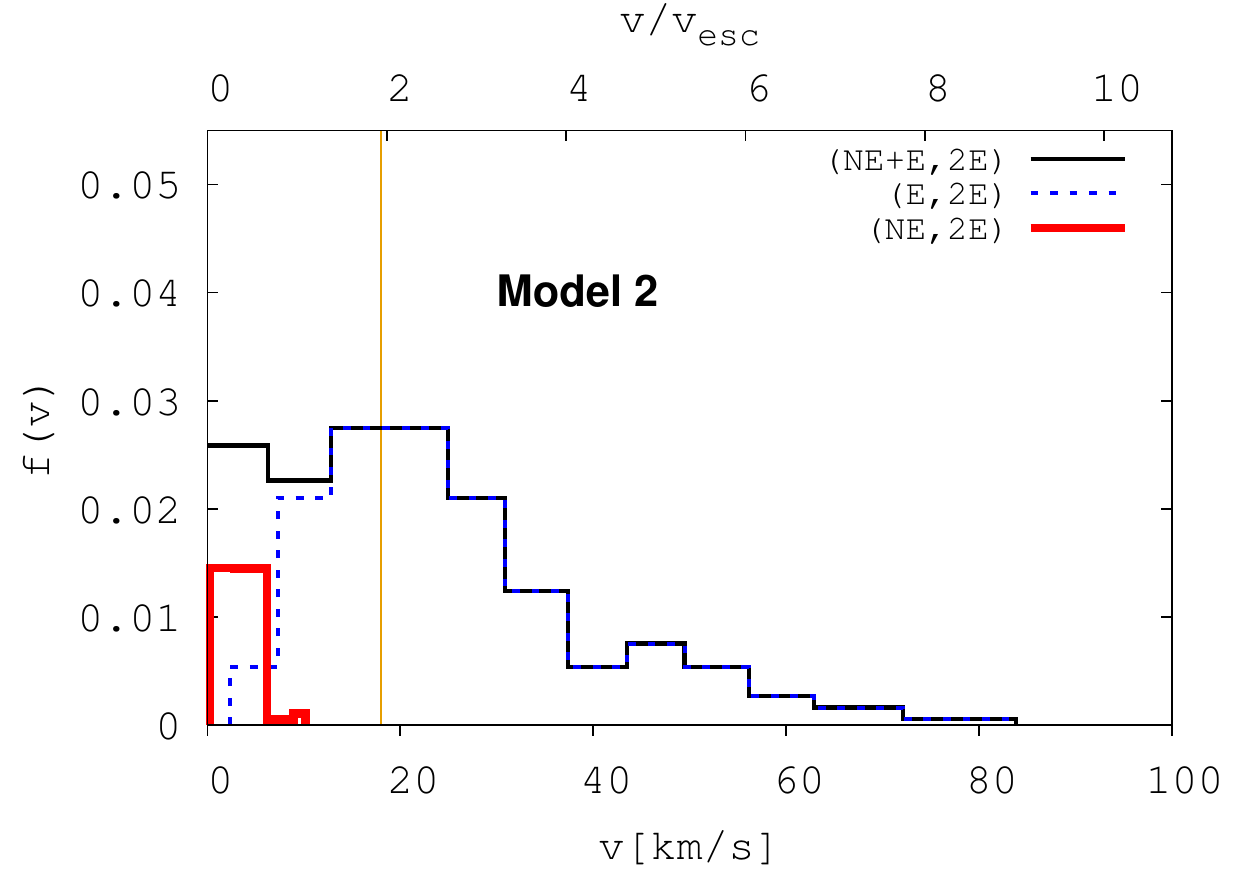}
	\includegraphics[width=8.5cm]{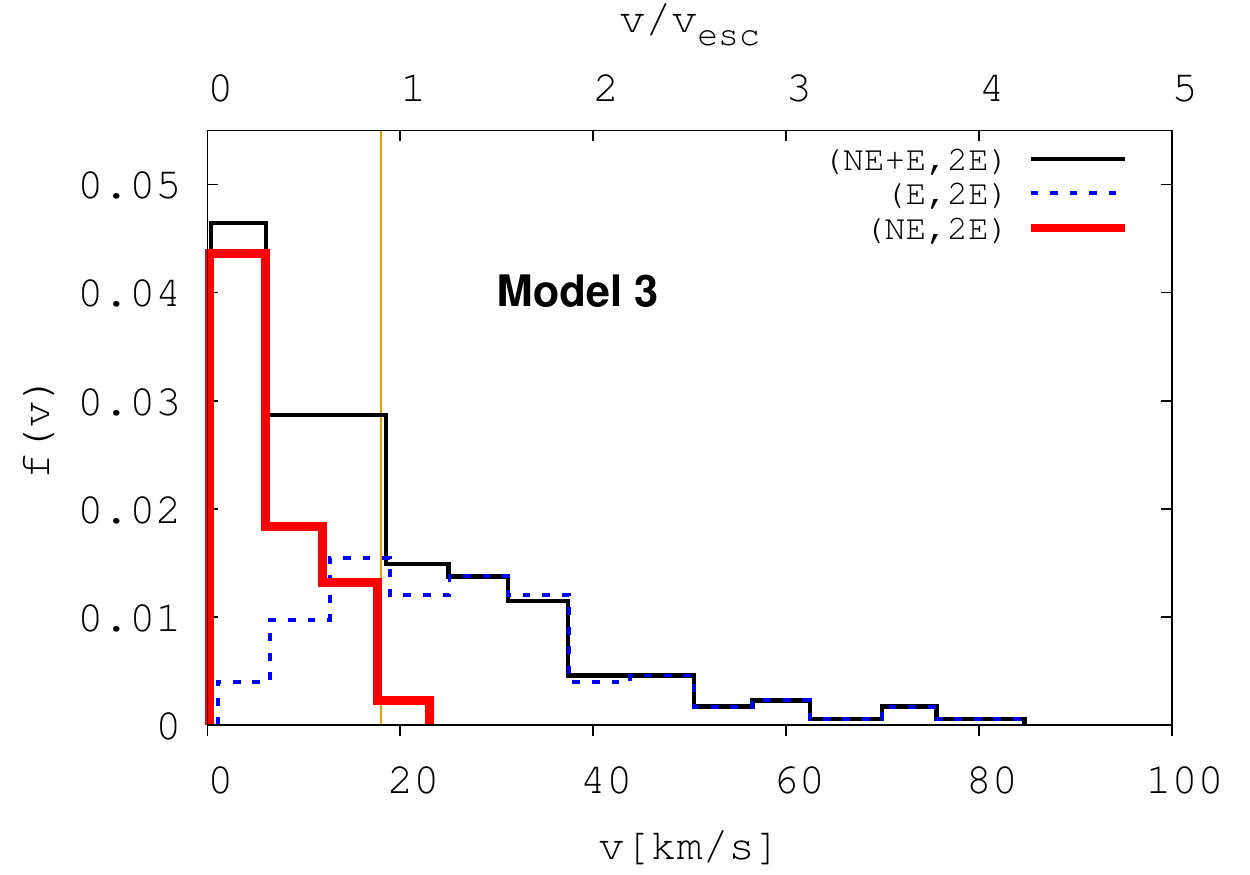}
	\caption{The distributions of the final velocities of the
		outcome binaries $v_{\rm binary}$, with vertical gold lines
		indicating an observed value of $v\simeq 18\km\s^{-1}$. The same
		plot formats and line colors are used as in Figure
		\ref{fig:semimajoraxis}.}
	\label{fig:velocity_binary}
\end{figure*}

\subsubsection{Escape fraction} 
\label{sec:escapefraction}

In contrast to the case with purely stellar interactions, ejected
stars could be trapped in the background potential. In fact, only when
the ejection velocities of the stars are sufficiently high can they
escape the background potential to spatial infinity. In
  our simulations, we check at every time step whether or not stars
  have escaped from the potential. We assume that stars have escaped
  when they are spatially outside the potential and they are unbound
  from the potential, i.e.,
\begin{equation}
	v\geq v_{\rm esc}=\sqrt{\frac{2 G M_{\rm en}(r)}{r}}$ and $r\geq r_{\rm bg}\;,
\end{equation}
where $M_{\rm en}(r)$ is the total mass enclosed in a spherical volume 
of radius $r$ and $v_{\rm esc}$ is the velocity required by a star at $r$ to escape
from the potential to spatial infinity.

For the 2+1+1 outcome, there are six possible scenarios depending on whether each 
binary and each single star could escape from the background potential,
\begin{enumerate}
	\item (E, 2E): both the binary (E) and the two single stars (2E) have escaped,
	\item (NE, 2E): the binary has not escaped (NE) whereas the two single stars have escaped,
	\item (N, 1E): the binary and only one single star (1E) have escaped,
	\item (NE, 1E): the binary has not escaped and only one single star (1E) has escaped,
	\item (E, NE): the binary has escaped whereas all single stars have not escaped (NE),
	\item (NE, NE): Neither the binary or the two single stars have escaped.
\end{enumerate}

The first/second slot in the parentheses indicates whether binaries/single stars have 
escaped from the potential at $t=4\Myr$. For notational succinctness, we also refer to
binaries which have (have not) escaped as E-binary (NE-binary). 

Figure \ref{fig:escapeprobability} and Table
\ref{tab:escapeprobability} show the escape fractions for the cases
listed above, and for each of the models with the background
potential.  The figure shows a clear trend in that as the escape velocity
increases, a larger number of binaries remain trapped (solid line with
hollow circles). In particular, in Model 3 with the largest escape
velocity (see Table \ref{tab:modelparameter2}), since the typical
ejection velocities of the binaries are comparable to the escape
velocity of the potential, around 40\% of the binaries could not
escape; this fraction becomes comparable to that of the escaped binaries
(E,2E). And it is also more likely in Model 3 that one single star stays
trapped in the background potential, i.e., (E,1E) and (NE,1E),
compared to the other models.  From this we can see that the escape
velocity may play a role as a good indicator to characterize the depth
of the potential instead of only $M_{\rm bg}$ or $\rho$.

In the following sections, we will present the distributions of the orbital parameters for 
the most massive binaries and the two single stars.

\subsection{Statistical properties of the most massive binaries and the two ejected single stars}
In this section we describe the statistical properties of the most
massive binaries (i.e. those formed from [$S_{11}+S_{21}$]), and the
two ejected stars produced by the scattering experiments.  Namely, we
provide the distributions of the semimajor axes and the eccentricities
for the binaries, as well as the distributions of the final velocities
for both the binaries and the ejected single stars.  In addition to
these, we also explore the ejection times and the relative angles
between the stars. Since we could not find any statistically
meaningful differences between Model 1 and Model 1-1, from now on, we
shall primarily focus on the analysis of the results for Model 0,
Model 1, Model 2 and Model 3.  All the distributions shown in the
following are made using the orbital parameters at $t=4\Myr$ for the
cases where two single stars have escaped, unless indicated
otherwise. We provide the cumulative distributions of
  radial distances from the system CM and the speeds at $t=1,~2,~3$
  and $4\Myr$ for the escaped binaries (E-binaries) in Figure
  \ref{fig:cumulative_binary} in Appendix \ref{appendix:cumulative}.

\subsubsection{The statistical properties of the final binaries}

\begin{table*}
	\centering
	\setlength\extrarowheight{0.4pt}
	\begin{tabulary}{0.1\linewidth}{c c c c c }
		\hline
		Model (Figure)																	&     	0&   1 & 2 & 3  \\ 
		\hline
		\hline
		$a~[{\rm AU}]$ (\ref{fig:semimajoraxis})								 & 1.27  & 1.73 / 1.17 & 1.62 / 1.17& 1.41 / 0.935 \\
		$v_{\rm binary}~[{\rm km~s}^{-1}]$ 	(\ref{fig:velocity_binary})	& 22.5  & 2.24 / 22.8 & 2.52 / 22.7& 5.81 / 25.5  \\
		$v_{\rm fast}~[{\rm km~s}^{-1}]$ (\ref{fig:velocity_singlestar}) & 84.1  & 35.2 / 92.7& 50.4 / 92.2& 64.7 / 113 \\
		$v_{\rm slow}~[{\rm km~s}^{-1}]$ (\ref{fig:velocity_singlestar})   &  40.9 & 28.6 / 46.5 & 31.1 / 41.8& 33.9 / 58.3 \\
		$v_{\rm fast}/v_{\rm slow}$  (\ref{fig:velocity_ratio})								&1.86   &1.20 / 1.78 & 1.39 / 1.96 &1.71 / 1.89 \\
		$v_{\rm fast}/v_{\rm binary}$ 	(\ref{fig:velocity_ratio})							&3.72   & 18.4 / 3.92& 21.7 / 4.14& 10.8 / 4.52\\
		$v_{\rm slow}/v_{\rm binary}$  	(\ref{fig:velocity_ratio})						   & 1.83 & 14.9 / 1.97& 14.4 / 1.98& 5.41 / 2.30\\
		$\xi_{{\rm fast},{\rm slow}}$ [$^{\circ}$]  ( \ref{fig:relativeangle_MODEL0}, \ref{fig:relativeangle_bg})  & 111  & 109 & 105& 89.1 \\
		$\xi_{{\rm fast},{\rm binary}}$ [$^{\circ}$]   (\ref{fig:relativeangle_MODEL0}, \ref{fig:relativeangle_bg})         &156  &154 &156 & 156 \\
		$\xi_{{\rm slow},{\rm binary}}$  [$^{\circ}$]  (\ref{fig:relativeangle_MODEL0},\ref{fig:relativeangle_bg}) &  105 & 108 & 112& 120 \\
		$\log[t_{\rm bg,~fast}/{\rm yr}] $ (\ref{fig:ejectiontime})														& -  & 3.88&4.04 &3.02 \\
		$\log[t_{\rm bg,~slow}/{\rm yr}]$  	(\ref{fig:ejectiontime})													&-  & 4.16  &4.37& 3.19\\
		$\log[t_{\rm bg,~binary}/{\rm yr}]$(\ref{fig:ejectiontime})														& - & 4.45 & 4.59& 3.39\\
		$\log[t_{\rm bg,~slow-fast}/{\rm yr}]$	(\ref{fig:ejectiontimedifference})													& -& 3.77  & 3.94 & 2.59 \\
		$\log[t_{\rm bg,~binary-fast}/{\rm yr}]$ (\ref{fig:ejectiontimedifference})		& - & 4.29& 4.44& 3.11\\
		\hline
	\end{tabulary}
	\caption{The median values of the distributions for the binaries and the 
		single stars. For the models with the background potential (Model 1 to Model 3), 
		we present the median values for (NE, 2E) before ``/" and (E, 2E) after ``/", separately. 
		Each row (from top to bottom) represents as follows:
		[Row 1] model name, 
		[Row 2] the semimajor axis $a$, 
		[Row 3-5] the speeds of the binaries, $S_{\rm fast}$ and $S_{\rm slow}$, 
		[Row 6-8] the speed ratios of $S_{\rm fast}$ to $S_{\rm slow}$, 
		$S_{\rm fast}$ to the binary and $S_{\rm slow}$ to the binary, 
		[Row 9-11] the relative angle $\xi$ between two single stars, 
		between $S_{\rm fast}$ and the binary and between $S_{\rm slow}$ and $S_{\rm fast}$, 
		[Row 12-16] the escape time $t_{\rm bg}$ in $\log_{10}$-scale of $S_{\rm fast}$, $S_{\rm slow}$ and the 
		binary and the escape time of $S_{\rm slow}$ and the binary with respect to $S_{\rm fast}$. }
	\label{tab:median}
\end{table*}

Figure \ref{fig:semimajoraxis} shows the distributions of the
semimajor axes $a$ for the [$S_{11}+S_{21}$] binaries. In the plots
for Model 1 to Model 3, we show three different distributions with
different line types: those for the binaries which have escaped from
the potential before $t=4\Myr$ (dotted blue line), those which have
not escaped and remain bound to the potential (thick red solid line),
and the overall distributions (thin black solid lines) as the sum of the two
distributions. The vertical gold lines indicate an observed value of
$a\simeq 0.7\AU$ or a period of 29 days \citepalias{Gualandris+2004}.

We can see how, in going from Model 1 to Model 4, the population
of final binaries splits into two separate populations, i.e.,
E-binaries and NE-binaries. The NE-binary population gradually
emerges.  This is because, as the background potential becomes deeper,
the escape velocity increases and hence it gets harder for the stars
to escape. The two populations become comparable in size in Model 3
(see \textit{right bottom} panel and Table
\ref{tab:escapeprobability}), hence contributing equally to the
overall distribution. Given the different values in the peaks of the
two populations, the overall distribution (thin black solid line) becomes
broader. The median values of $a$ for Model 0 are $a_{\rm
	median}=1.27\AU$. Those for (NE,2E) case (from Model 1 to Model 3) 
are $a_{\rm
	median}=1.17\AU,~1.17\AU$ and $0.94\AU$ and 
for (E,2E) case and $a_{\rm
	median}=1.73\AU,~1.62\AU$ and $1.41\AU$. We present the 
all of the median values for the distributions for the binaries 
and the single stars in Table \ref{tab:median}.

We also notice that the distributions of NE-binaries (red thick solid
line) are located at larger $a$ than those of E-binaries. This can be
understood in terms of conservation of energy along with the
escape velocity.  In general, when two single stars are ejected, the
recoiled binary carries some kinetic energy. As the recoil velocity of
the binary increases, given a fixed total energy, a larger reservoir
of negative energy is left for the binary itself, implying a tighter
binary.  Correspondingly, the fact that binaries could not escape from
the background potential means that the instantaneous velocities (or the
kinetic energies) of the binaries at the last ejection event were not
sufficiently high. Therefore, with smaller energy reservoirs given to
the binaries, their semimajor axes are distributed at larger $a$.

\begin{figure}
	\centering
	\includegraphics[width=8.5cm]{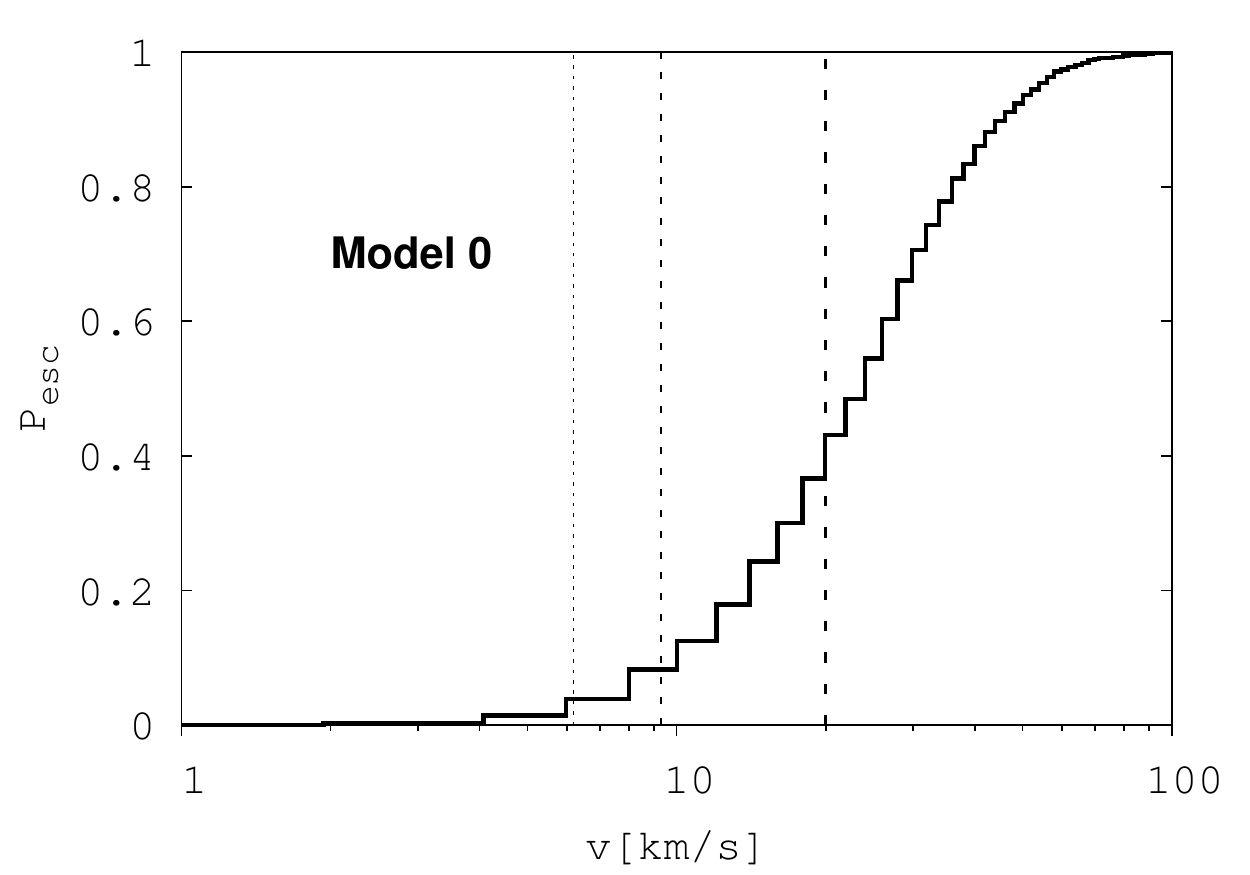}
	\caption{The escape probability $P_{\rm esc}$ calculated using
		Equation \ref{eq:P_esc} using the velocity distribution
		$f(v)$ for Model 0. The three vertical dotted lines indicate
		the escape velocities for Model 1 to Model 3 (from left to
		right). The probability $P_{\rm esc}$ at each escape
		velocity agrees well with the fractions given in Table
		\ref{tab:escapeprobability}.  }
	\label{fig:cumulative_velocity}
\end{figure}

\begin{figure}
	\centering
	\includegraphics[width=8.5cm]{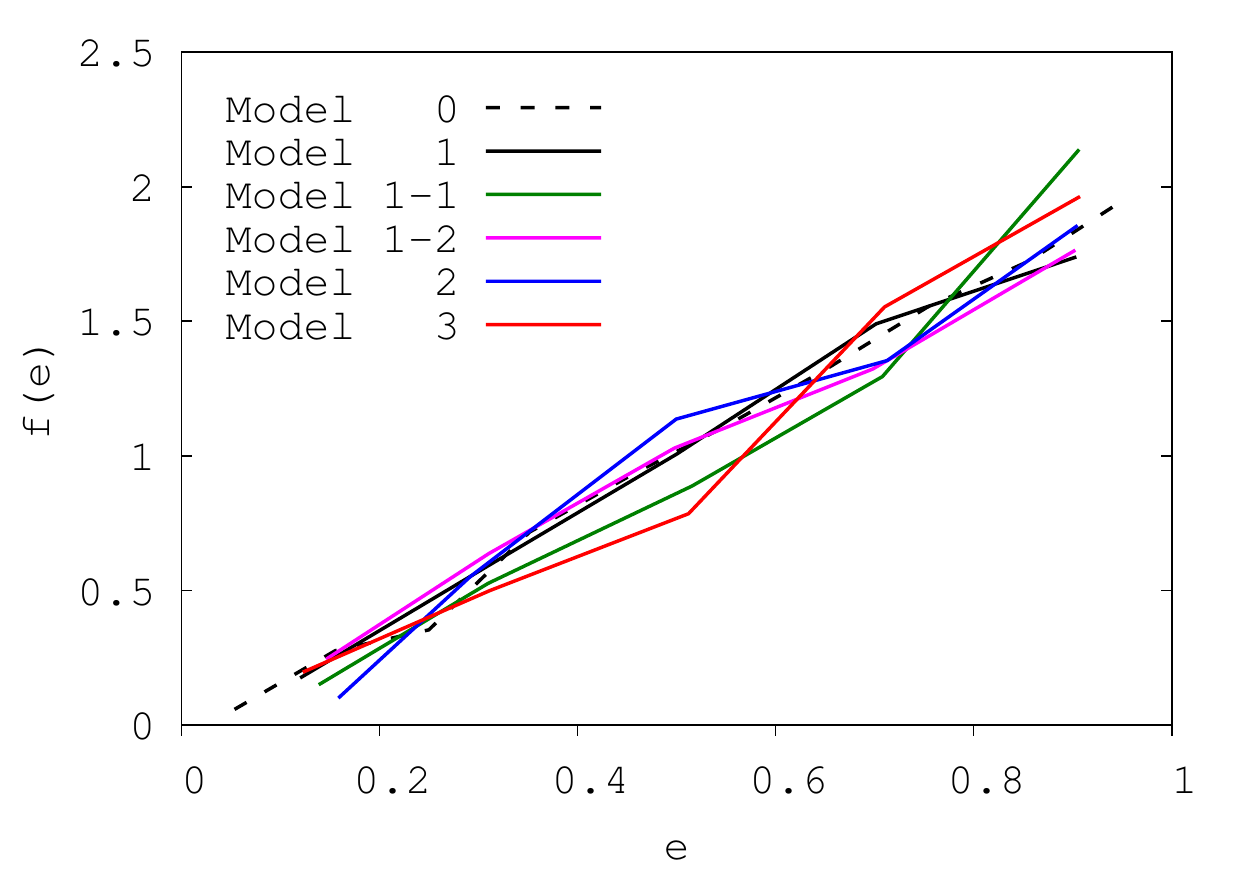}	
	\caption{The distributions of the eccentricities for our 
		models. All distributions follow a thermal distribution function, $f(e)\sim 2e$.}
	\label{fig:eccentricity}
\end{figure}

\begin{figure*}
	\centering
	\includegraphics[width=8.2cm]{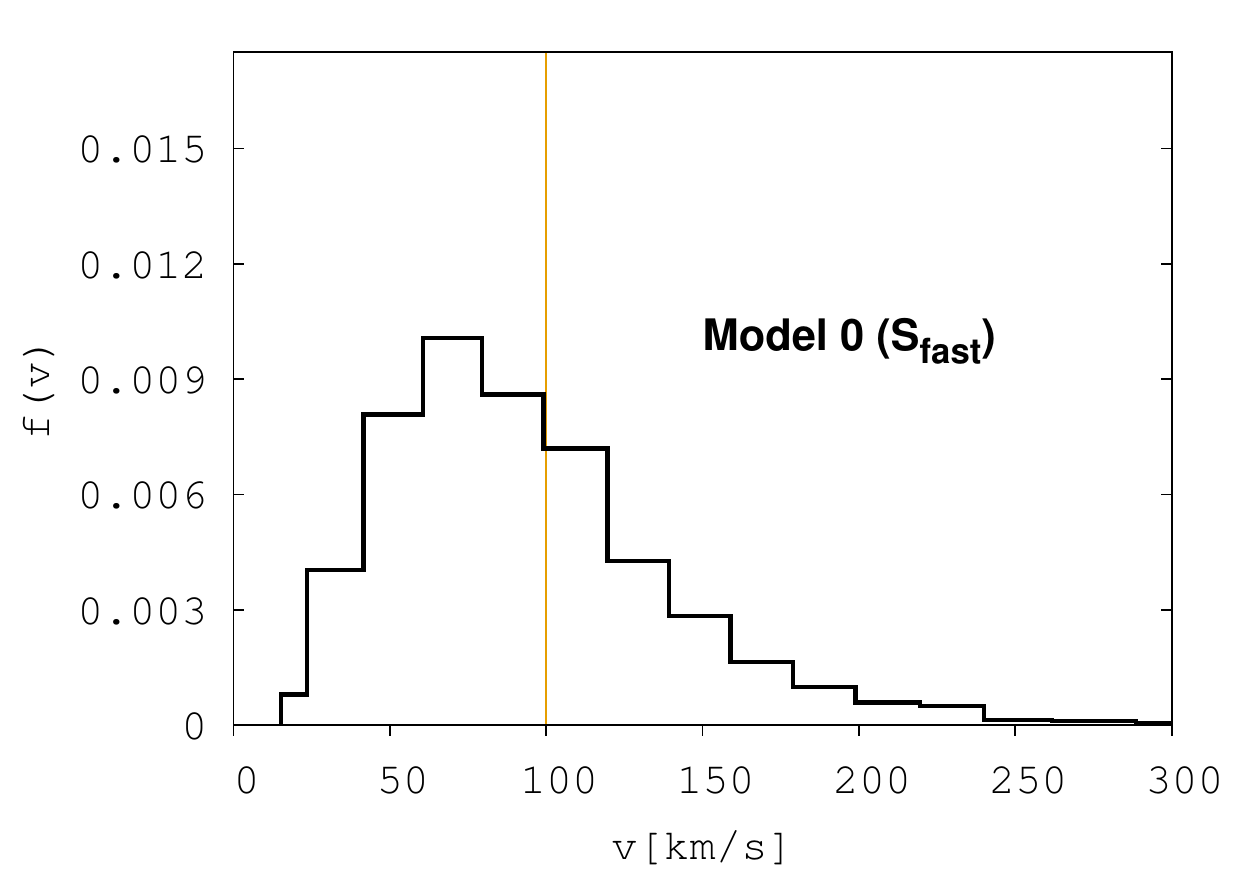}			
	\includegraphics[width=8.2cm]{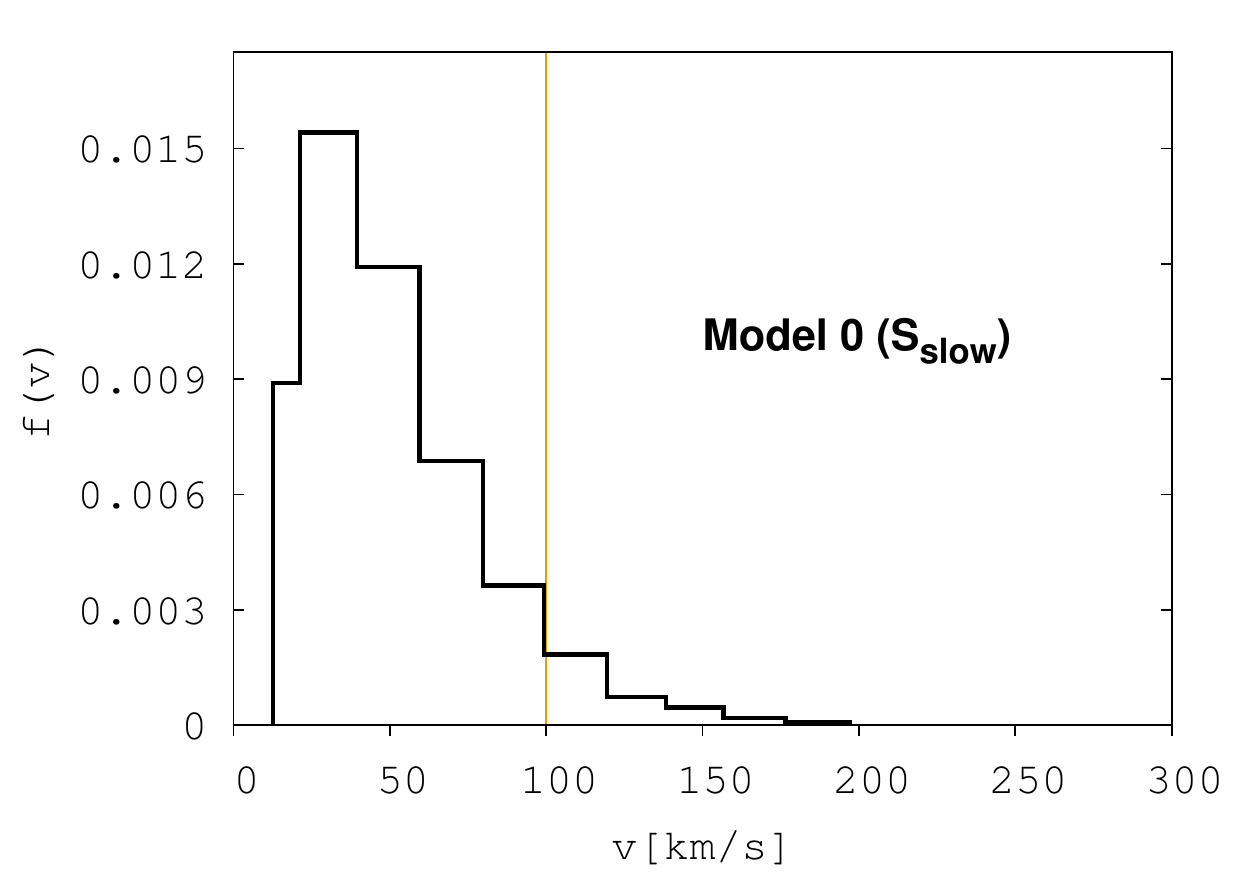}					\includegraphics[width=8.2cm]{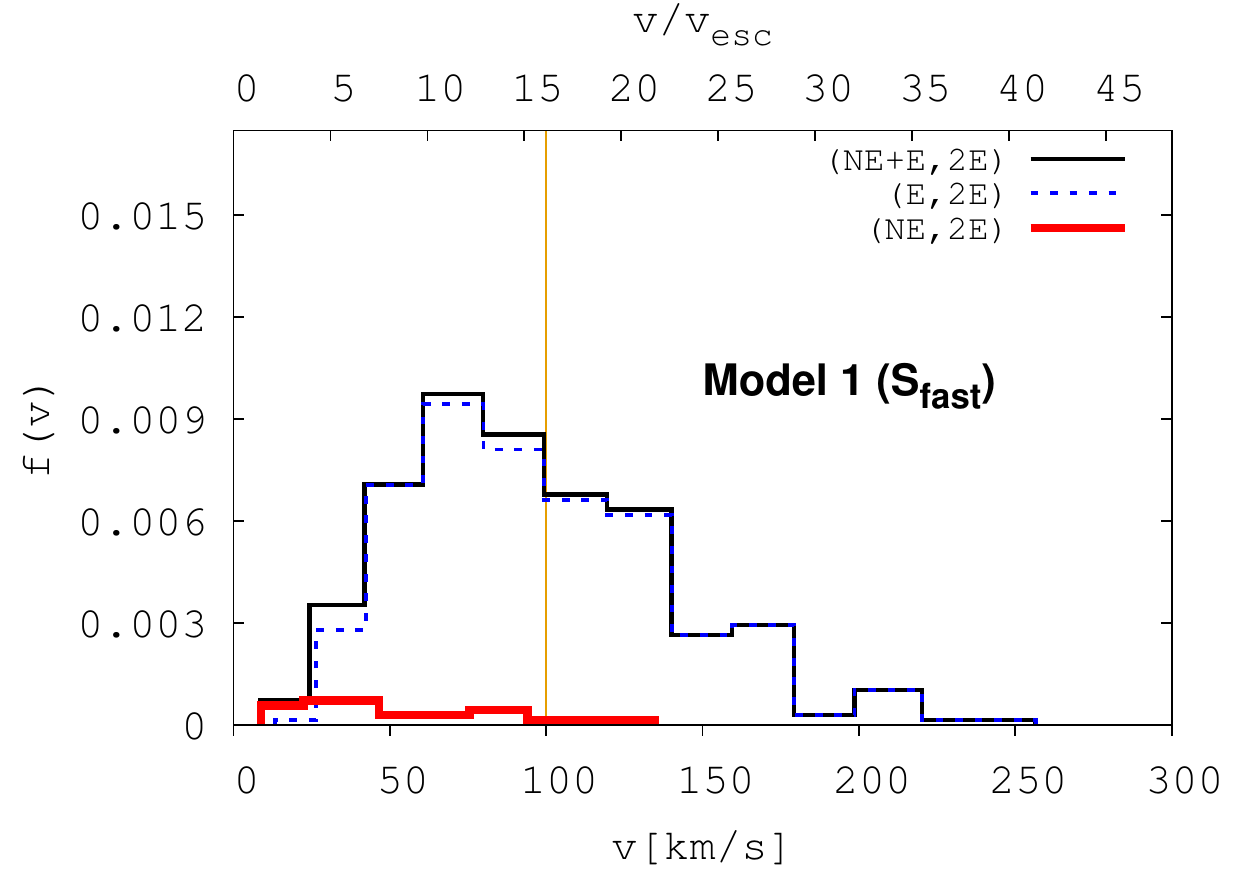}			
	\includegraphics[width=8.2cm]{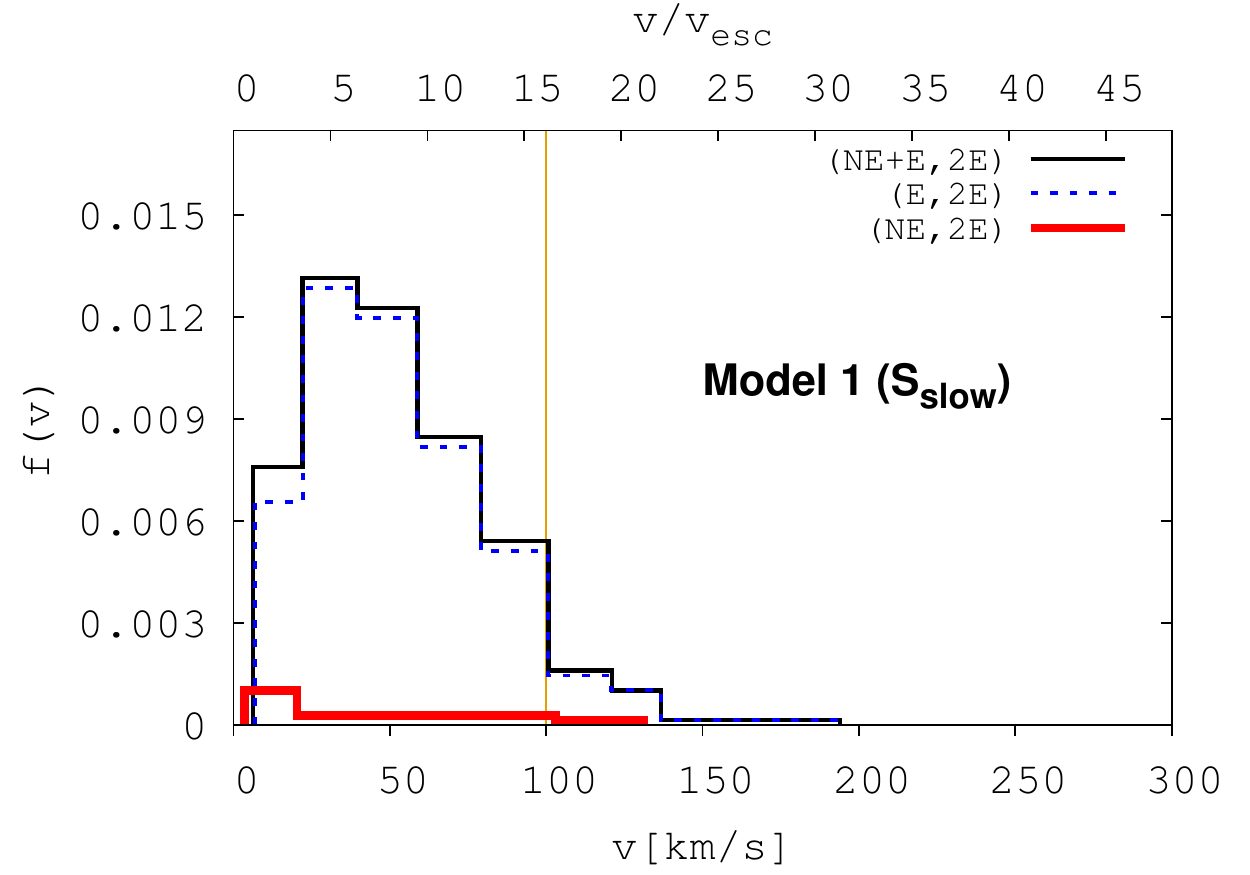}					\includegraphics[width=8.2cm]{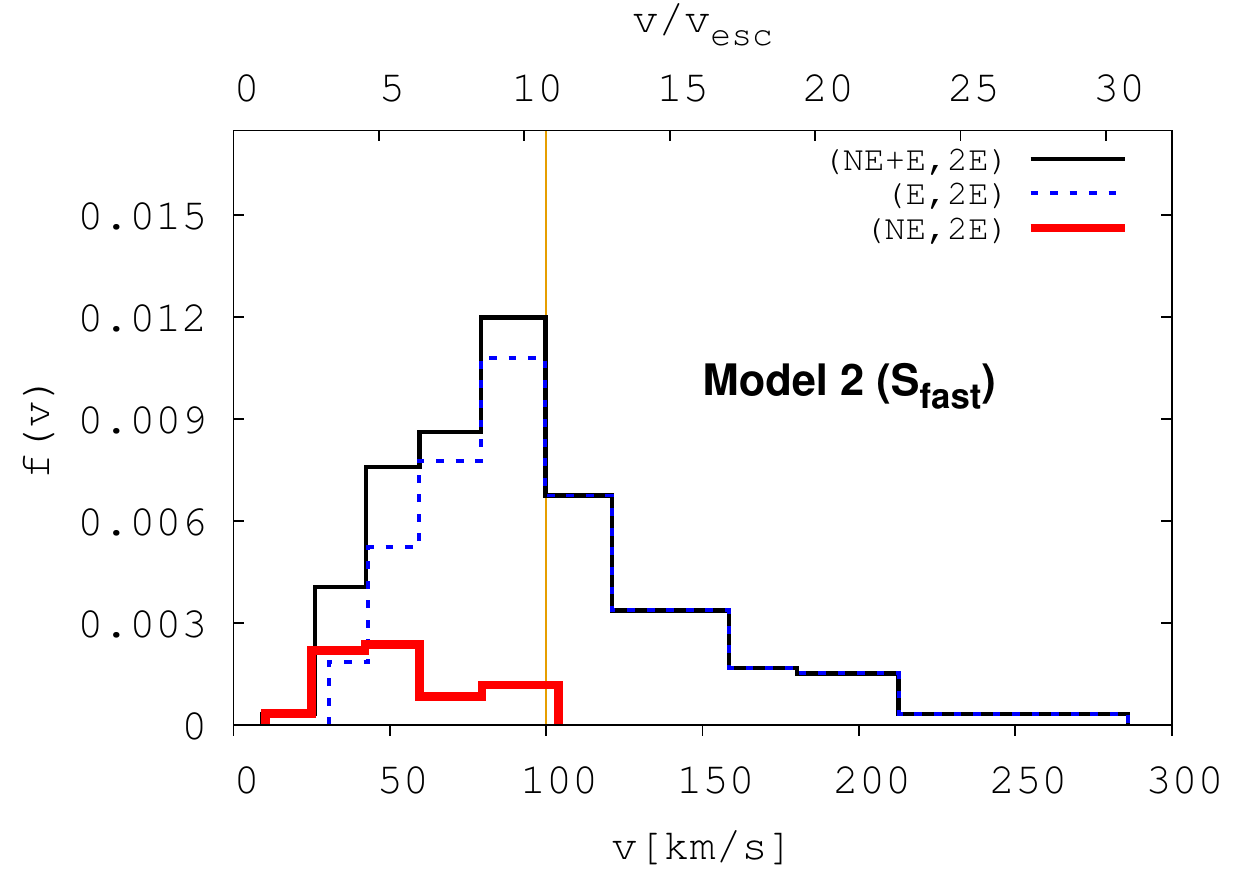}			
	\includegraphics[width=8.2cm]{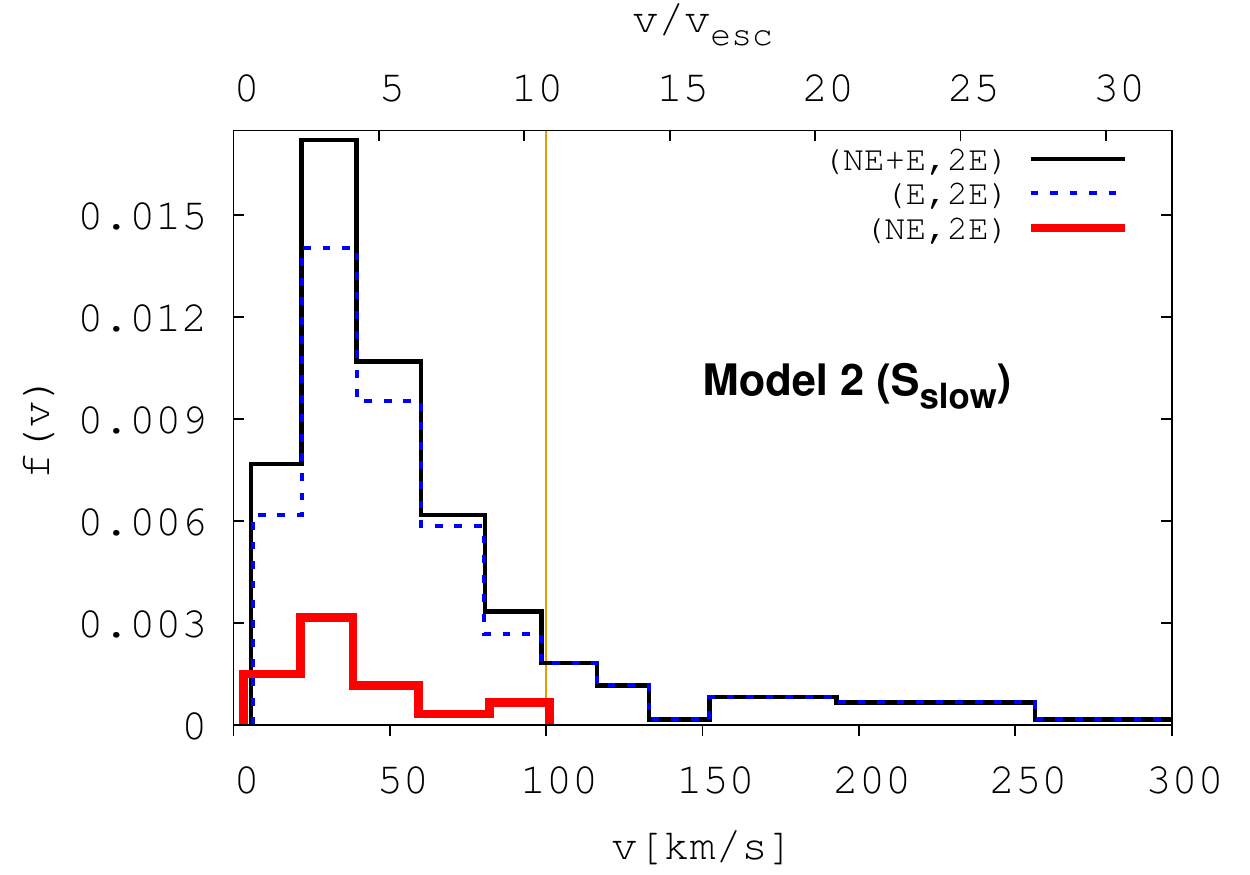}					\includegraphics[width=8.2cm]{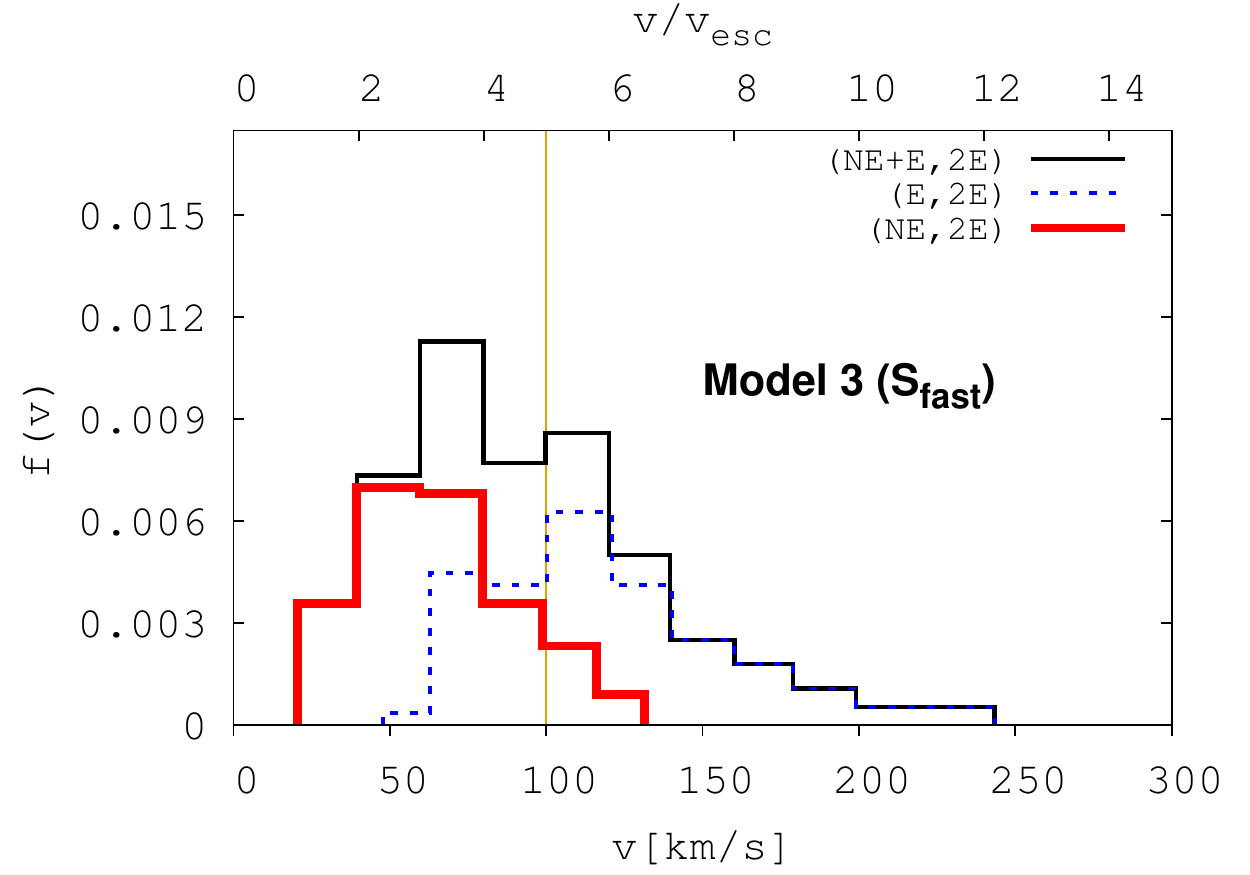}			
	\includegraphics[width=8.2cm]{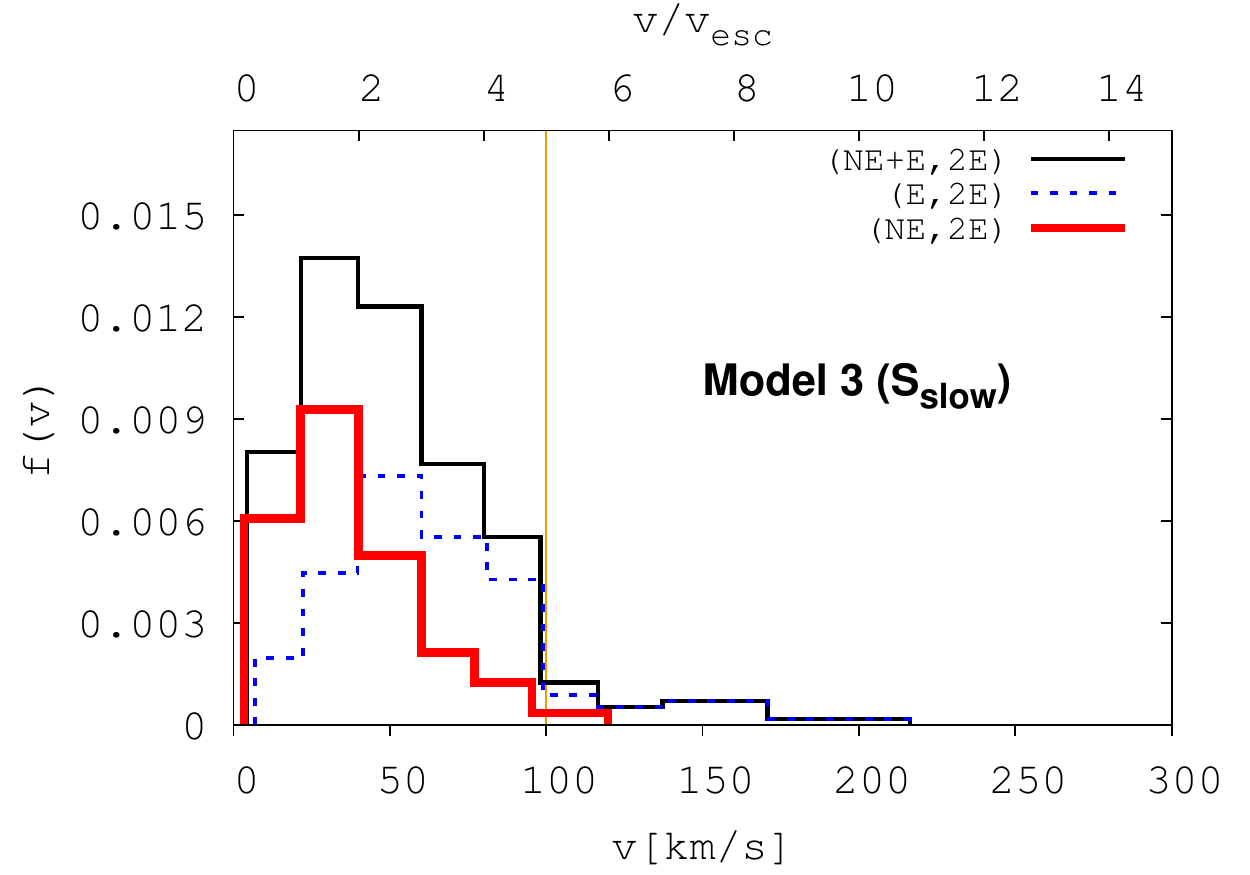}				
	\caption{The distributions of the final velocities for $S_{\rm
			fast}$ (\textit{left} column) and $S_{\rm slow}$
		(\textit{right} column) for all the models (from
		\textit{top} to \textit{bottom}), with the vertical gold
		lines indicating an observed value of $v\simeq
		100\km\s^{-1}$. In all panels, the same colors are
		used as in Figure \ref{fig:semimajoraxis}. }
	\label{fig:velocity_singlestar}
\end{figure*}

We find similar trends for the distributions of the final velocities
$v_{\rm binary}$ shown in Figure \ref{fig:velocity_binary}, where the
vertical gold lines indicate the observed value of $v\simeq
18\km\s^{-1}$. However, the velocity distributions of NE-binaries
start to appear at small $v$. Furthermore, we can see the overall
shift of the distributions for E-binaries (blue dotted line) toward
lower $v$ due to the gravitational pull from the potential as the binaries
move outward.  These two effects (the emergence of NE-binary
populations at small $v$ and the overall shift of the velocity distribution for
E-binaries to lower $v$) become more substantial as $v_{\rm esc}$
increases; finally, in Model 3, the shapes of the overall distributions
have completely changed. The peak of the overall distribution for
Model 3 has disappeared, turning into a more power-law-like distribution.

Using the velocity distributions for Model~0, we can understand the
escape fractions shown in Section~\ref{sec:escapefraction}. If
we make the rough assumption that the velocities at $r=r_{\rm bg}$
follow the distribution $f(v)$ for Model 0, the escape probability
$P_{\rm esc}$ for a background potential with $v_{\rm esc}$ can be
written in terms of the cumulative distribution function of the
velocities as,

\begin{equation}
\label{eq:P_esc}
P_{\rm esc}=\frac{\int_{0}^{v_{\rm esc}}f(v)dv}{\int_{0}^{v_{\rm max}}f(v)dv}\,.
\end{equation}
where we have adopted $6.2\km\s^{-1}$ (Model 1), $9.3\km\s^{-1}$
(Model 2) and $20 \km\s^{-1}$ (Model 3) for $v_{\rm esc}$ and the
maximum value of the final velocities of the binaries in Model 0 ($v\sim
100\km\s^{-1}$) for $v_{\rm max}$. 
Figure~\ref{fig:cumulative_velocity} shows $P_{\rm esc}$ calculated
using Equation \ref{eq:P_esc} with the velocity distribution $f(v)$
for Model 0. The three vertical dotted lines indicate the escape
velocities for Model 1 to Model 3 (from left to right). The
probability $P_{\rm esc}$ at each escape velocity agrees well with the
fractions given in Table \ref{tab:escapeprobability}.\footnote{Note
  that for more precise calculations, we have to consider the equation
  of motion in a harmonic potential with an upper bound of the
  integration in the numerator of $\sqrt{3/2}v_{\rm esc}$ to account
  for the energy required to travel from the centre to the potential
  boundary.  We refer to \citet{RLP2017} for a more comprehensive
  analytical formulation of the 3-body outcomes.}  
We
  show that the escape fractions are well explained using the results
  from Model 0, but this does not necessarily mean that all other
  results can be inferred from simulations without the background
  potential (Model 0). We find that in some cases stars cannot
  escape at their first ejections, but return back and go through multiple encounters
  with the remaining stellar systems in the potential. On the one
  hand, the escape fractions are not significantly affected by those
  cases because the single stars can manage to escape as they gain
  kinetic energy. On the other hand, this implies that the statistical
  properties of such single stars and binaries could be different
  compared to those from Model 0. Furthermore, for different choices
  of binary parameters (e.g. wide binaries) or different depths of the
  potential, the presence of the background potential could
  significantly change the properties of the final outcomes
  \citep{Ryu+2017}.

In Figure \ref{fig:eccentricity}, we present the distributions of the eccentricities for all 
models. The eccentricity distributions approximately follow a thermally-averaged 
density function \citep{Heggie1975}, or:
\begin{equation}
f(e)\sim 2e\,,
\label{e_thermal}
\end{equation} 
which is typically found for eccentricity distributions in three-body scattering problems.

\begin{figure}
	\centering
	\includegraphics[width=8.3cm]{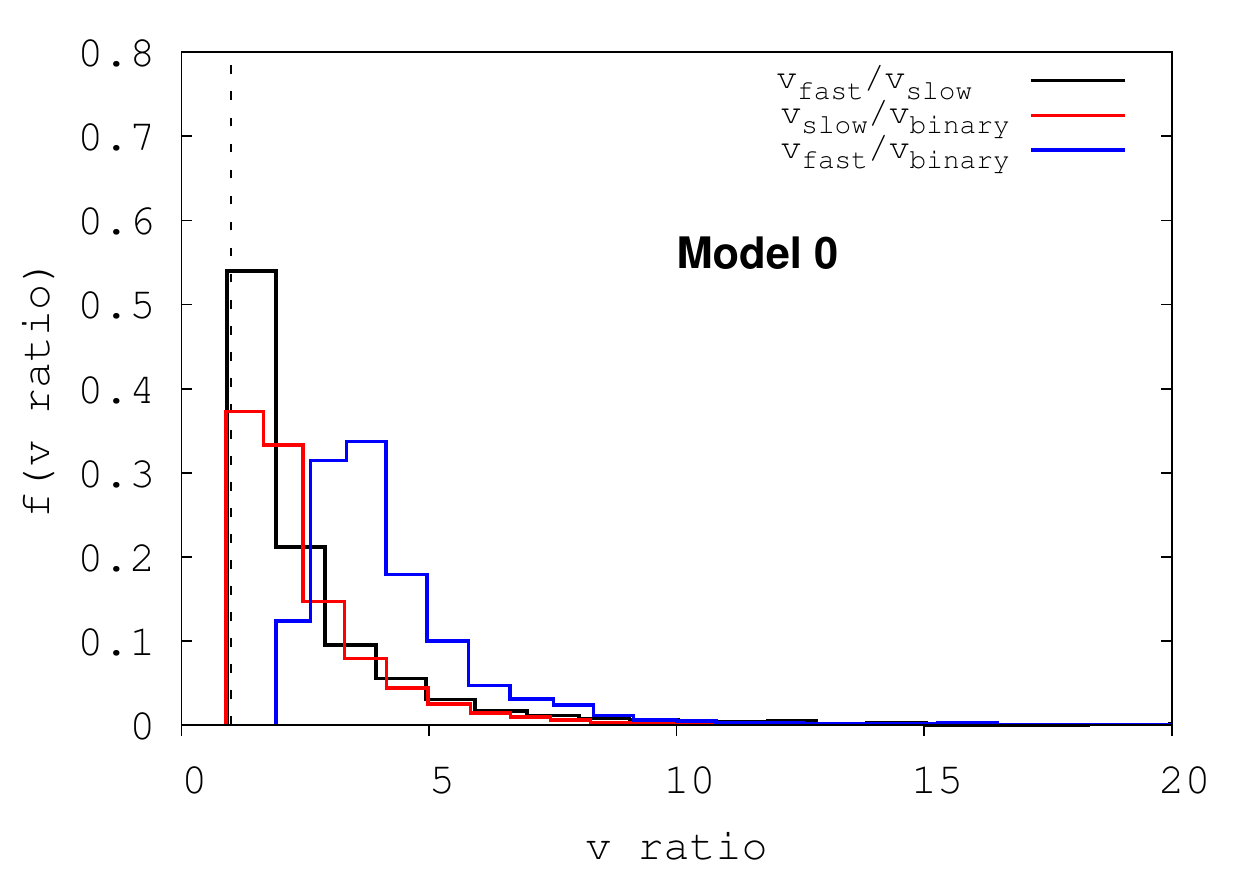}			
	\includegraphics[width=8.3cm]{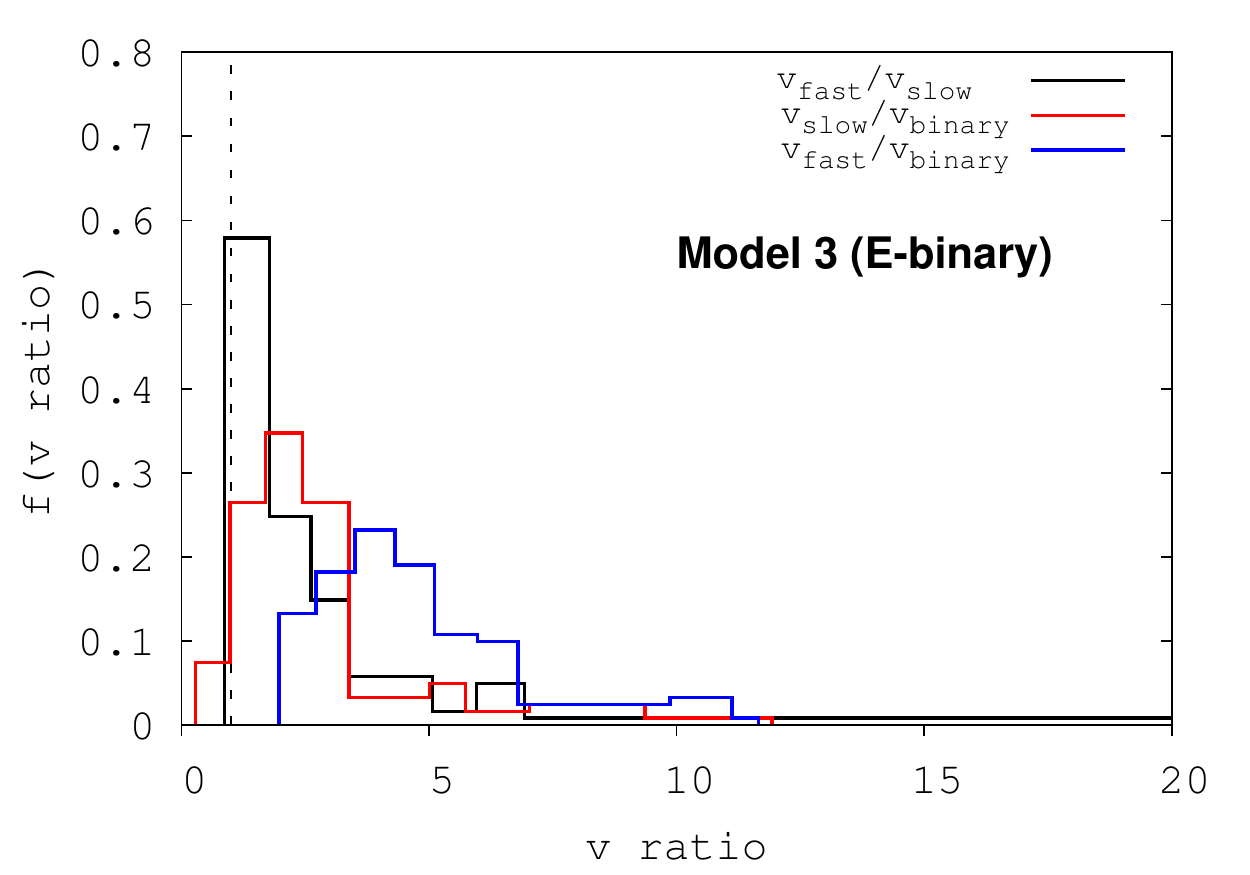}	
	\includegraphics[width=8.3cm]{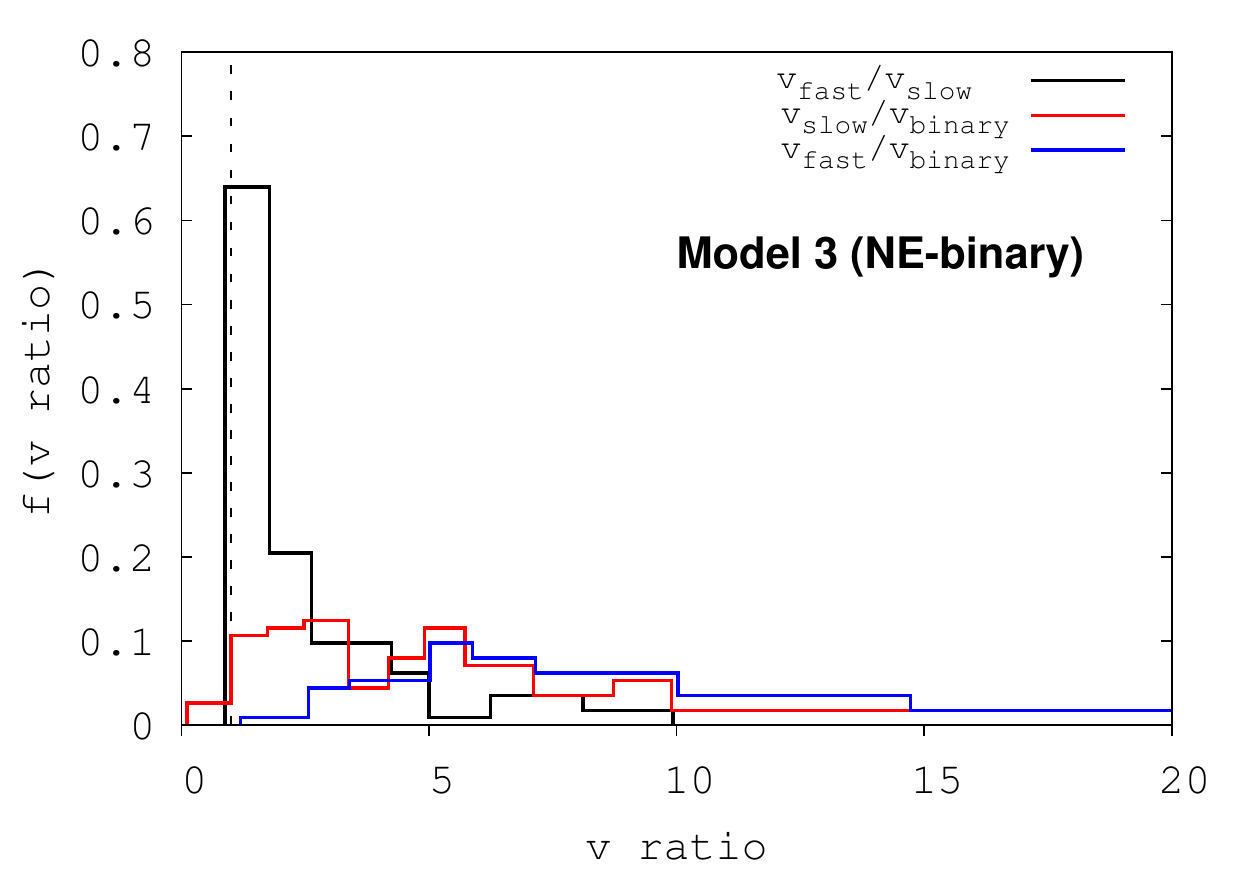}						
	\caption{The ratio of $v_{\rm fast}$, $v_{\rm slow}$ and
		$v_{\rm binary}$ for Model 0 (\textit{upper} panel) and
		Model 3 (E-binaries in the \textit{middle} panel and NE-binaries
		in the \textit{lower} panel). We only provide those for Model 3
		because the shapes and the median values are not
		significantly different from those for Model 0, Model 1 and
		Model 2. Notice that $v_{\rm fast}/v_{\rm slow}$ does not
		vary much. The median values of $v_{\rm
			fast}/v_{\rm slow}$ for all models are $\sim1.2-2.0.$}
	\label{fig:velocity_ratio}
\end{figure}

\begin{figure}
	\centering
	\includegraphics[width=8.3cm]{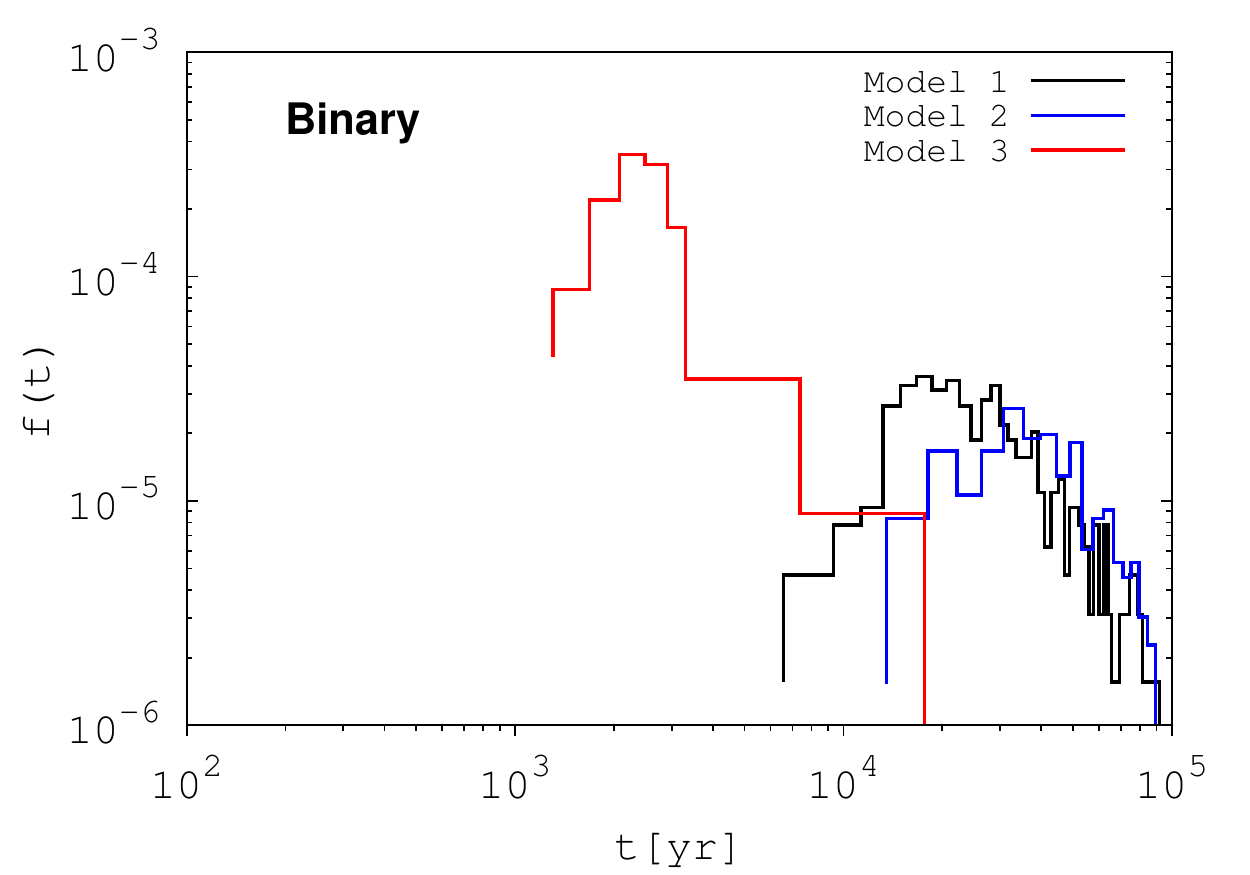}			
	\includegraphics[width=8.3cm]{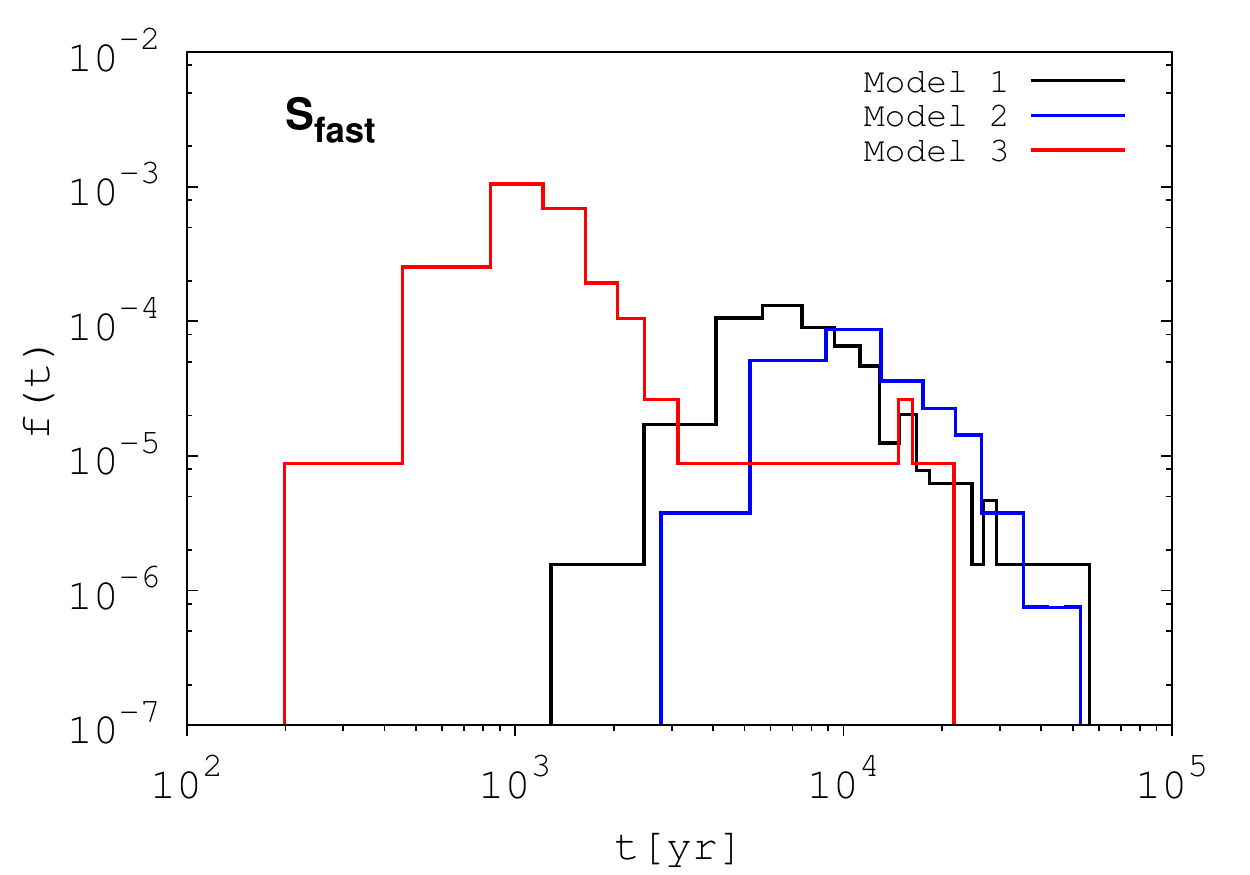}			
	\includegraphics[width=8.3cm]{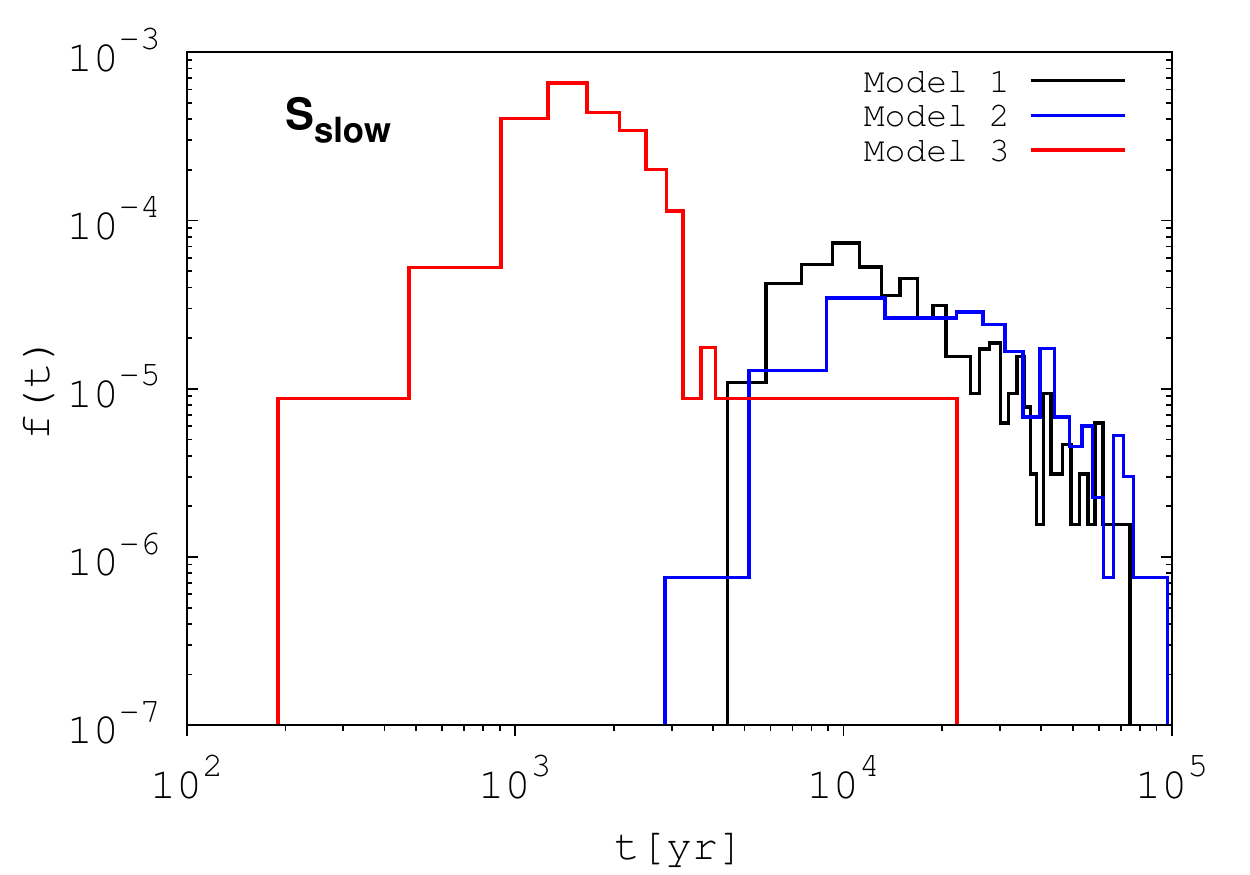}			
	\caption{The distributions of the escape time $t_{\rm bg}$ for
		the final binaries (\textit{upper} panel), $S_{\rm fast}$
		(\textit{middle} panel) and $S_{\rm slow}$ (\textit{bottom}
		panel) for our background potential models.  The escape time
		$t_{\rm bg}$ is defined as the time taken for stars to cross
		the outer boundary with speeds higher than the escape velocity
		$v_{\rm esc}$ from the drop-in time of the simulations. We
		only consider the cases where both of the two single stars could
		escape (2E).}
	\label{fig:ejectiontime}
\end{figure}

\subsubsection{The statistical properties of the ejected single stars}

\begin{figure}
	\centering
        \includegraphics[width=8.7cm]{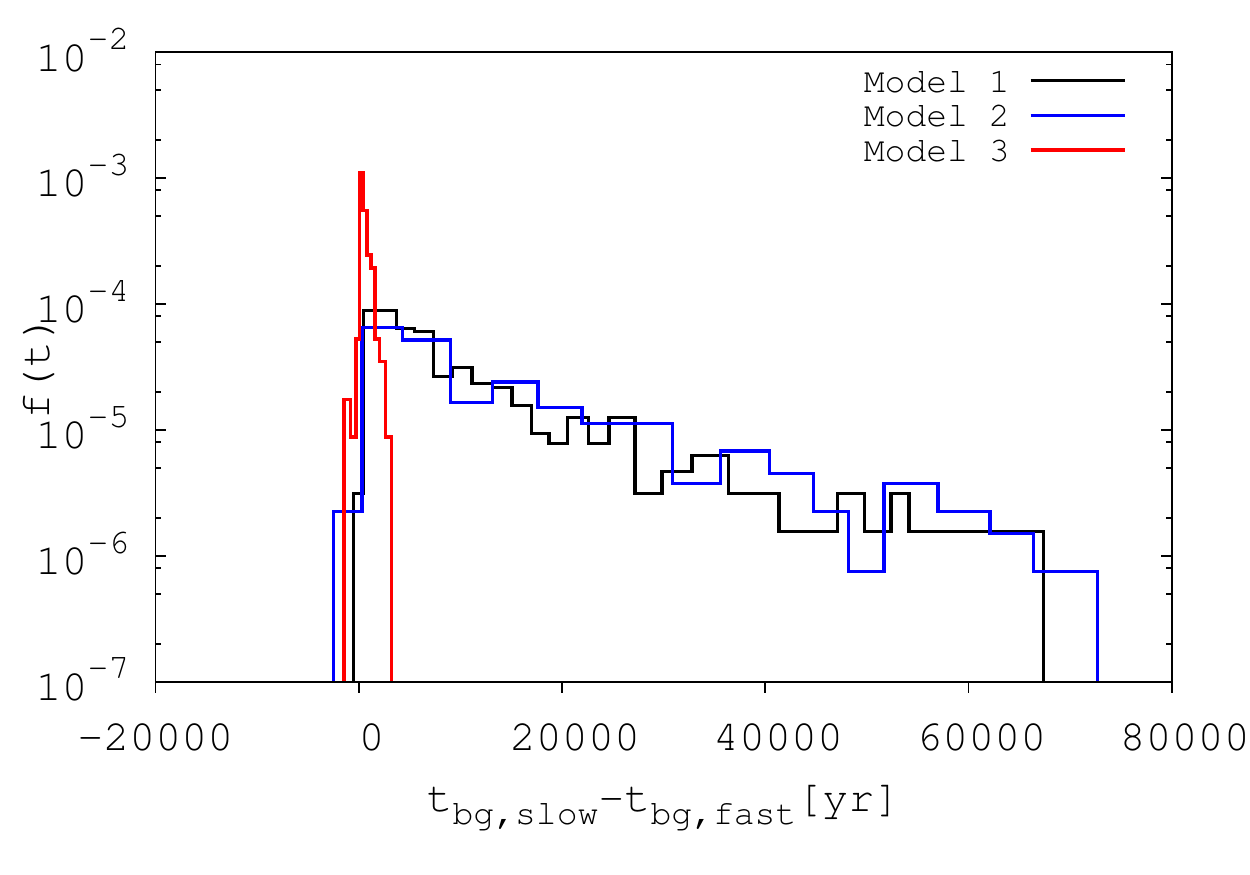} 
        \caption{The
          distributions of $t_{\rm bg}$ for $S_{\rm slow}$ with
          respect to those for $S_{\rm fast}$, or $t_{\rm bg,
            slow}-t_{\rm bg, fast}$. With similar $t_{\rm bg}$
          distributions for $S_{\rm fast}$ and $S_{\rm slow}$, the
          resulting distributions for $t_{\rm bg, slow}$ and $t_{\rm
            bg,fast}$ continue down to slightly negative $t_{\rm
            bg}$. }
	\label{fig:ejectiontimedifference}
\end{figure}
\begin{figure*}
	\centering
	\includegraphics[width=5.5cm]{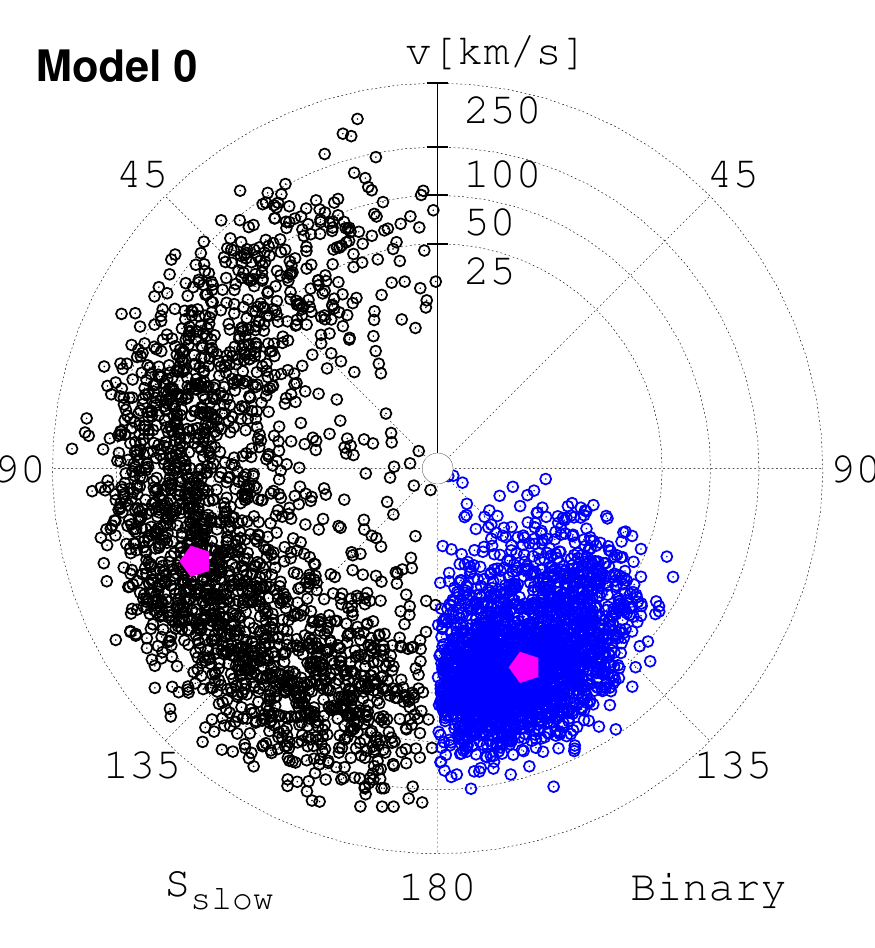}
	\includegraphics[width=5.5cm]{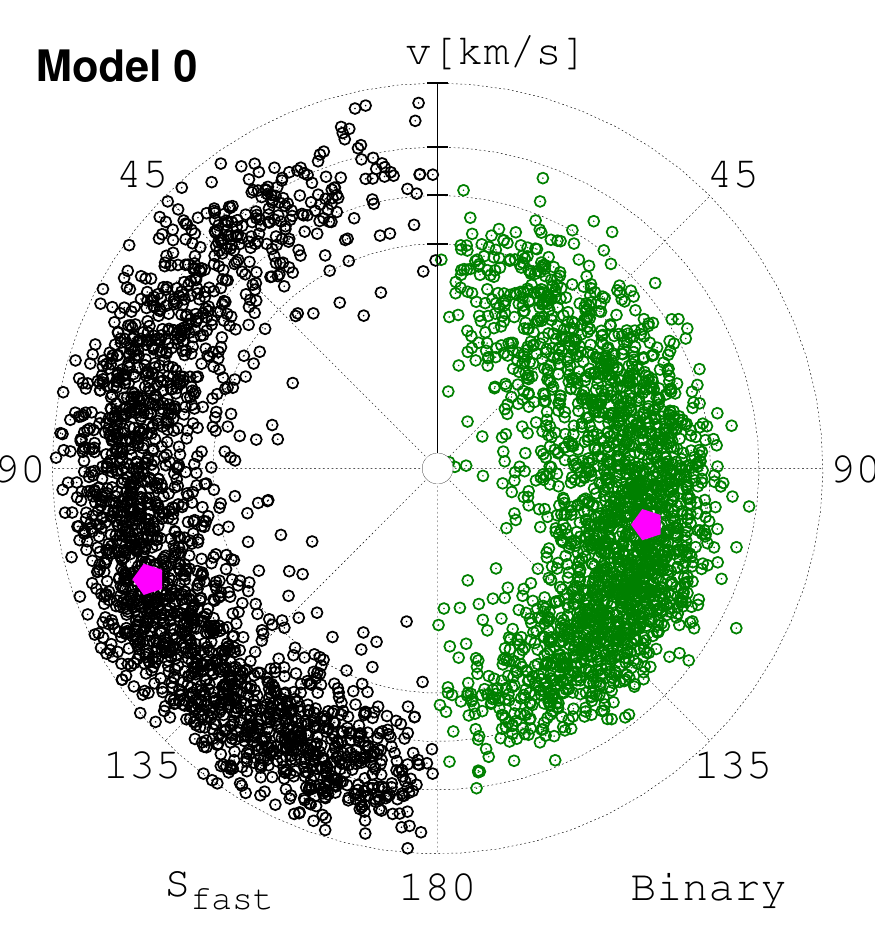}
	\includegraphics[width=6.5cm]{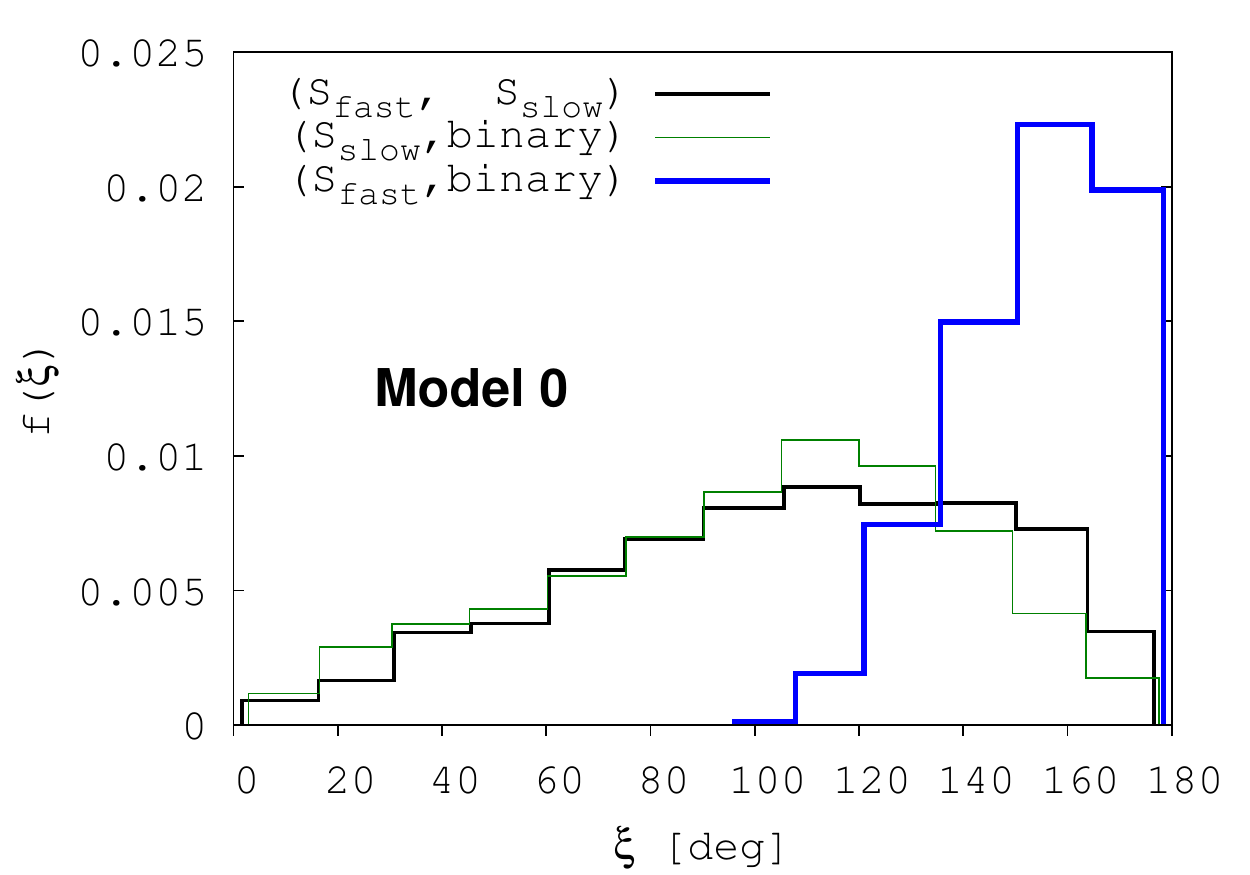}
	\caption{The distributions of the angle $\xi$ between the two
          stars (single/binary stars) as a function of the speeds $v$
          for Model 0 (see the definition in the text). In the first two
          circular plots, the distributions are projected on to the plane
          ($v,~\xi$) with their median values
          (magenta pentagon dots). In particular, in the \textit{left} panel, we
          present $\xi$ of $S_{\rm slow}$ and the binaries with
          respect to $S_{\rm fast}$ (so $v$ represents the speeds of
          $S_{\rm slow}$) while in the \textit{middle} panel, we present
          $\xi$ of $S_{\rm fast}$ and the binaries with respect to
          $S_{\rm slow}$. And $\xi$ of the single stars (the binaries)
          are marked in the left-half (right-half) panel. Each dotted
          circular grid (from the inner circle to the outer circle)
          indicates velocities of 25, 50, 100 and
          250$\km\s^{-1}$. The radial grids (from north to south)
          show $\xi$ separated by $45^{\circ}$ from $0^{\circ}$ to
          $180^{\circ}$. In the \textit{right} panel, we show the
          distribution functions $f(\xi)$ for $\xi$.  We have used the
          same colors for the same type of angles in all three panels.
          For example, the blue dots in the \textit{left} panel and the blue
          line in the \textit{right} panel refer to $\xi$ between $S_{\rm
            fast}$ and the binaries. }
	\label{fig:relativeangle_MODEL0}
\end{figure*}

\begin{figure*}
	\centering
	\includegraphics[width=5.4cm]{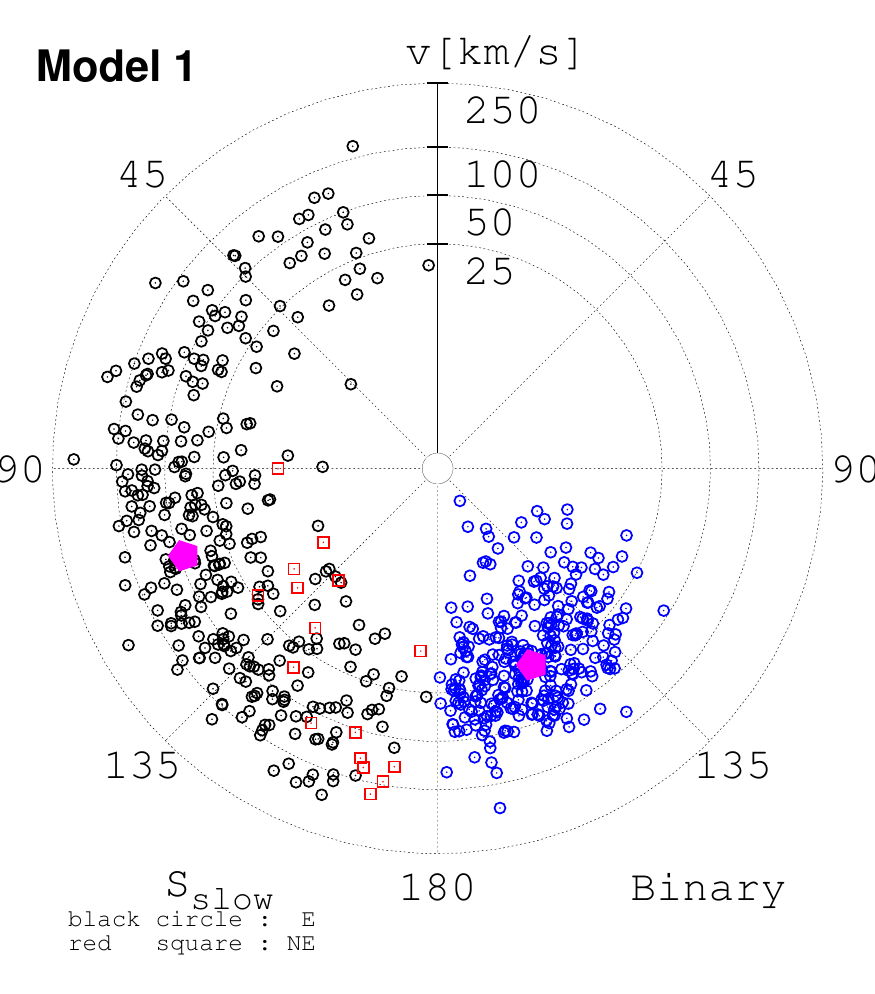}
	\includegraphics[width=5.4cm]{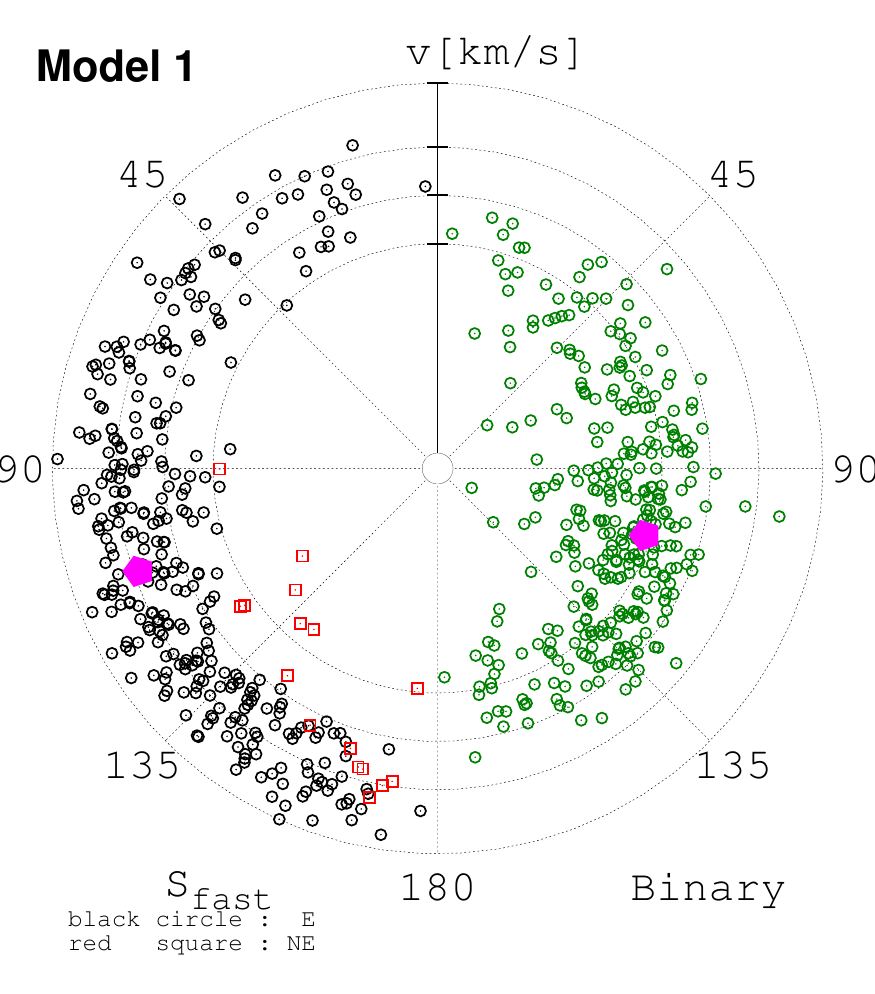}
	\includegraphics[width=6.4cm]{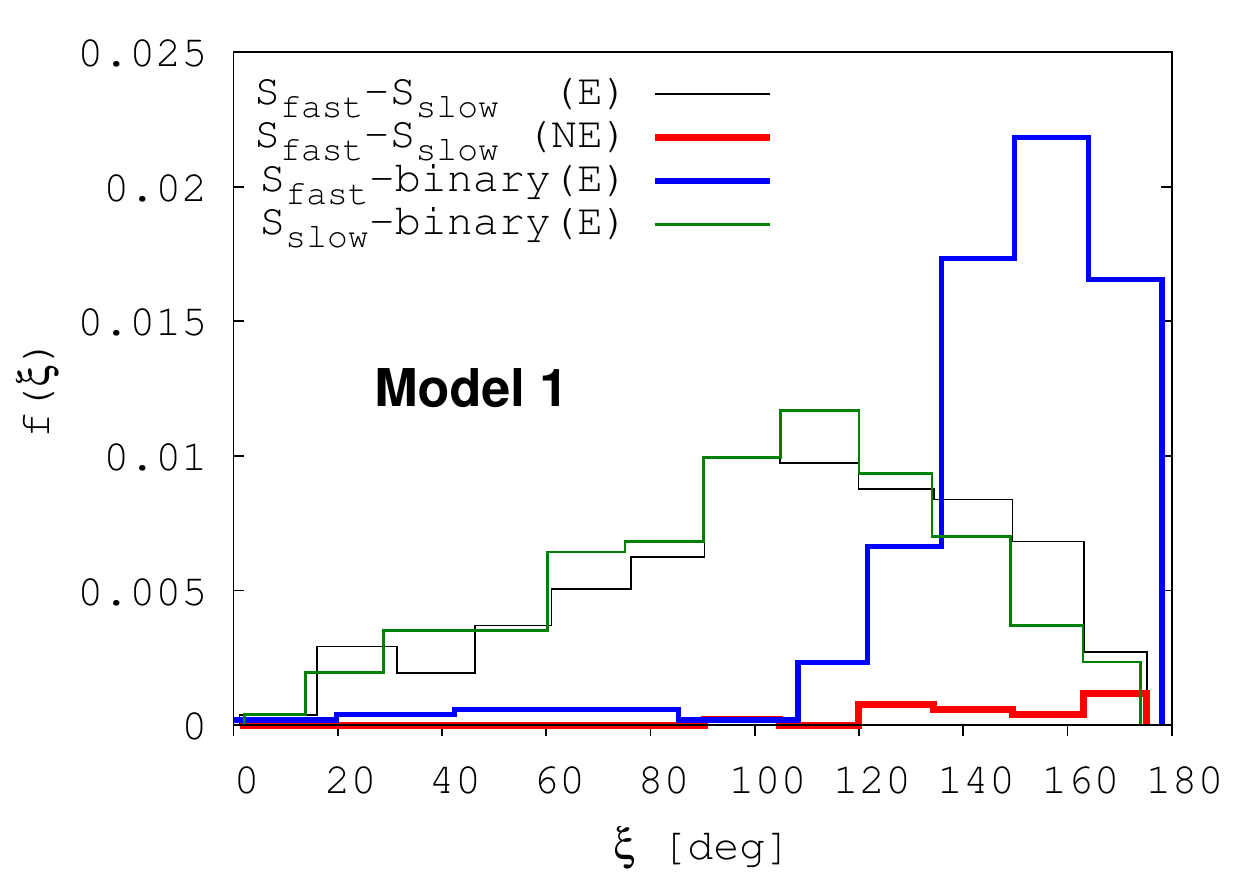}
	\includegraphics[width=5.4cm]{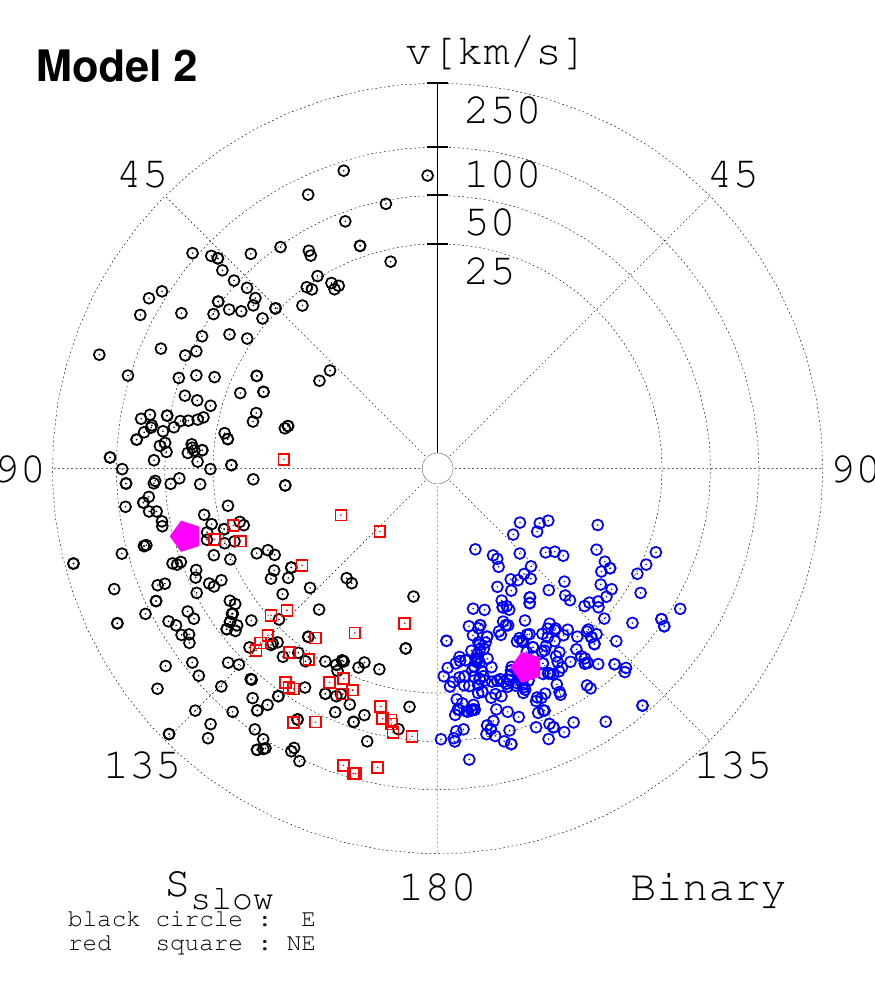}
	\includegraphics[width=5.4cm]{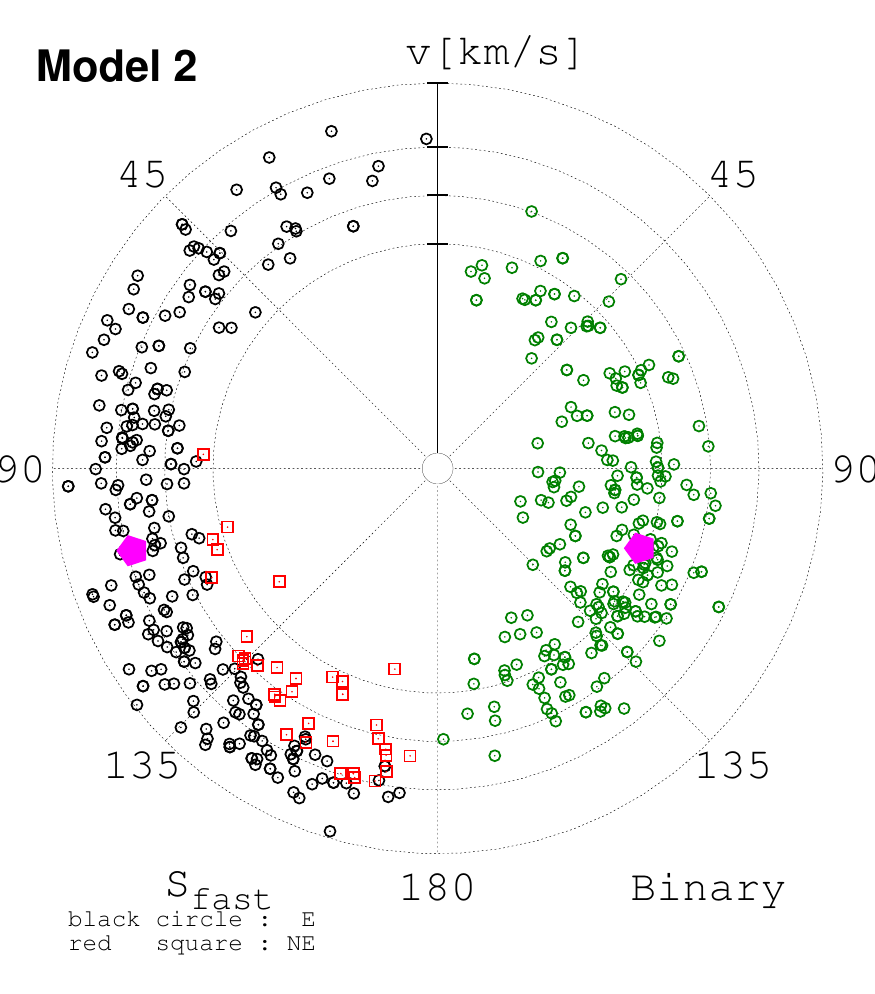}
	\includegraphics[width=6.4cm]{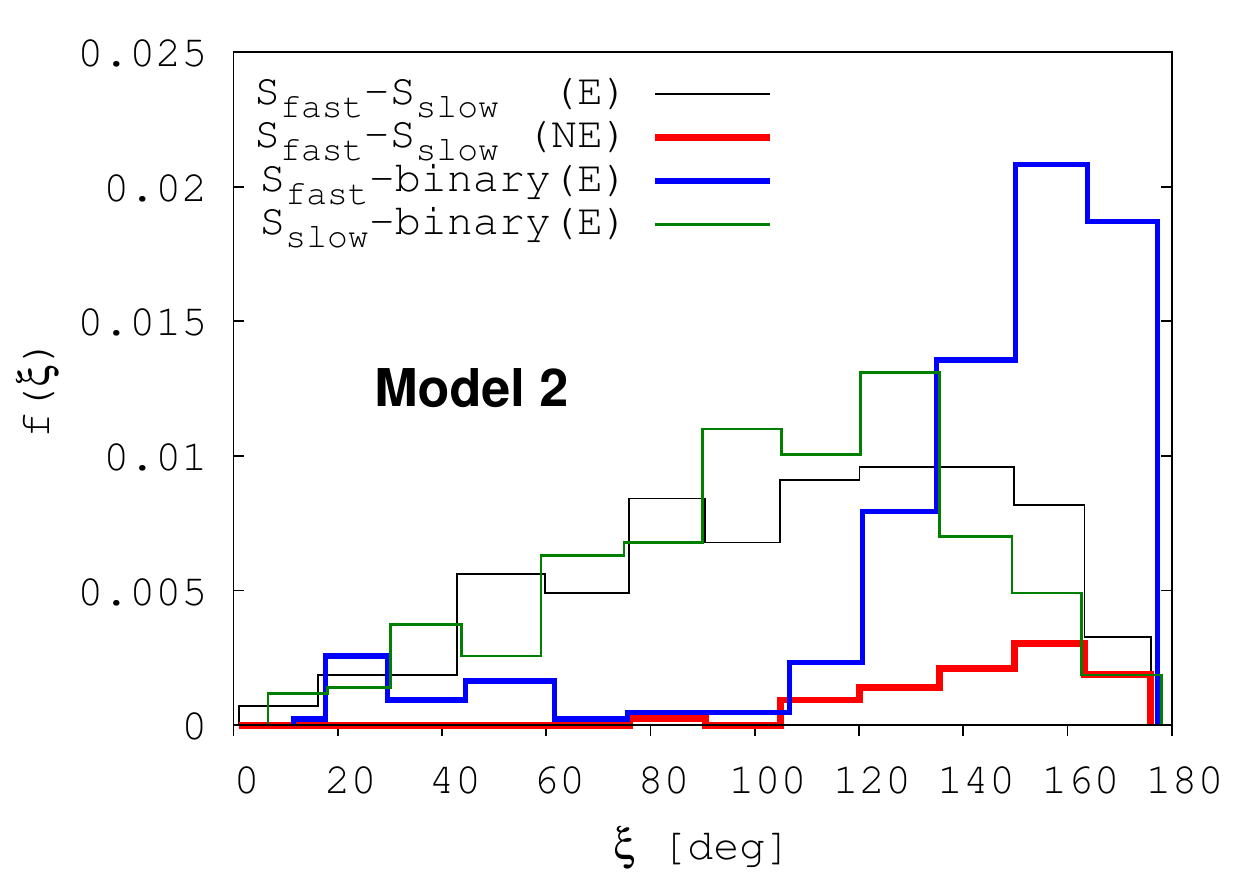}
	\includegraphics[width=5.4cm]{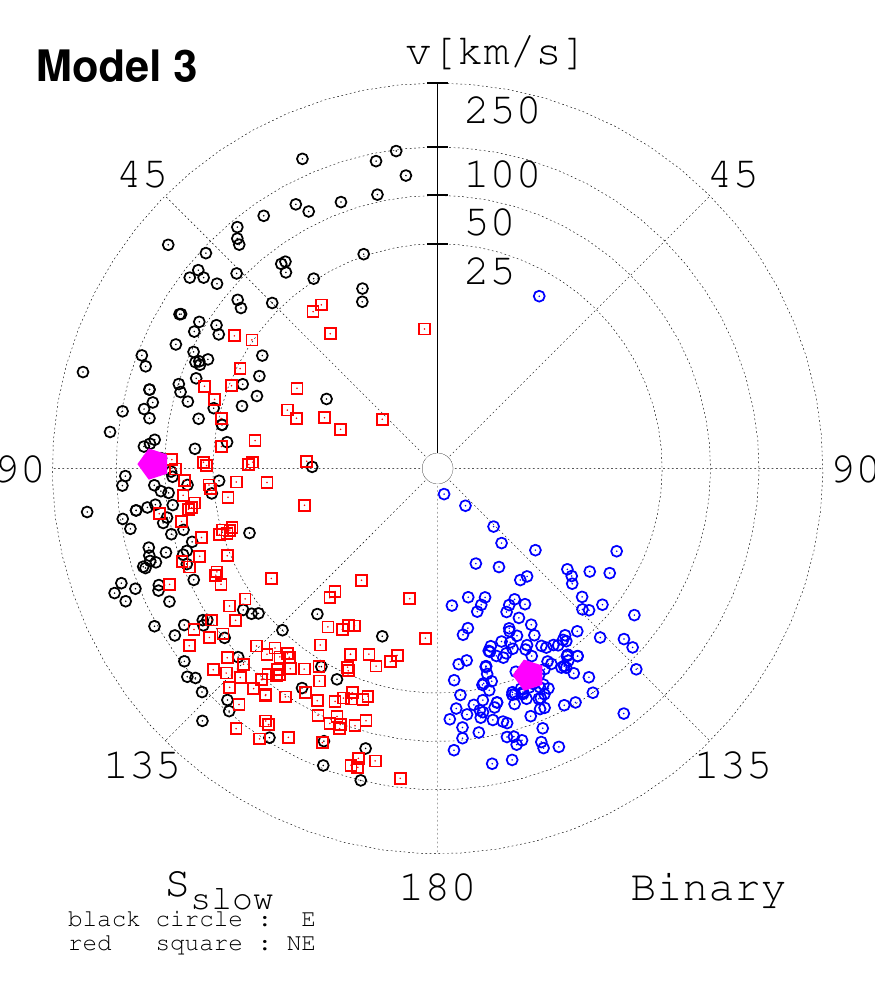}
	\includegraphics[width=5.4cm]{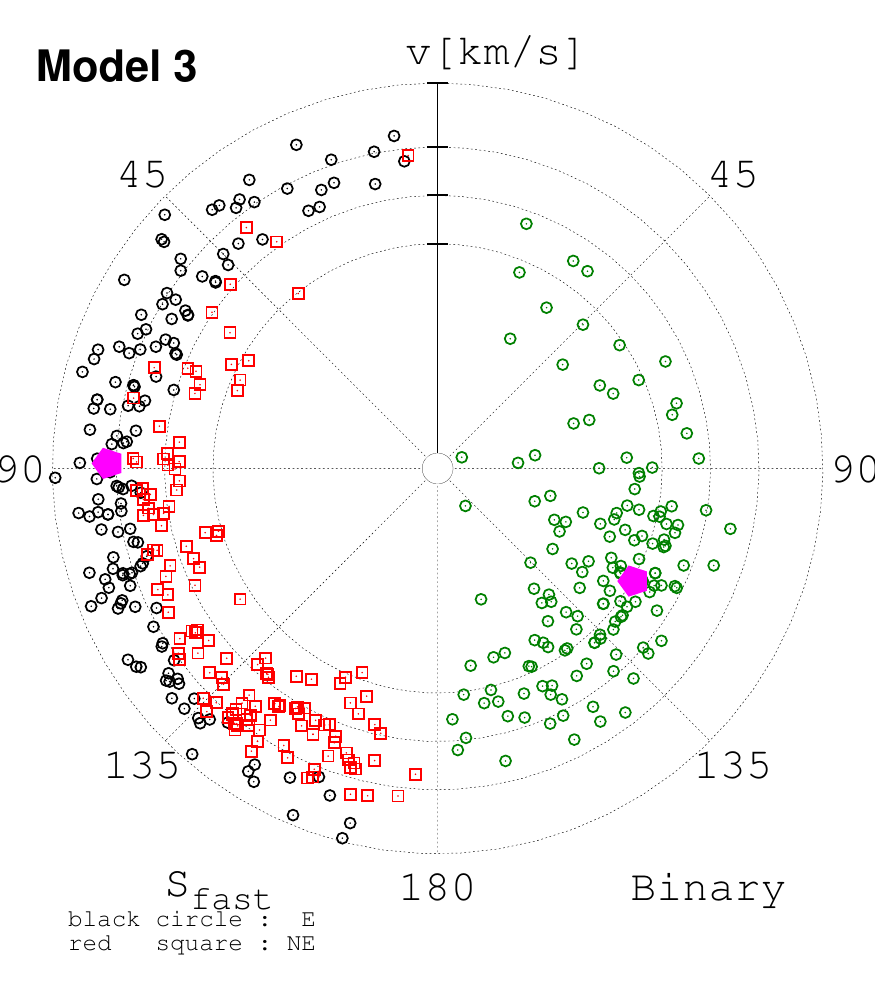}
	\includegraphics[width=6.4cm]{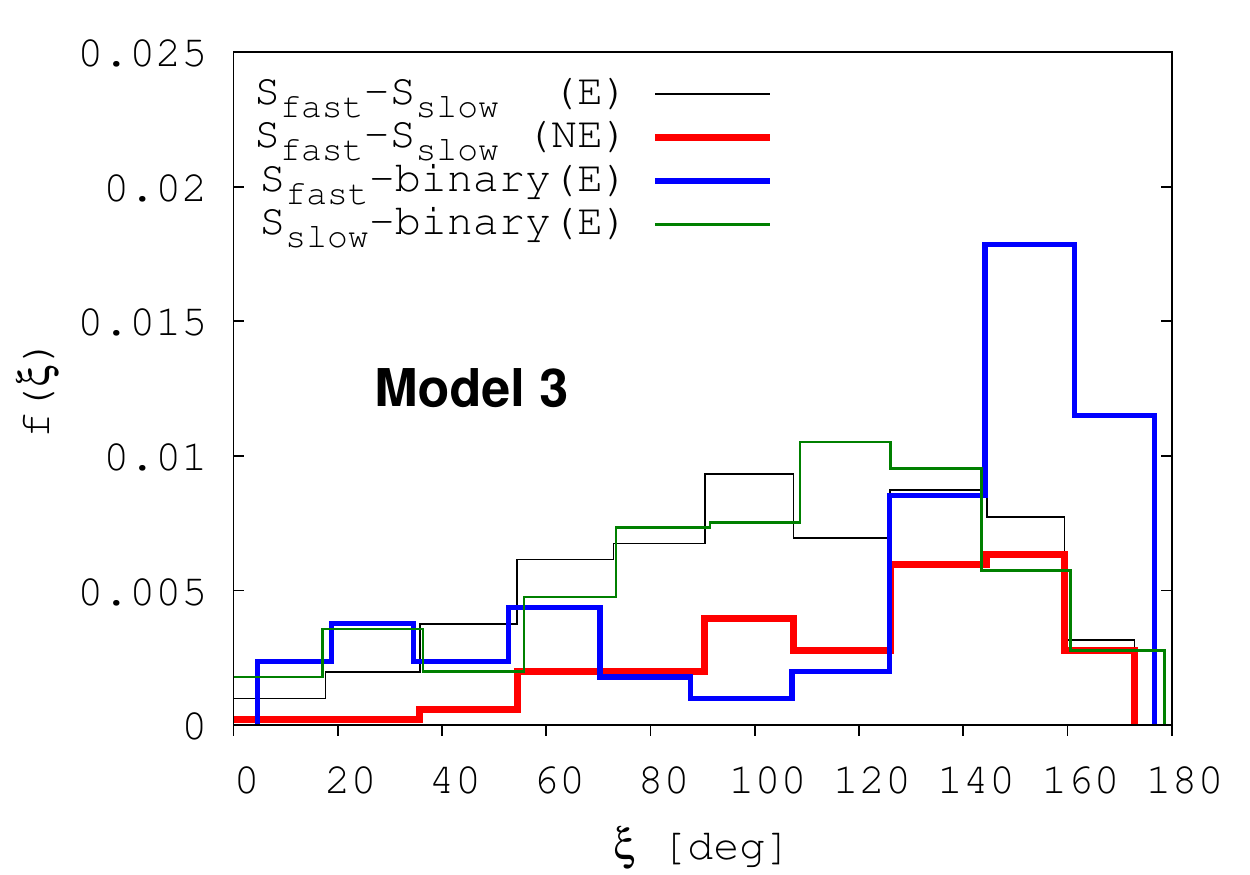}
	\caption{The distributions of the angle $\xi$ for models
          with a background potential. The same layout and plot formats
          are used as in Figure \ref{fig:relativeangle_MODEL0}. The magenta pentagon dots represent the median values of $\xi$ and $v$. In
          these plots, unlike the one for Model 0, we make a distinction
          between the cases where the binaries have escaped (E, marked
          with the black circles) and where the binaries have not
          escaped (NE, marked with the red
          squares). }
	\label{fig:relativeangle_bg}
\end{figure*}

In this section we present the statistical properties of the single
stars, and in particular the distributions of the final
velocities. As stars undergo chaotic interactions, all
  memory of the initial binaries (i.e., which objects comprised the initial
  binaries $S_{12}$ and $S_{22}$) may be lost. Therefore, given their
  identical masses, it is more appropriate to differentiate the single
  stars by their final speeds.  Hence we refer to the ejected single
  stars with higher (lower) final speeds at the end of our simulations as
  $S_{\rm fast}$ ($S_{\rm slow}$). We also use $v_{\rm fast}$ and
$v_{\rm slow}$ for their speeds, respectively.

In Figure \ref{fig:velocity_singlestar}, we present the distributions
of the final velocities for $S_{\rm fast}$ (\textit{left} column) and
$S_{\rm slow}$ (\textit{right} column) for all models (from
\textit{top} to \textit{bottom}), with the vertical gold lines
indicating the observed value of $v\simeq 100\km\s^{-1}$. In all 
panels, the same colors are used as in Figure
\ref{fig:semimajoraxis}. In Figure \ref{fig:velocity_ratio}, we
compare the ratio of $v_{\rm fast}$, $v_{\rm slow}$ and $v_{\rm
  binary}$ for Model 0 (\textit{upper} panel) and Model 3 (E-binaries
in the \textit{middle} panel and NE-binaries in the \textit{lower} panel). We
only show those for Model 3 because the shapes and the median values
are not significantly different from those for Model 1 and
Model 2. We find from the figures that:

\begin{enumerate}
	\item The changes in shape of the distributions are not
          significant compared to the large changes in $a$ or $v_{\rm
            binary}$. But we can still see that, as the background
          potential gets deeper (from \textit{top} to \textit{bottom}), the
          overall distributions become skewed toward low
          velocities due to the increasing contribution from the population
          of NE-binaries (red solid lines).  Their
          median values for $S_{\rm fast}$ and $S_{\rm slow}$ for the models
          are given in Table \ref{tab:median}.
	
	\item The velocities of $S_{\rm fast}$ and $S_{\rm
          slow}$ for the NE-binary cases are distributed at small
          $v$. In general, for a single ejection event between a
          binary and a single star, a small kick of the binary
          (corresponding to the NE-binary case) translates into a small ejection
          velocity for the single star, because they are proportional
          to the mass ratio.  However, for two ejection events (like
          2+1+1), we have to take into account the vector sum of the
          two kicks to estimate the final kick of the binary (and
          hence the relative angles between them).  There can be
          situations with two ejection events with wide relative
          angles where, even though each kick is very strong, the
          binary left behind has a smaller final kick velocity and
          remains bound to the potential. This could be one of
          the reasons for the extended high-end tail seen in the (NE,2E)
          distribution.

	\item The ratio of the speeds between $S_{\rm fast}$ and
          $S_{\rm slow}$ does not vary much (see Figure
          \ref{fig:velocity_ratio}). Even for the NE-binary case in
          Model 3, the distribution for $v_{\rm fast}/v_{\rm slow}$
          remains the same compared to the other models. 
            The median values for all models are $\sim1.2-2.0$
             (See Table \ref{tab:median}).
	 
	\item It is clearly seen that, as the potential becomes
          deeper, the distributions for the E-binary cases (blue
          dotted lines in Figure \ref{fig:velocity_singlestar}) 
          decrease in size as the population of 
          NE-binary cases emerges, but their shapes remain similar
          compared to the noticeable transformation in the velocity
          distributions of the binary (see (N,2E) distributions for
          Model 3 in Figure \ref{fig:velocity_binary}). This, along
          with the above, may imply that the motions of the single
          stars are not substantially affected by the background
          potential once they leave the potential. This is
          not surprising in light of the fact that the velocities of
          the rapidly-moving stars barely change because their kinetic
          energies are much larger than the background potential energy.

\end{enumerate}
We show cumulative distributions of radial distances from
  the system CM and the speeds at $t=1,~2,~3$ and $4\Myr$ for $S_{\rm
    fast}$ (Figure \ref{fig:cumulative_S_fast}) and $S_{\rm slow}$
  (Figure \ref{fig:cumulative_S_slow}) in Appendix
  \ref{appendix:cumulative}.

\subsection{The escape time}
\label{sec:escapetime}

In Figure \ref{fig:ejectiontime}, we compare the distributions of the
escape time $t_{\rm bg}$ for the final binaries (\textit{upper}
panel), for the faster star $S_{\rm fast}$ (\textit{middle} panel),
and for the slower star $S_{\rm slow}$ (\textit{bottom} panel), for
our background potential models.  We define the escape time $t_{\rm
  bg}$ as the time (from the start of each simulation)
taken by the stars to pass the outer boundary with
speeds higher than the escape velocity $v_{\rm esc}$. 
We only consider cases where both of the two single stars
could escape (2E). For binaries which have not escaped until
$t=4\Myr$, we take $t_{\rm bg}=0$. The distributions for such
binaries are not shown in the figure.

The \textit{upper} panel of Figure~\ref{fig:ejectiontime} shows that
the escape time for Model 3 is more concentrated at small times
compared to the other models. In order to gain a quantitative understanding of this,
we introduce a characteristic time $\tau_{\rm bg}$ for the potential
defined as follows,
\begin{align}
\tau_{\rm bg}=\frac{r_{\rm bg}}{\sqrt{3/2}v_{\rm esc}}=
\frac{1}{\sqrt{4\pi G\rho}}=\frac{1}{\sqrt{3}w_{\rm bg}}\,,
\end{align}
where $w_{\rm bg}$ is the frequency of oscillation in a harmonic
potential.  This provides a good approximation to the "maximum" time taken for
a star to arrive at the outer boundary from the origin with a velocity
equal to the escape velocity. Notice that $\tau_{\rm bg}$ is a function
of only $\rho$, not $M_{\rm bg}$ or $v_{\rm esc}$. Since in our
simulations the last ejection events occur at $t$ shorter than $t_{\rm
  bg}$, $t\leq1000\yr$, we may ignore the duration of the stellar
interactions for this analysis. We estimate that $\tau_{\rm bg}$ for
Model 3 is shorter than for the other models by two orders of magnitude
($\tau_{\rm bg}\simeq 3000\yr$ for $\rho=10^{3}$ and $\tau_{\rm
  bg}\simeq 10^{5}\yr$ for $\rho=1$). From these estimates, we expect
that the escape times for Model 3 with high $\rho$ are shorter, which
is consistent with what is shown in the \textit{upper panel}. In
addition to that, $\tau_{\rm bg}$ indeed represents the peak value for
Model 3 since typically the velocities of the binaries at $r=r_{\rm bg}$ 
\footnote{Note that it may be hard to infer $v$ at $r=r_{\rm bg}$ from
  the velocity distributions for Model 3 in Figure
  \ref{fig:velocity_binary} since the velocities of the escaped
  binaries keep decreasing due to the gravitational pull from the
  potential (the distributions are for $v$ at $t=4\Myr$).} are
comparable to $v_{\rm esc}\simeq20\km \s^{-1}$ for
$\rho=10^{3}$. However, when the typical velocities of the binaries at
$r=r_{\rm bg}$ are higher than the escape velocities, $\tau_{\rm bg}$
may overestimate the actual escape times. The distribution of $t_{\rm
  bg}$ for Model 2 is shifted towards longer $t_{\rm bg}$ compared
to that for Model 1, which may appear contradictory to the estimates
based on $\tau_{\rm bg}$ given their same densities. But $v_{\rm esc}$
for these two models is $6-9\km \s^{-1}$, lower than typical binary
speeds at $r=r_{\rm bg}$. For this case, the trend arises simply because the
width of the potential is larger (larger $r_{\rm bg}$) for Model
3. See Table \ref{tab:median} for the median values of $t_{\rm bg}$.

In the \textit{middle} and \textit{bottom} panels, we can see that the
single stars show similar trends in the distributions to those of the
binaries. But it is clear that, given the higher speeds of the
single stars, $t_{\rm bg}$ for $S_{\rm fast}$ and $S_{\rm slow}$ are
distributed at shorter times than that for the binaries, roughly
shorter by the mass ratios between the single stars and the binary,
namely, $\simeq (39+19)/16\simeq3.6$.

In Figure \ref{fig:ejectiontimedifference}, we further show the
distributions of $t_{\rm bg}$ for $S_{\rm slow}$ with respect to those
for $S_{\rm fast}$. Given the similar $t_{\rm bg}$ distributions for
$S_{\rm fast}$ and $S_{\rm slow}$, the resulting distributions for
$t_{\rm bg, slow}$ and $t_{\rm bg,fast}$ extend down to $t_{\rm
  bg}\simeq0$ (even to negative values).  As shown in the figure, it is more
likely that stars moving at higher speeds have escaped first.

\subsection{The relative angles between the ejected objects}
\label{sec:relativeangle}

In this section, we study the relative angles between objects, i.e., single-single, 
single-binary. We expect that the statistical properties of the angles can provide 
observationally useful insight to identify runaway objects with a common origin.

In Figure \ref{fig:relativeangle_MODEL0}, we present the distributions
of the relative angle $\xi$ as a function of the speeds $v$ for Model
0. Here, we define the relative angle $\xi$ as the relative angle
between the velocity vectors of the two stars. In the first two circular
plots, the distributions for $\xi$ are projected on to the ($v,~\xi$)
plane with the median values (magenta pentagon dots). The $\xi$ values for the single
stars (the binaries) are marked in the left-half (right-half) of the 
circle. In particular, in the \textit{left} panel, we present $\xi$ for
$S_{\rm slow}$ and the binaries with respect to $S_{\rm fast}$ (so $v$
represents the speeds for $S_{\rm slow}$ and the binaries) 
while in the \textit{middle}
panel, we present $\xi$ for $S_{\rm fast}$ and the binaries with
respect to $S_{\rm slow}$ (so $v=v_{\rm fast}$ and $v_{\rm binary}$). 
Each dotted circular grid (from the inner circle to the outer
circle) indicates velocities of 25, 50, 100 and
250$\km\s^{-1}$. The radial grids (from north to south) show $\xi$
separated by $45^{\circ}$ from $0^{\circ}$ to $180^{\circ}$. In the
\textit{right} panel, we show the distribution function $f(\xi)$ for
$\xi$. We have used the same colors for the same types of angles in all
three panels. For example, the blue dots in the \textit{left} panel refer
to $\xi$ between $S_{\rm fast}$ and the binaries, which corresponds to
the distribution with the blue line in the \textit{right} panel.

In Figure \ref{fig:relativeangle_bg}, we show the same distributions
for the models with the background potential (Model 1 to Model 3). The
same plot format and layout are used as in Figure
\ref{fig:relativeangle_MODEL0}. In these plots, different from Model
0, we make a distinction between cases where the binaries have
escaped (E, marked with the black circles) and cases where the binaries have
not escaped (NE, marked with the red squares).

There are several characteristic features found from the distributions for $\xi$ in 
both Figures~\ref{fig:relativeangle_MODEL0} and \ref{fig:relativeangle_bg}. 
\begin{enumerate}
		\item \label{feature1}The relative angles between
                  $S_{\rm fast}$ and the binaries (blue solid lines
                  and blue dots) are densely distributed at
                  $\xi\gtrsim120^{\circ}-135^{\circ}$ with peaks near
                  $\xi\simeq160^{\circ}$.

	\item \label{feature2}For cases with NE-binaries in Figure
	\ref{fig:relativeangle_bg}, the
          relative angles between the two single stars are more
          concentrated at higher $\xi$ (red square dots).  In addition, as the
          escape velocity $v_{\rm esc}$ increases (from Model 1 to
          Model 3), the red dots spread out over a wider range in $\xi$.

		\item \label{feature3}The relative angles $\xi$ for
                  $S_{\rm fast}$ and the binaries with respect to
                  $S_{\rm slow}$ are broadly distributed (black/green
                  dots and lines) compared to the distribution for
                  $\xi$ between $S_{\rm fast}$ and the binaries (blue
                  lines and dots), but more biased toward higher
                  values of $\xi$ with peaks at
                  $\xi\simeq110^{\circ}-120^{\circ}$.

\end{enumerate}

Those features above can be understood in terms of the relation
between the recoil velocities of the binaries and the relative angle
between the two single stars.  If the two single stars are ejected
with a wider angle (large $\xi$), it is more likely that the binary
gets a smaller recoil kick. As an extreme case, when two single stars are
ejected at the same speed in opposite directions (or
$\xi=180^{\circ}$), the final recoil velocity of the binary sums
to zero (as a result of the two kicks in opposite directions) so
that it remains where the last ejection event occurred.

We find feature \ref{feature1} in our simulations as a result of the
fact that kicks from more rapidly-moving stars tend to contribute more
to the final recoil velocity of the binary than slowly moving
stars. In order to better understand this feature
quantitatively, we use Equation 27 from \citet{RLP2017}
 (with $r_{\rm ej}\simeq0$), which relates the speeds of the binary to the relative
angle between the two single stars in a harmonic potential. The
equation gives the relative angles $\xi_{\rm fast,slow}$ between
$v_{\rm fast}$ and $v_{\rm slow}$ at the outer boundary of the
potential as follows,
\begin{align}
\label{eq:solution_vbrest_overall}
\mathcal{P}_{\rm binary}=\sqrt{\mathcal{P}_{\rm fast}^{2}+
	\mathcal{P}_{\rm slow}^{2}+2\mathcal{P}_{\rm fast}
\mathcal{P}_{\rm slow}\cos\xi_{\rm fast,slow}}\;,
\end{align}
where $\mathcal{P}_{\rm i}=m_{\rm i}\sqrt{(1/2) v_{\rm esc}^2+v_{i}^{2}}$ ($i$=binary, $S_{\rm fast}$ or $S_{\rm slow}$)
\footnote{Here, $v_{i}$ is the speed at $r=r_{\rm bg}$.}. 
However, in deriving this equation, the binary and the ejected single star are 
only distinguished by their masses. Therefore, the relative angles between $S_{\rm fast}$ 
and the binary $\xi_{\rm fast,binary}$ can be estimated by exchanging $\mathcal{P}_{\rm binary}$ 
with $\mathcal{P}_{\rm slow}$. Expressing $\xi$ in terms of the momenta, 
 \begin{align}
 \label{eq:angle_fast_binary}
 \cos\xi_{\rm fast,binary}=\frac{\mathcal{P}_{\rm slow}^{2}-
\left[\mathcal{P}_{\rm fast}^{2}+\mathcal{P}_{\rm binary}^{2}\right]}
{2\mathcal{P}_{\rm fast}\,\mathcal{P}_{\rm binary}}\;.
 \end{align}
We can first see that $\xi\geq90^{\circ}$ since the numerator is negative. 
Note that the case for $v_{\rm esc}=0$ corresponds to Model~0. 
In particular, for binaries and single stars which have escaped from the 
background potential, and given the velocity distributions for the binaries 
(Figure \ref{fig:velocity_binary}) and the single stars (Figure \ref{fig:velocity_singlestar}) 
and their ratio (Figure \ref{fig:velocity_ratio}), we can roughly estimate that 
$\xi_{\rm fast,binary}\simeq 140^{\circ}$ and $\geq90^{\circ}$.

In order to interpret feature \ref{feature2}, we introduce the maximum
angle $\xi_{\rm max}$ between two single stars required to give a
sufficiently high recoil kick to a binary that the binary can
escape to infinity.  In other words, if two single stars are ejected
with $\xi>\xi_{\rm max}$ at their given speeds, the final binary
will not gain enough kinetic energy to completely
escape. Therefore, we expect that $\xi_{\rm fast,slow}$ for the
NE-binary cases should be distributed at $\xi>\xi_{\rm max}$.  We note that
$\xi_{\rm max}$ is a function of the speeds of the two single stars ($v_{\rm
  fast}, v_{\rm slow}$) and the escape velocity $v_{\rm esc}$, namely,
$\xi_{\rm max}=\xi_{\rm max} (v_{\rm fast}, v_{\rm slow}, v_{\rm
  esc})$.

 Using $\xi_{\rm max}$, we can understand feature \ref{feature2} and try to provide 
 useful insight from an observational perspective. Using Equation 
 \ref{eq:solution_vbrest_overall} (or Equation 36 in \citealt{RLP2017} 
 with $r_{\rm ej}\simeq0$), 
 we impose the condition that at $r=r_{\rm bg}$ and $\xi=\xi_{\rm max}$, 
 $v_{\rm binary}= v_{\rm esc}$. More explicitly,
\begin{equation}
			\label{eq:esc_condition}
\sqrt{\mathcal{P}_{\rm fast}^{2}+\mathcal{P}_{\rm slow}^{2}+2\mathcal{P}_{\rm fast}
\mathcal{P}_{\rm slow}\cos\xi_{\rm max}}=\sqrt{\frac{3}{2}}m_{\rm binary}v_{\rm esc}\,.
\end{equation}
Again, the term on the left-hand side represents 
the final momentum of the binary at the last ejection event and the term on the 
right-hand side the minimum momentum of the binary necessary to travel from the origin 
to the outer boundary of the potential, and subsequently to spatial infinity. 

\begin{figure}
	\centering
	\includegraphics[width=8.5cm]{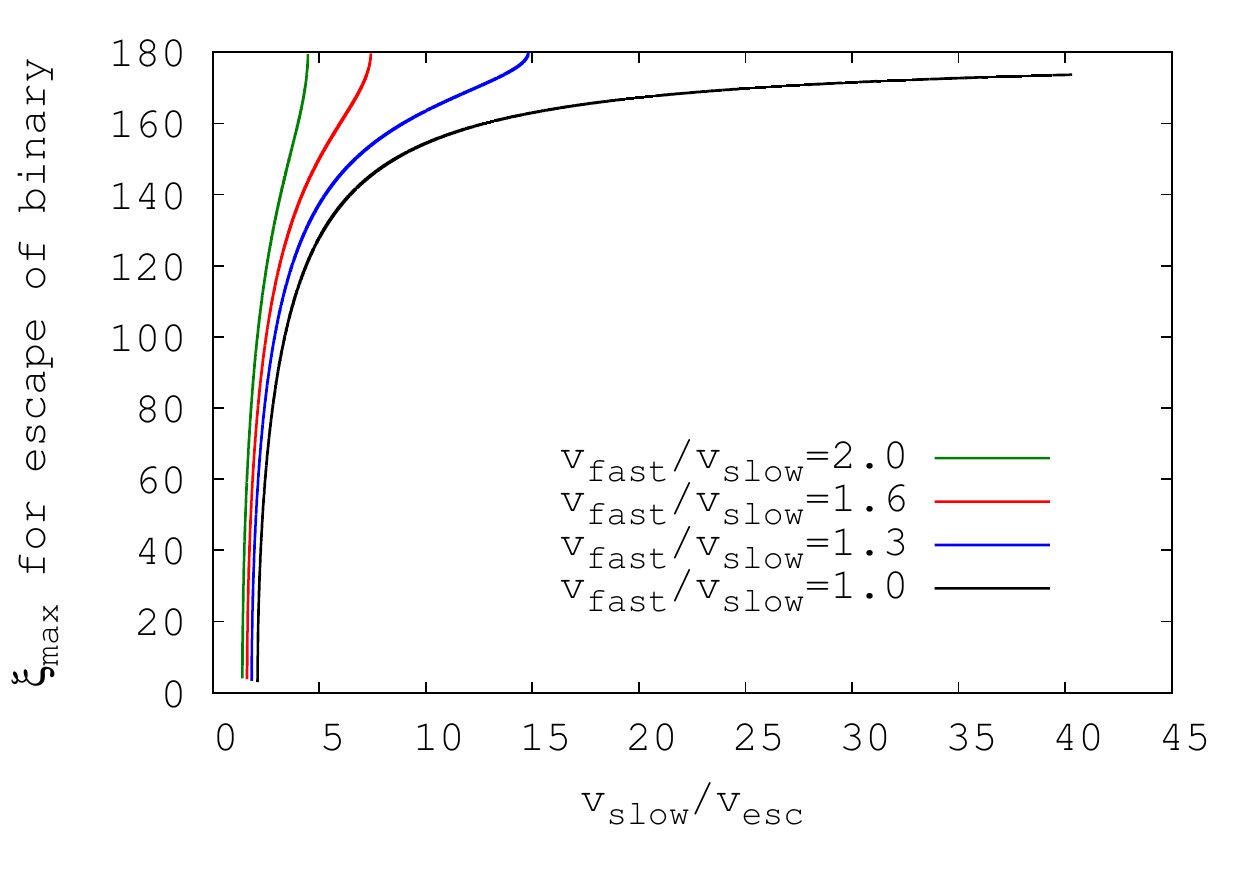}
	\includegraphics[width=8.5cm]{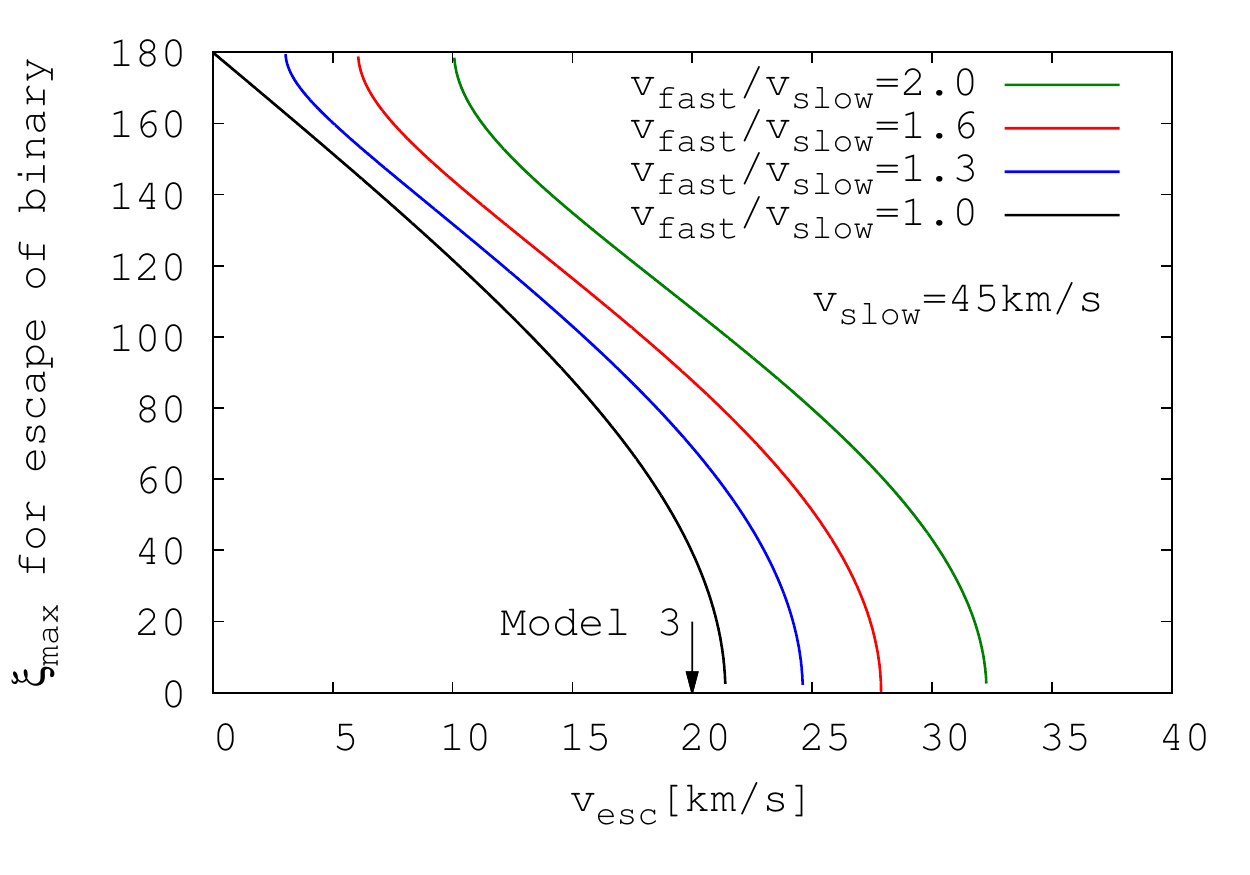}
	\caption{$\xi_{\rm max}$ estimated using Equation \ref{eq:esc_condition}.
		\textit{Upper} panel: $\xi_{\rm max}$ is expressed as a function 
		of $v_{\rm slow}$ in units of $v_{\rm esc}$ for various ratios 
		of $v_{\rm fast}/v_{\rm slow}$ (see Figure \ref{fig:velocity_ratio}). $\xi_{\max}$
		 increases rapidly up to $120^{\circ}-140^{\circ}$ at typical velocities for 
		 $S_{\rm slow}$ for Model 1 and 2. In the \textit{lower} panel, 
		  $\xi_{\rm max}$ is described in terms of $v_{\rm esc}$ assuming $v_{\rm slow}\sim 40\km\s^{-1}$. $\xi_{\rm max}$ decreases down to $50^{\circ}-90^{\circ}$ for
		   $v_{\rm esc}=20\km\s^{-1}$ in Model 3 indicated by the downward arrow 
(feature \ref{feature2}). }
	\label{fig:esc_minangle}
\end{figure}

In the \textit{upper} panel of Figure \ref{fig:esc_minangle}, we
estimate $\xi_{\max}$ (using Equation \ref{eq:esc_condition}) as a
function of $v_{\rm slow}$ in units of $v_{\rm esc}$ for various
ratios of $v_{\rm fast}/v_{\rm slow}$ (see Figure
\ref{fig:velocity_ratio}).  As shown in the figure, $\xi_{\max}$
increases rapidly at $v_{\rm slow}/v_{\rm esc}\lesssim5$. It
reaches $120^{\circ}-140^{\circ}$ at typical velocities for $S_{\rm
  slow}$ found in Model 1 and 2 (median velocities of $v_{\rm
  slow}/v_{\rm esc}\sim3$) for all speed ratios, which explains
well the feature \ref{feature2}. However, considering smaller values
for $v_{\rm slow}/v_{\rm esc}\lesssim2$ in Model 3, it is hard to read
$\xi_{\rm max}$ from the \textit{upper} panel. Hence we additionally provide
a plot (\textit{bottom} panel), which shows $\xi_{\rm max}$ as a
function of $v_{\rm esc}$ assuming $v_{\rm slow}\sim 45\km\s^{-1}$. We
can see that for $v_{\rm esc}=20\km\s^{-1}$ in Model 3 (indicated with
the downward arrow), $\xi_{\rm max}$ decreases down to
$40^{\circ}-120^{\circ}$, which accounts for the wide range of 
dots in Figure~\ref{fig:relativeangle_bg}, or feature \ref{feature2}.

From an observational perspective, we expect that the analytic
relations above (as derived in the analytic paper) and the statistical
properties of the relative angles found from our simulations can help
restrict the region in parameter space where we need to look in order to 
find an unknown related object given observations of some other runaway stars. Note
that, in the derivation of the above relation, the three bodies are
only identified by their masses, meaning that the relation can be
applied to any kind of isolated system that evolves to produce 3-body outcomes 
(two ejected systems and one left-over system), not
necessarily only to the 2+1+1 outcome.

\section{Discussion and Summary}
\label{sec:discussion_summary}

We have investigated the formation of runaway stars during binary -
binary encounters in a background potential via dynamical
ejections. To this end, we have performed numerical scattering
experiments for various potential models, under the assumption that the
potential remains static in time.  In order to understand the effects
of the background potential on the statistical properties of the
outcomes, we explored a range of different depths for the potential:
($\rho,~M_{\rm bg},~v_{\rm esc}$) = (0, 0, 0) for Model 0, (1,
$3000\Msol,~6.3\km\s^{-1}$) for Model 1, (1,
$10000\Msol,~9.2\km\s^{-1}$) for Model 2 and (1000,
$3000\Msol,~20\km\s^{-1}$) for Model 3. The parameters of the models
are summarized in Table \ref{tab:modelparameter2}.

We have found that, for encounters of two tightly bound binaries, the
background potential does not have an appreciable effect on the
outcome formation probabilities. However, in the presence of the potential,
two distinctive populations of binaries, with different orbital
properties, emerge depending on whether they could escape or
not. Upon analyzing our results, we have shown that the escape velocity of
the potential can play a role as a good indicator to identify these two
populations. We also found that the relative angles between stars have a
dependence on the depth of the potential, or the escape velocity.

We summarize the effects of the background potential as follows:

\begin{enumerate}
	
	\item We find that for the encounters of two tightly bound
          binaries, the outcome
          fractions have no appreciable dependence on the depth of the
          background potential. This is because the stellar
          interactions take place where the mutual gravitational
          potentials between stars are dominant over the gravitational
          background potential. Additionally, during the interactions,
          the ejection velocities of single stars are typically higher
          than the escape velocities from the background potential so
          they easily escape. However, for encounters involving wide
          binaries with low ejection velocities, they may experience
          subsequent interactions while they are oscillating inside
          the potential \citep{Ryu+2017}.

	\item As the background potential deepens, a larger number of
          binaries cannot escape (NE-binaries), resulting in the
          emergence of two distinct populations of binaries. When the
          escape velocities are comparable to the typical velocities
          of the binaries (see Figure \ref{fig:velocity_binary}) as in
          Model 3, $v_{\rm binary}\simeq v_{\rm esc}\simeq 20\km
          \s^{-1}$, around 40\% of the binaries cannot escape, almost
          the same fraction as that of the escaped binaries
          (E-binaries). More importantly, these two populations have
          different orbital parameters: the semimajor axes of the
          escaped binaries tend to be shorter. Their velocity
          distributions extend up to $v\simeq80\km \s^{-1}$
          whereas the velocities for $S_{\rm slow}$ are limited by the
          escape velocities. Additionally, as the potential gets
          deeper, the gravitational pull from the potential more
          significantly slows down the binaries, resulting in a 
          shift of the velocity distributions for the escaped binaries
          towards lower $v$.  Here, we emphasize that the escape
          velocity plays a role as a good measure to gauge the effects
          of the background potential.

	\item In contrast with the binaries, we could not find
          any noticeable effects of the background potential on the
          velocities of the ejected single stars. This is essentially
          due to the fact that the velocities of the rapidly-moving
          stars barely change because their kinetic energies are much
          larger than the background potential energy.

	\item We find that the escape times (the time taken for stars
          to cross the potential boundary from the onset of the
          encounter) for Model 1 and Model 2 are $\sim10^{4}\yr$ and
          that for Model 3 is around a few $1000\yr$.  From the
          difference in the escape times between the two single stars, we
          confirm that it is more likely that stars ejected at higher
          speeds are the ones to escape first.
	
	\item The relative angles between stars ($S_{\rm
          fast}$-$S_{\rm slow}$, $S_{\rm fast}$ - binary and $S_{\rm
          slow}$ - binary) are generally large, $\xi\simeq
          110^{\circ}-120^{\circ}$ independent of the presence of the
          background potential. More interestingly, the angle $\xi$ between
          $S_{\rm fast}$ and the binary (blue circles in Figure
          \ref{fig:relativeangle_MODEL0} and Figure
          \ref{fig:relativeangle_bg}) is mostly concentrated at
          $\xi>120-135^{\circ}$, while the $\xi$ values between $S_{\rm fast}$ and
          $S_{\rm slow}$ when the final binaries remain bound to the
          potential (red squares in in Figure
          \ref{fig:relativeangle_MODEL0} and Figure
          \ref{fig:relativeangle_bg}) are distributed at larger
          $\xi$. Moreover, the distributions stretch out over a wider
          range of $\xi$ as the potential gets deeper. We have explained
          these trends using the term $\xi_{\rm max}$ defined in Equation
          \ref{eq:esc_condition}.

	\item In summary, we have discussed the effects of a
          background potential on the formation of runaway stars
          and the statistical properties of the orbital parameters. We
          find that in the presence of the background potential, the
          outcomes have different orbital properties depending on
          whether or not they have escaped from the potential. From
          our study, we find that stars moving faster than the escape
          velocities, as is the case for runaway stars, are less affected by the
          background potential. Furthermore, we find some constraints
          on the relative angles between the ejected objects, which we expect to be
          observationally useful for identifying related runaway stars.
\end{enumerate}

\vspace{0.5cm}

\section*{Acknowledgements}

Results in this paper were obtained using the
high-performance LIred computing system at the Institute for Advanced
Computational Science at Stony Brook University, which was obtained
through the Empire State Development grant NYS \#28451.




\bibliographystyle{mnras}




\appendix
\section{The Cumulative distributions at $t=1,~2,~3$ and $4\Myr$.}
\label{appendix:cumulative}

\begin{figure*}
	\centering
	\includegraphics[width=7.8cm]{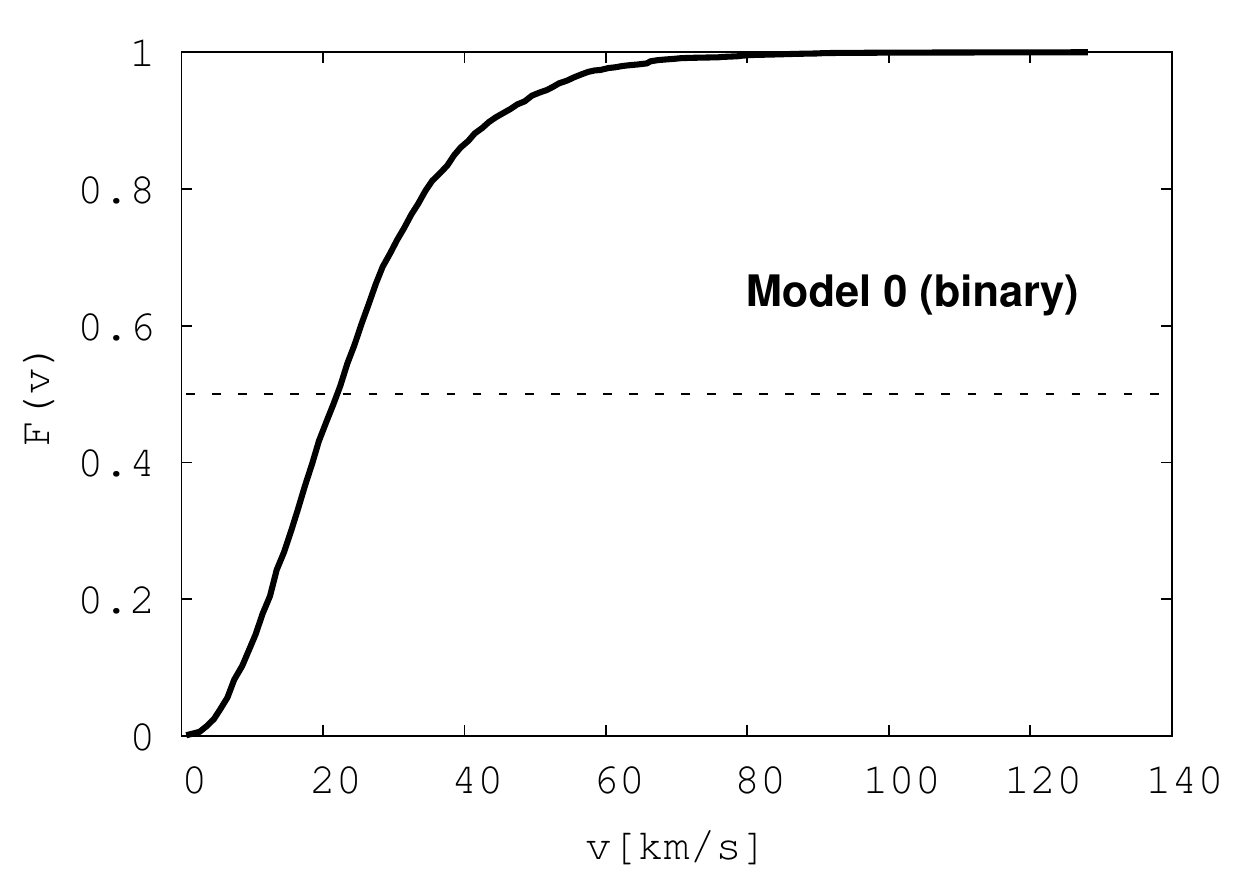}			
	\includegraphics[width=7.8cm]{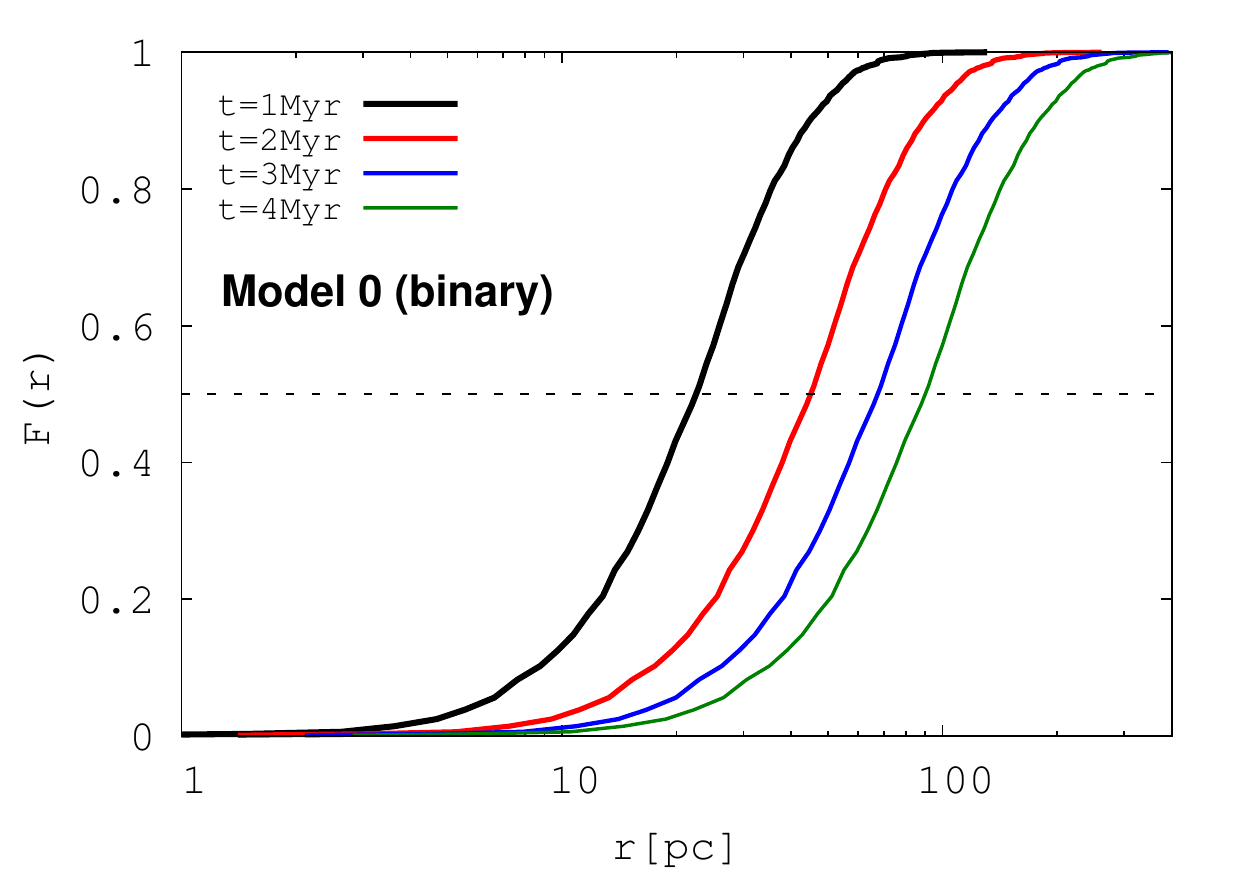}					\includegraphics[width=7.8cm]{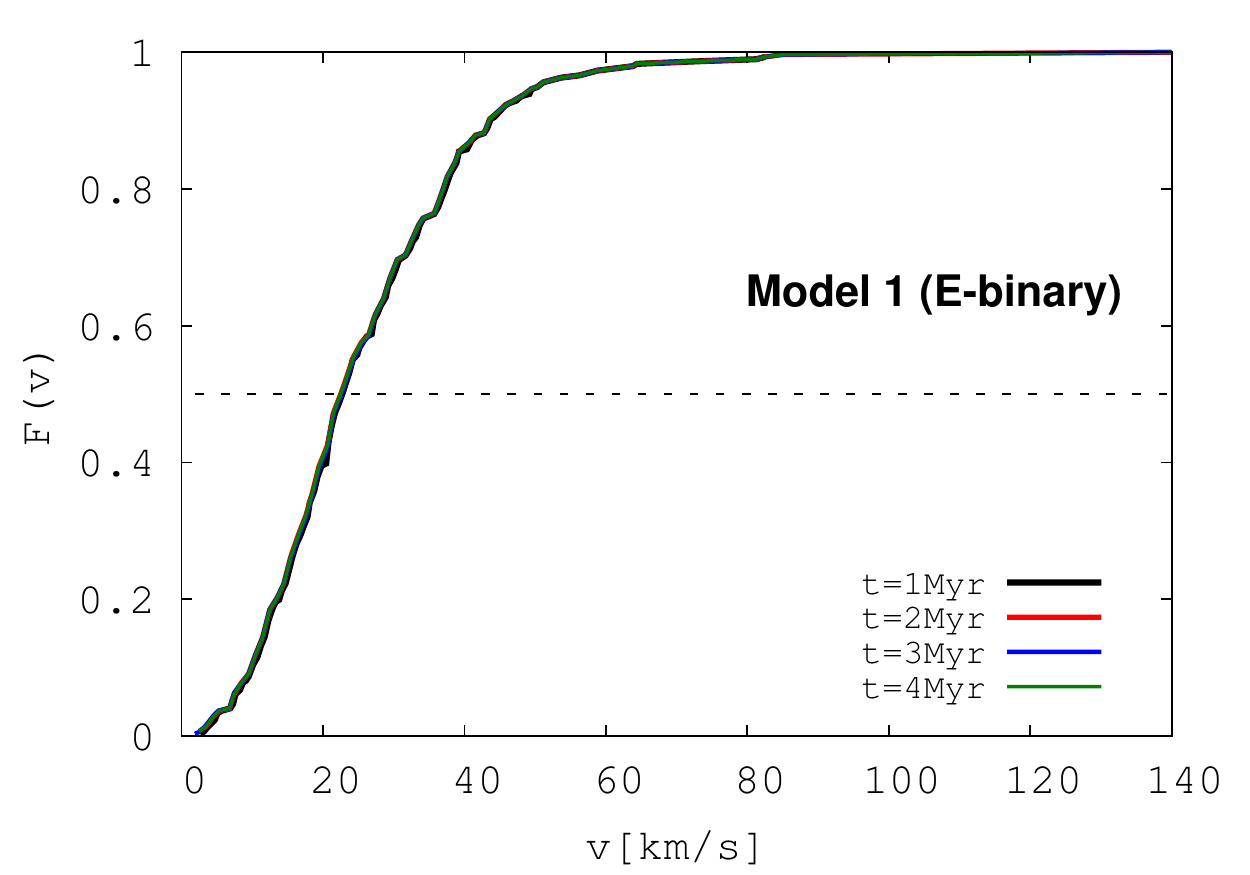}			
	\includegraphics[width=7.8cm]{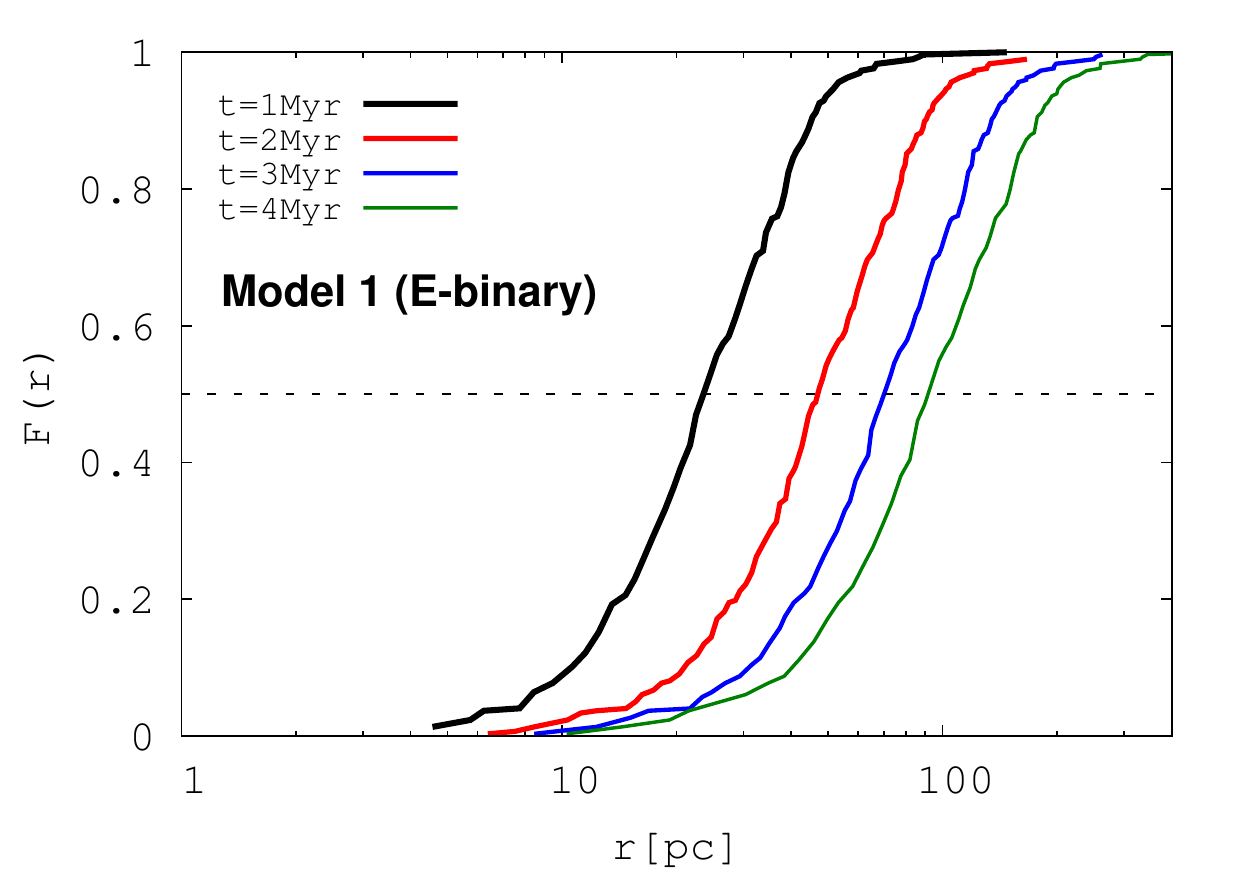}					\includegraphics[width=7.8cm]{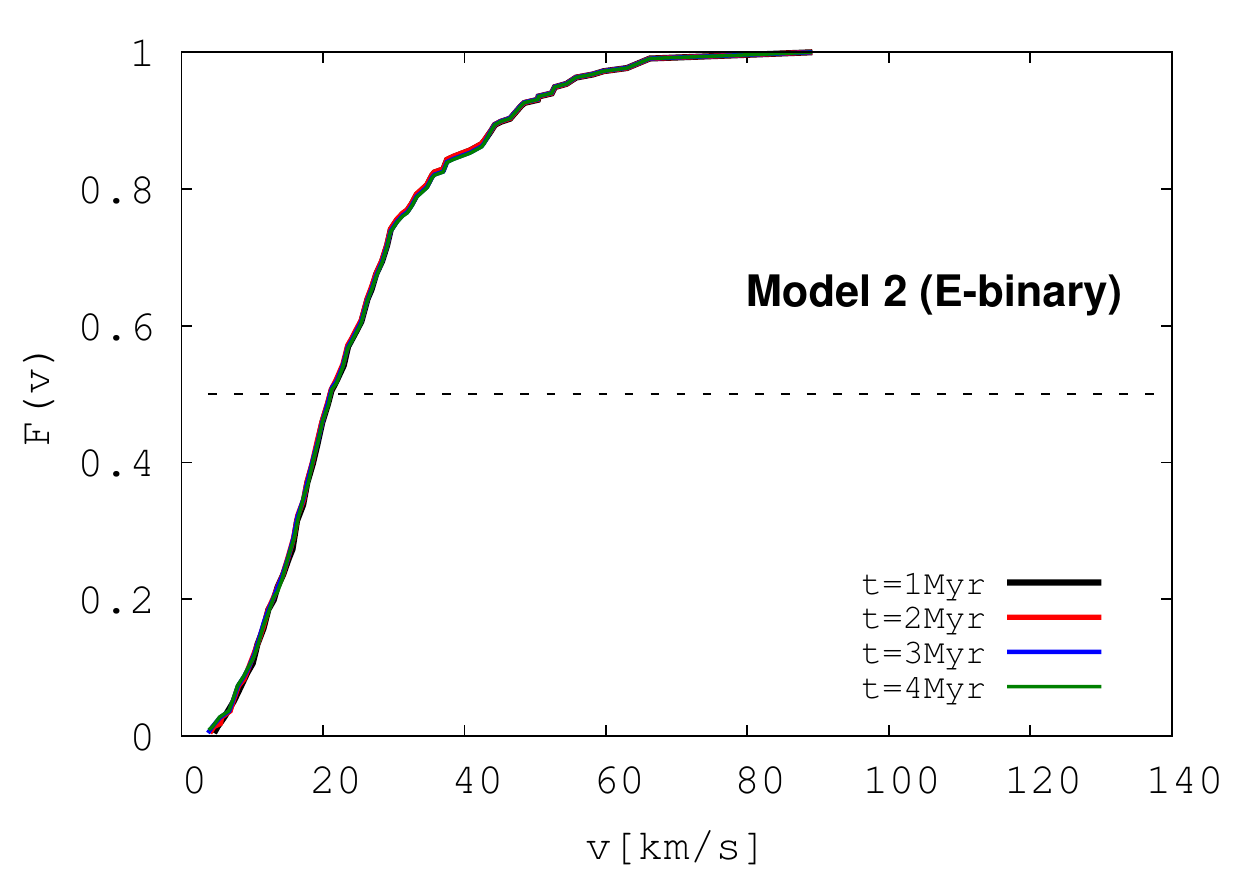}			
	\includegraphics[width=7.8cm]{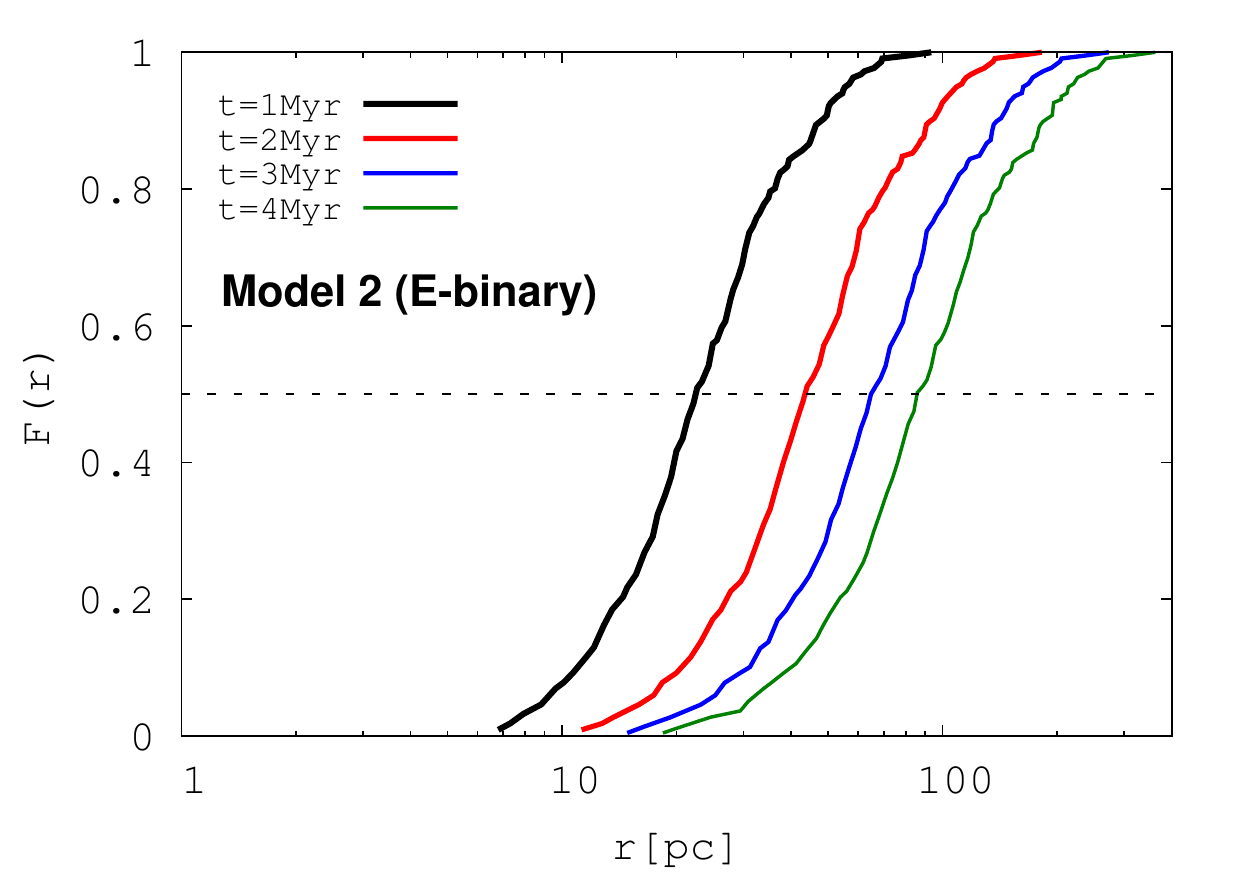}					\includegraphics[width=7.8cm]{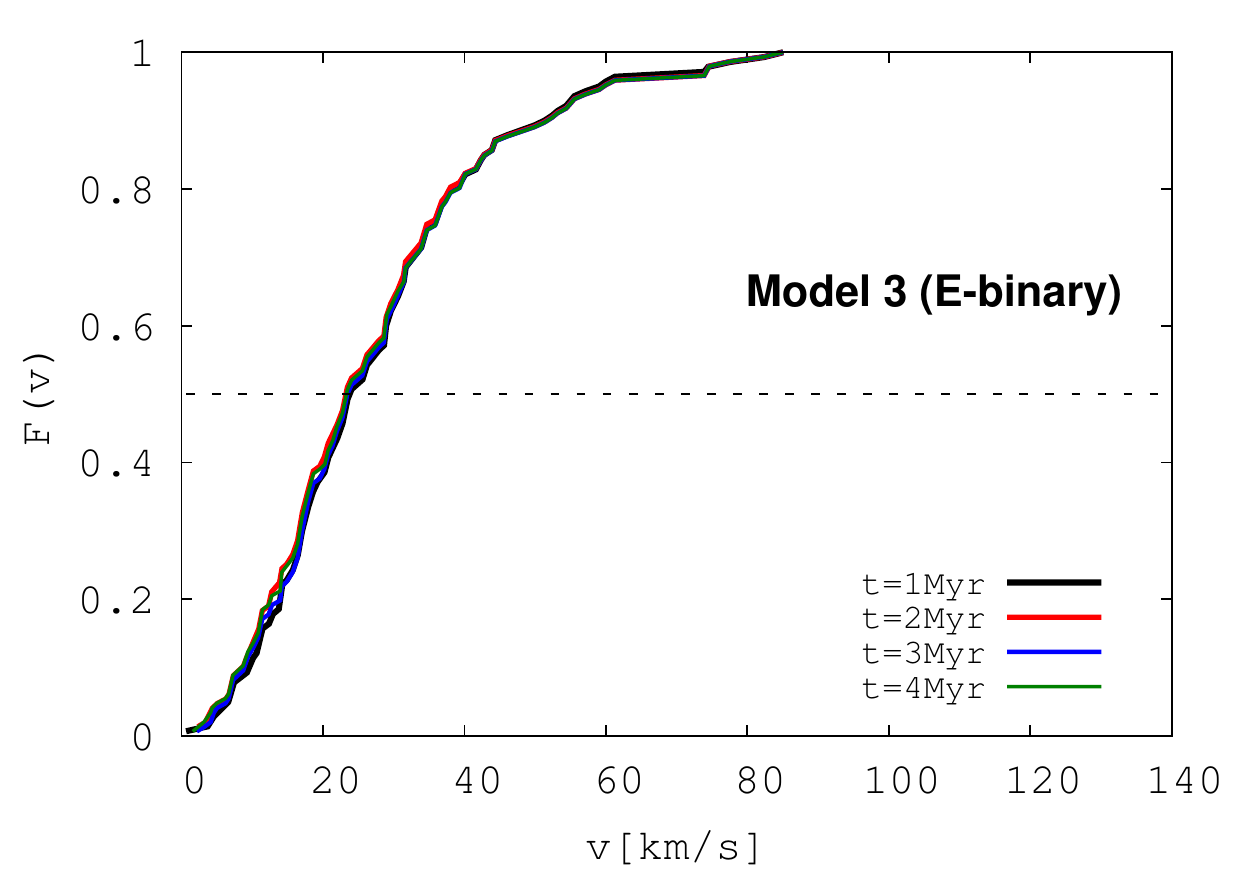}			
	\includegraphics[width=7.8cm]{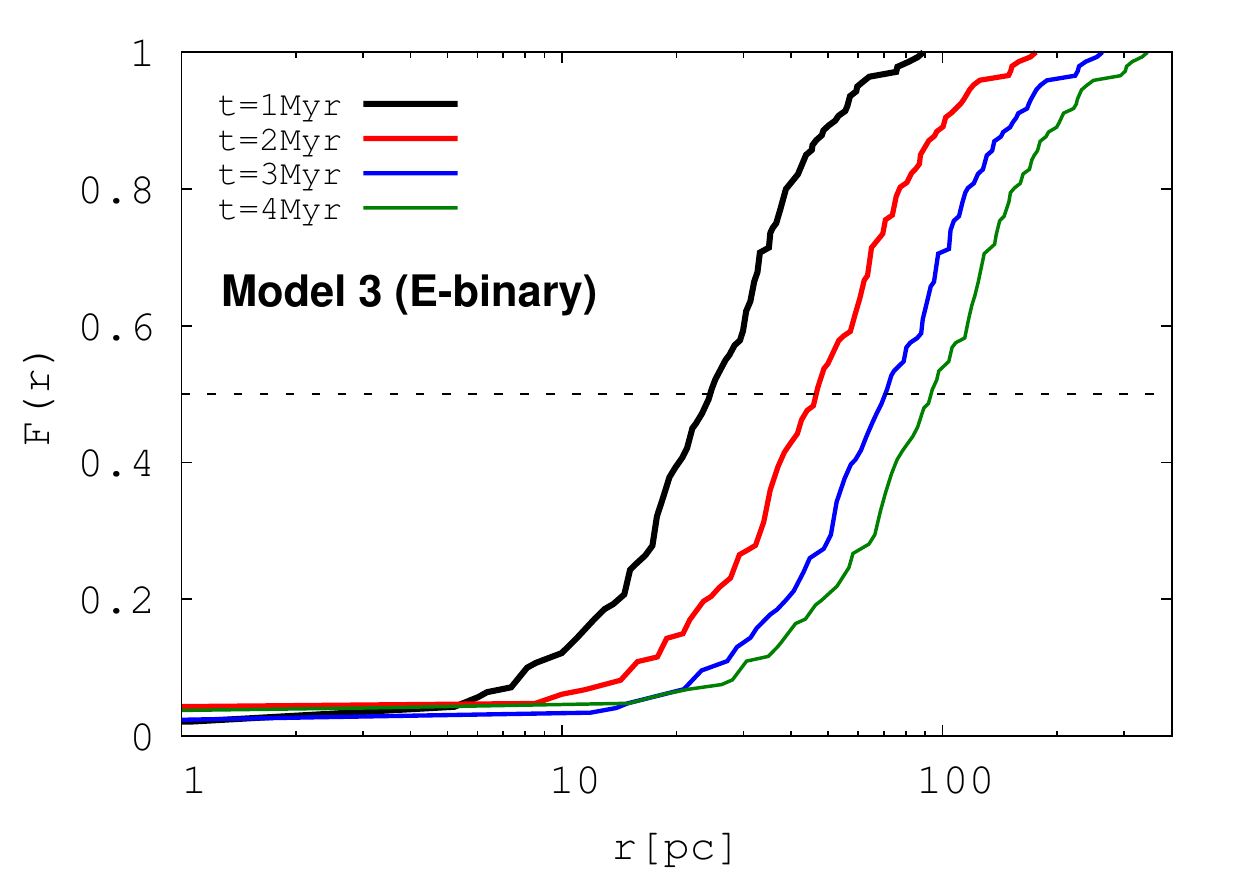}				
	\caption{\textit{Binary} : The cumulative distributions of the final velocities (\textit{left} column) and the radial distances from the system CM (\textit{right} column) for the escaped binaries (E-binaries)
		for all the models at $t=1\Myr$ (black), $2\Myr$ (red), $3\Myr$ (blue) and $4\Myr$ (green) from the drop-in time. The escape time scales (see Figure \ref{fig:ejectiontime}) are shorter than $\sim$Myr. For Model 0, we assume that the final binaries would keep propagating at the final velocities $v_{\rm final, binary}$ outward from the system CM after being isolated from other substellar systems, i.e., $r(t)=v_{\rm final,binary}t$. The horizontal dotted line indicate the median values.}
	\label{fig:cumulative_binary}
\end{figure*}

\begin{figure*}
	\centering
	\includegraphics[width=7.8cm]{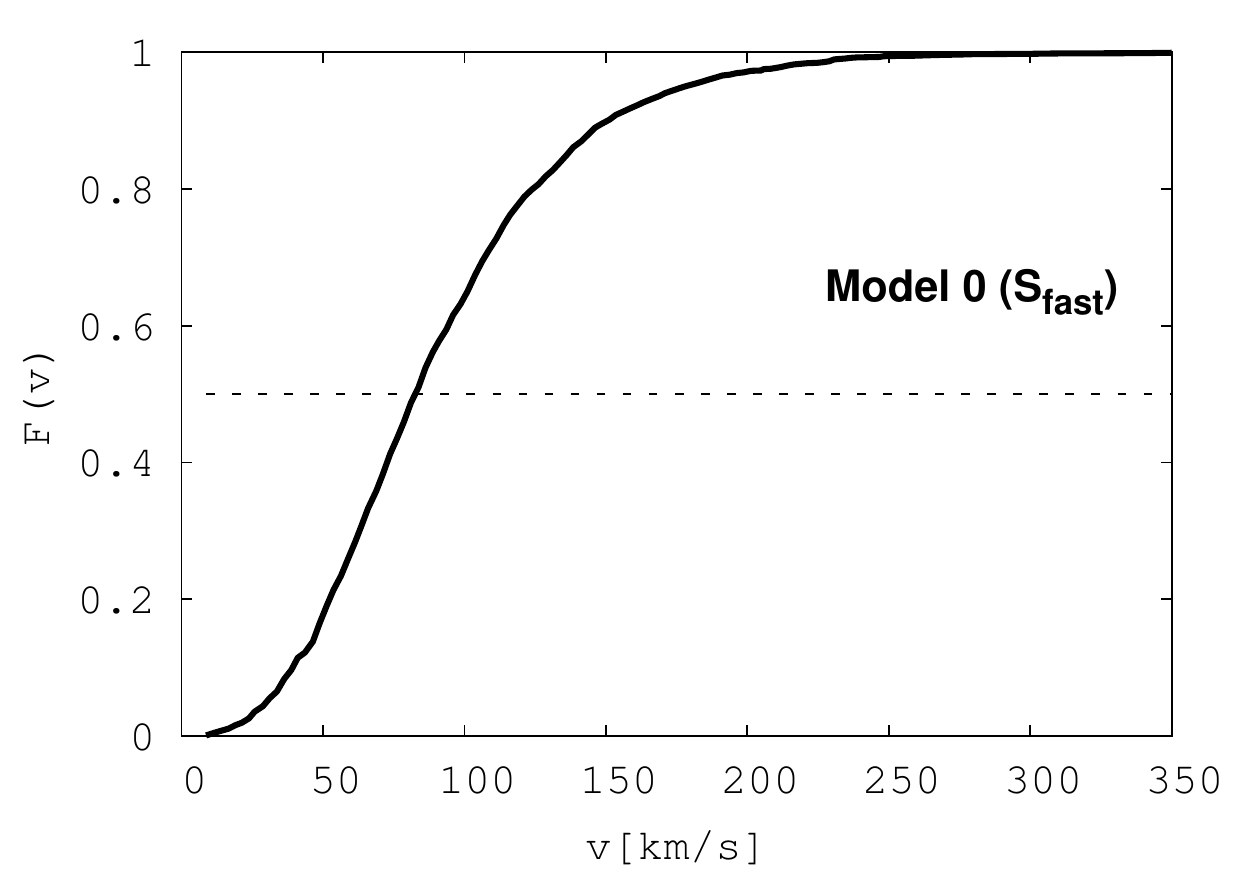}			
	\includegraphics[width=7.8cm]{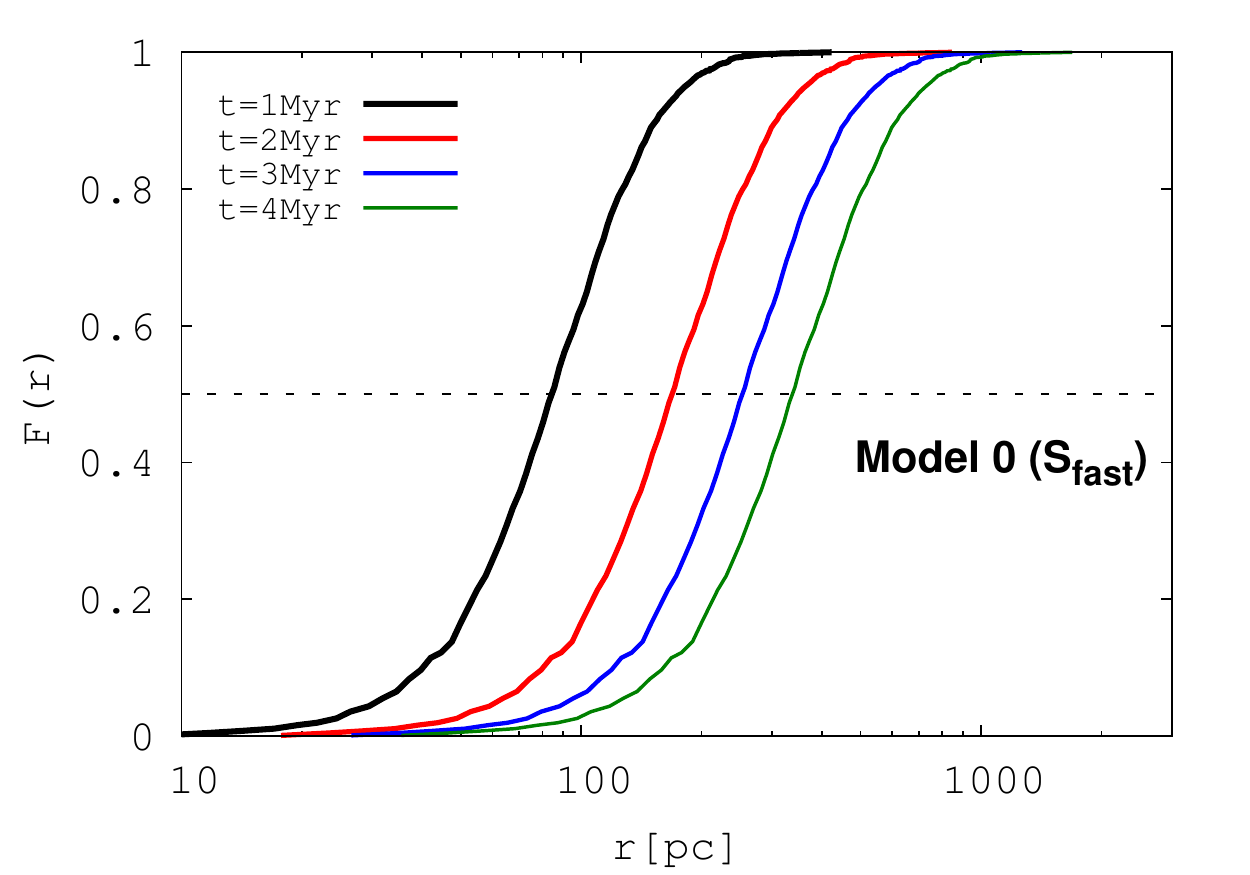}					\includegraphics[width=7.8cm]{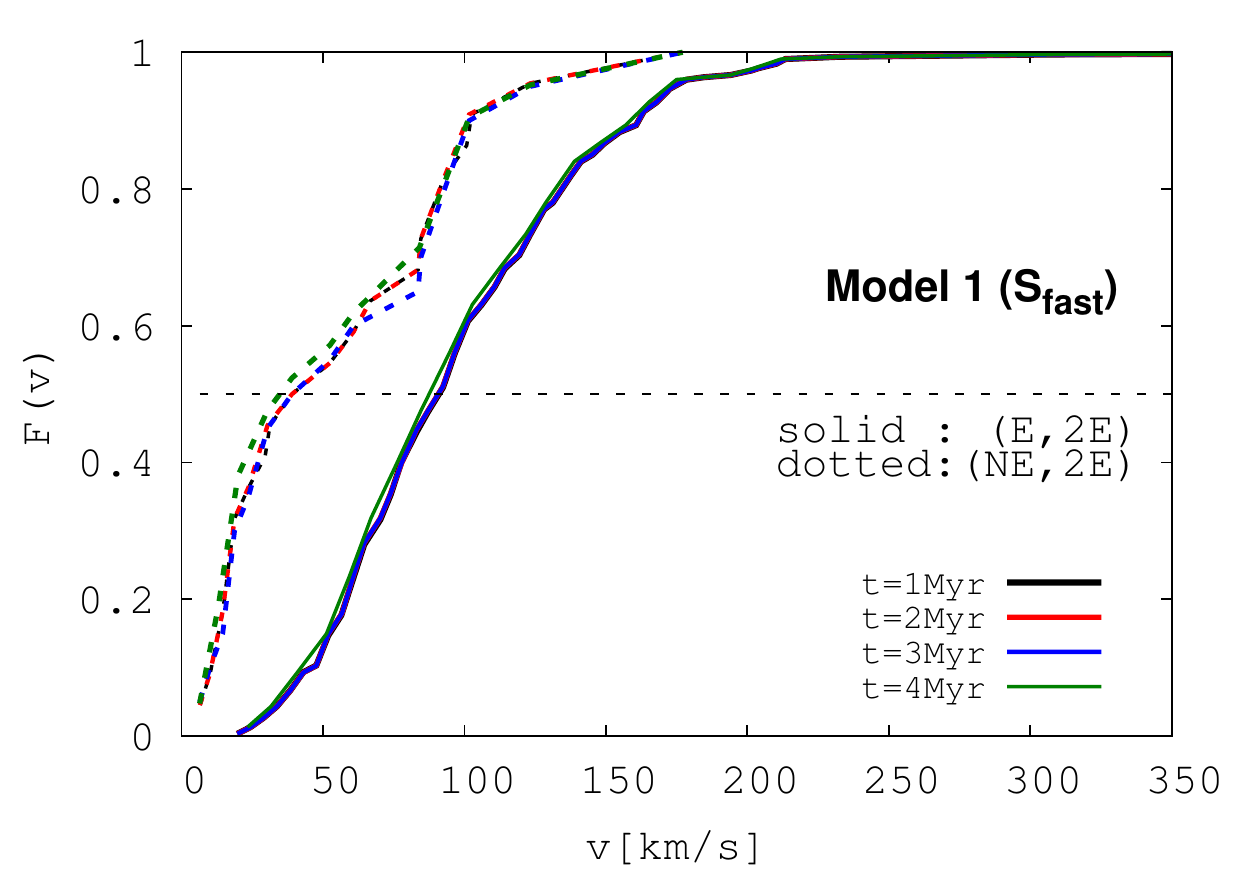}			
	\includegraphics[width=7.8cm]{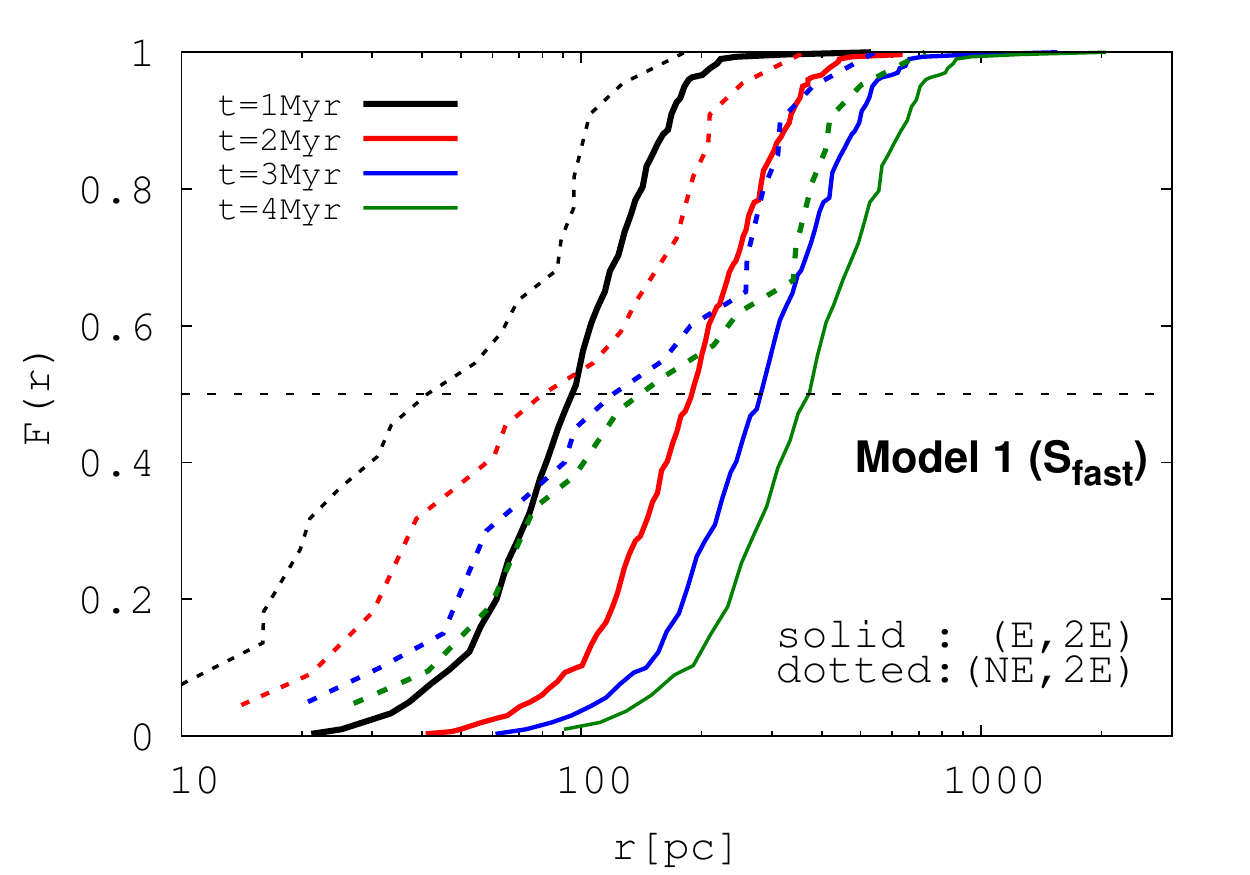}					\includegraphics[width=7.8cm]{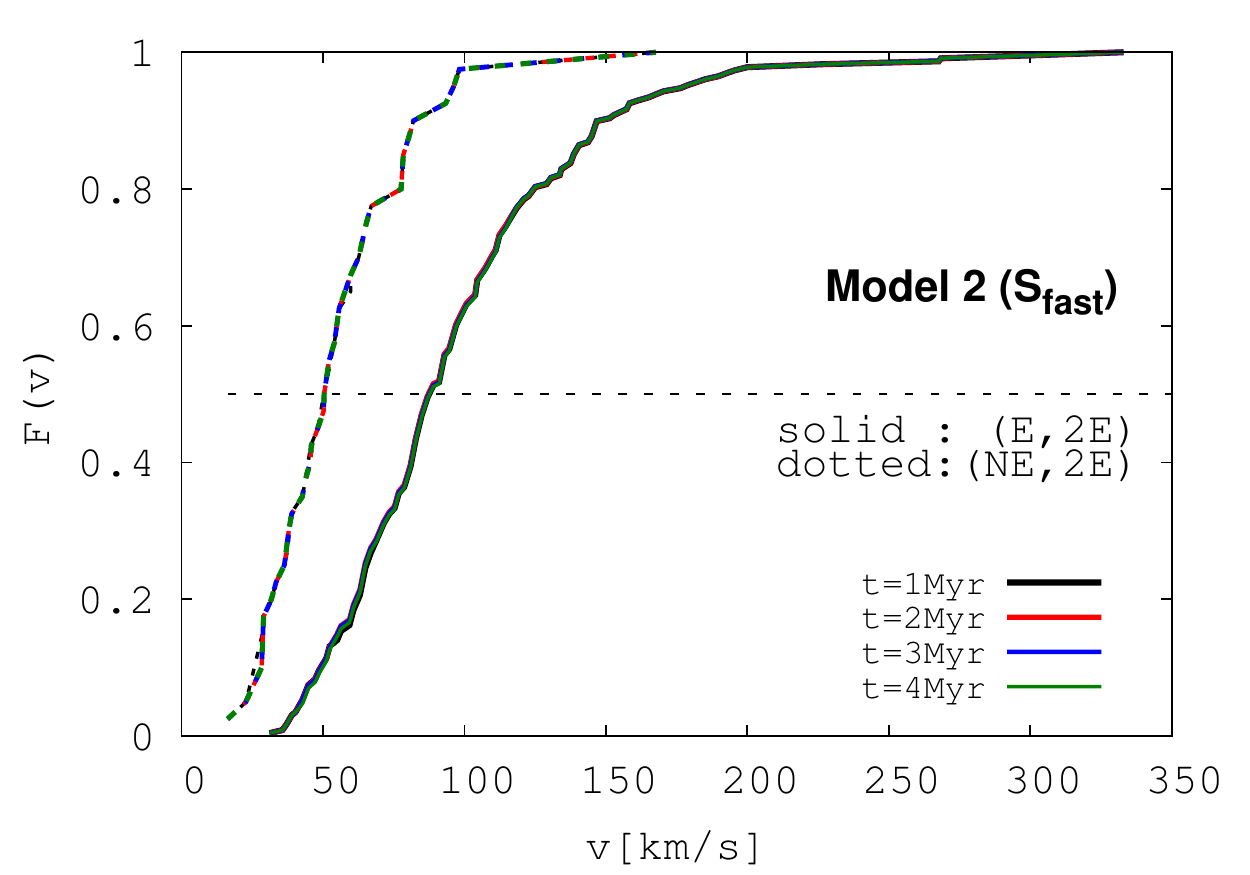}			
	\includegraphics[width=7.8cm]{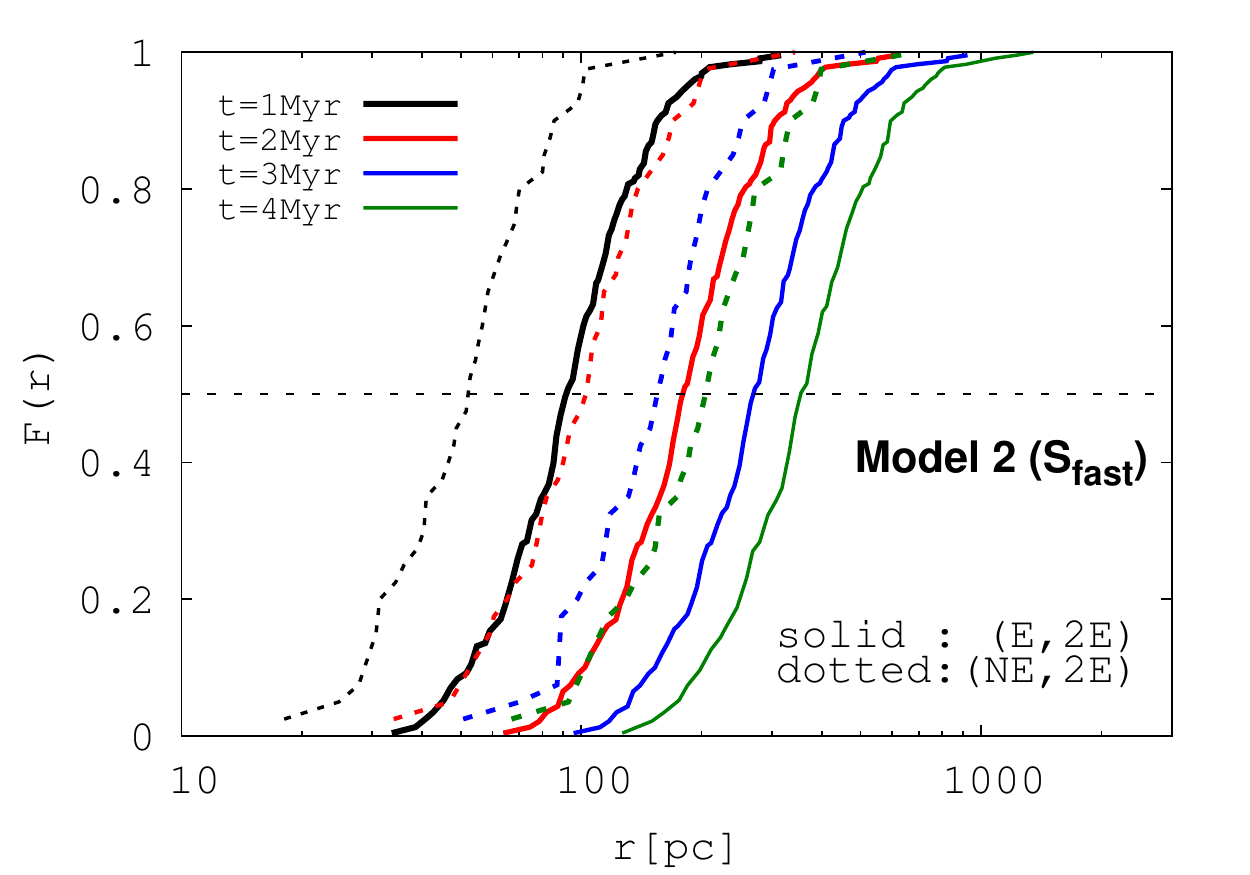}					\includegraphics[width=7.8cm]{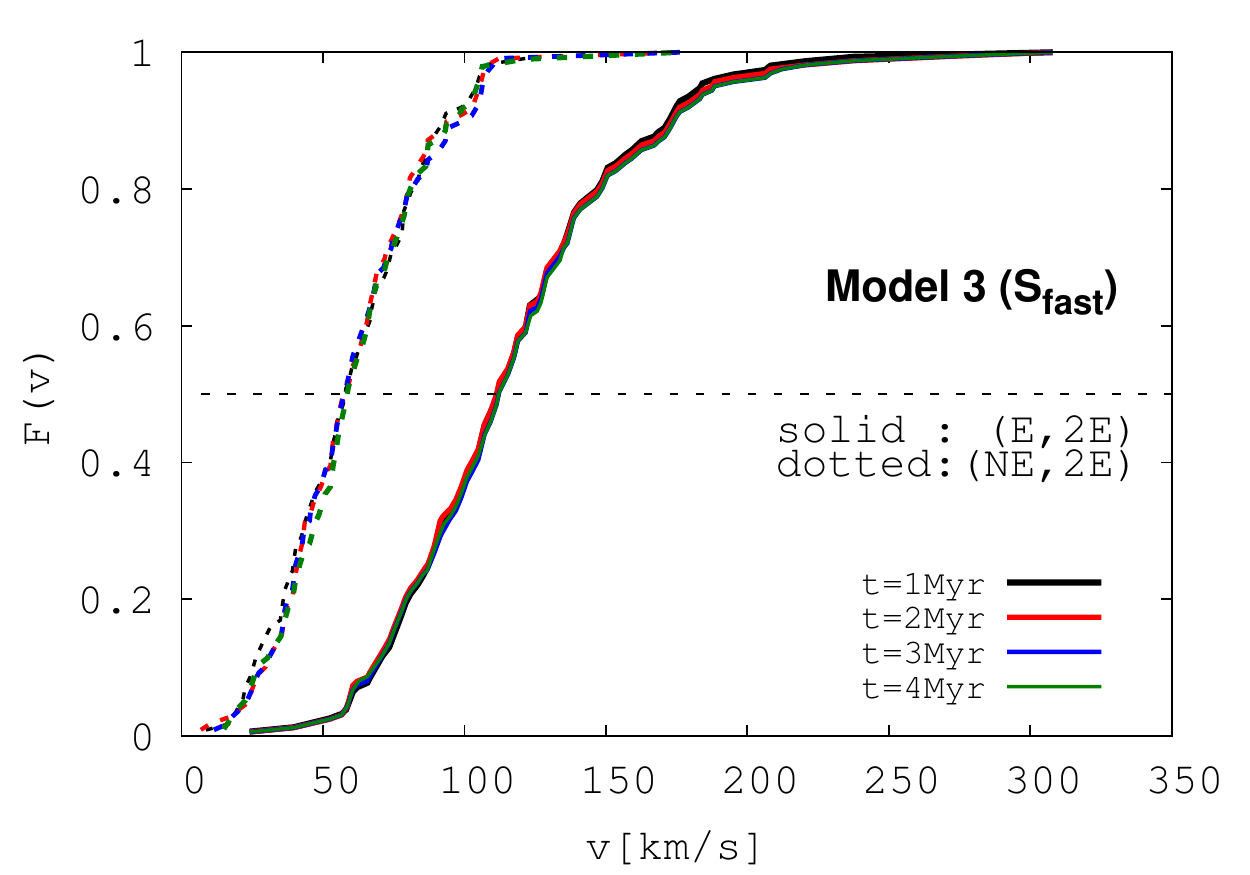}			
	\includegraphics[width=7.8cm]{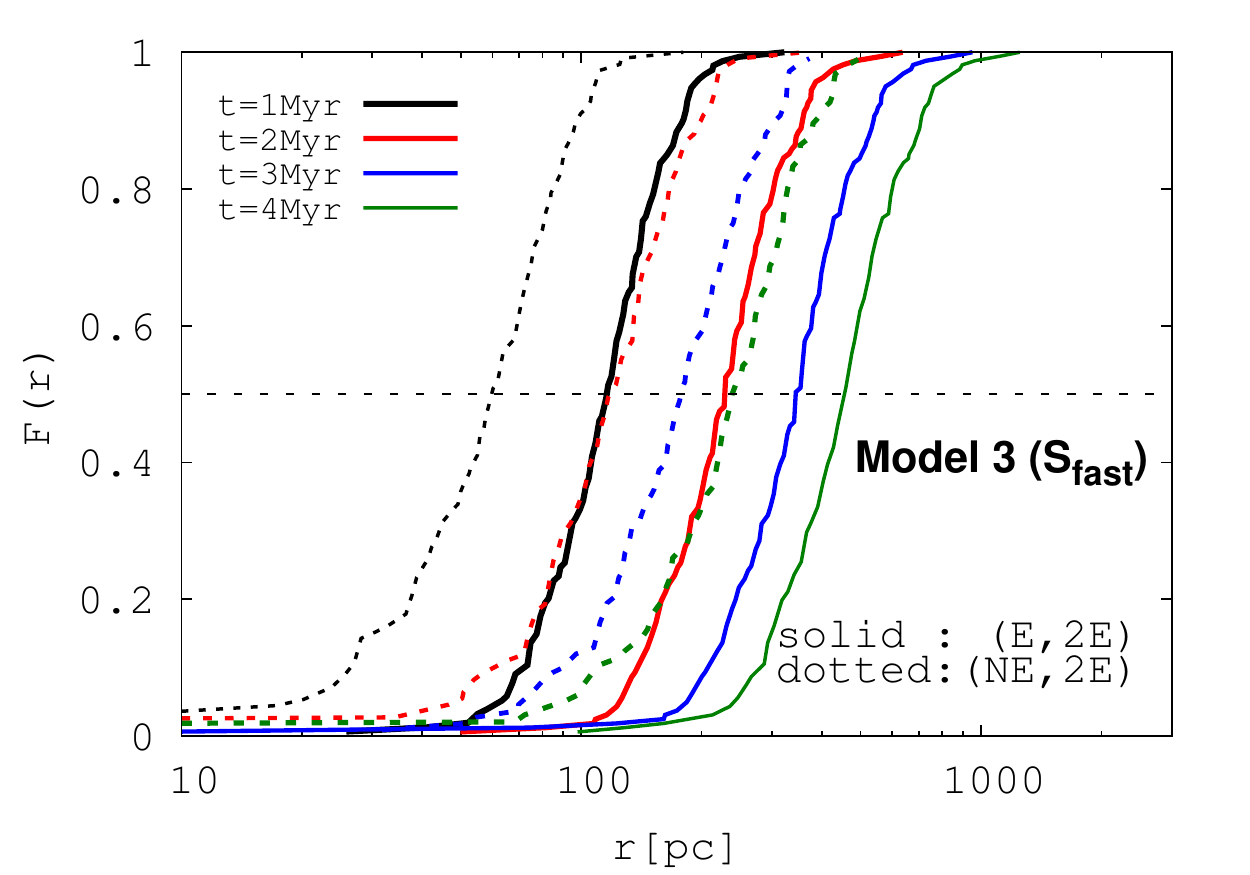}				
	\caption{\textit{faster single star} : The cumulative distributions of the final velocities (\textit{left} column) and the radial distances from the system CM (\textit{right} column) for the faster single stars ($S_{\rm fast}$) for all the models at $t=1\Myr$ (black), $2\Myr$ (red), $3\Myr$ (blue) and $4\Myr$ (green) from the drop-in time. We only consider the case where two single stars have completely escaped from the potential (2E). We use different line types to make a distinction between the cases with the E-binaries (solid lines) and NE-binaries (dotted lines). We use same line colors for different $t$ as in Figure \ref{fig:cumulative_binary}. The escape time scales (see Figure  \ref{fig:ejectiontime}) are shorter than $\sim$Myr. Same as Figure \ref{fig:cumulative_binary}, we estimate the spatial distance with $r(t)=v_{\rm final,fast}t$. The horizontal dotted line indicate the median values.}
	\label{fig:cumulative_S_fast}
\end{figure*}

\begin{figure*}
	\centering
	\includegraphics[width=7.8cm]{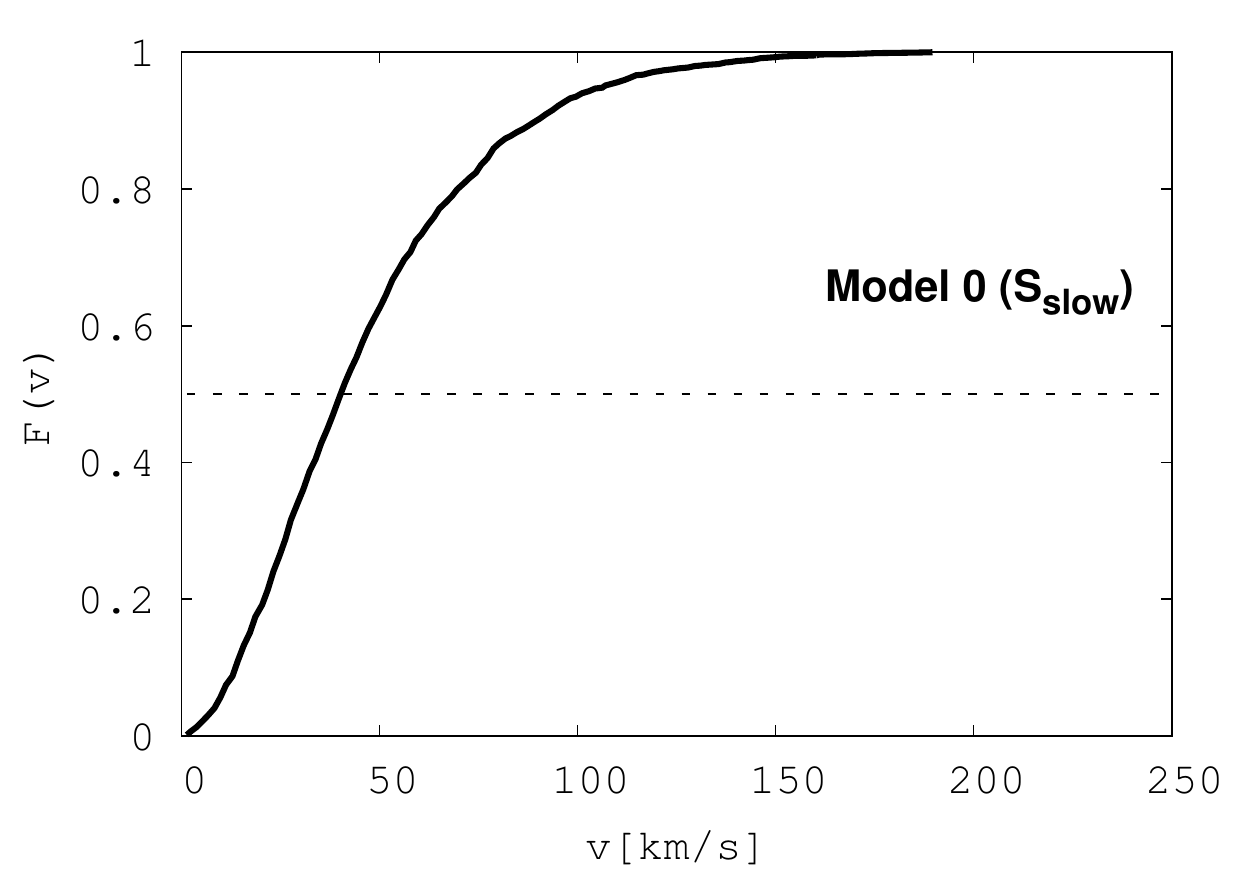}			
	\includegraphics[width=7.8cm]{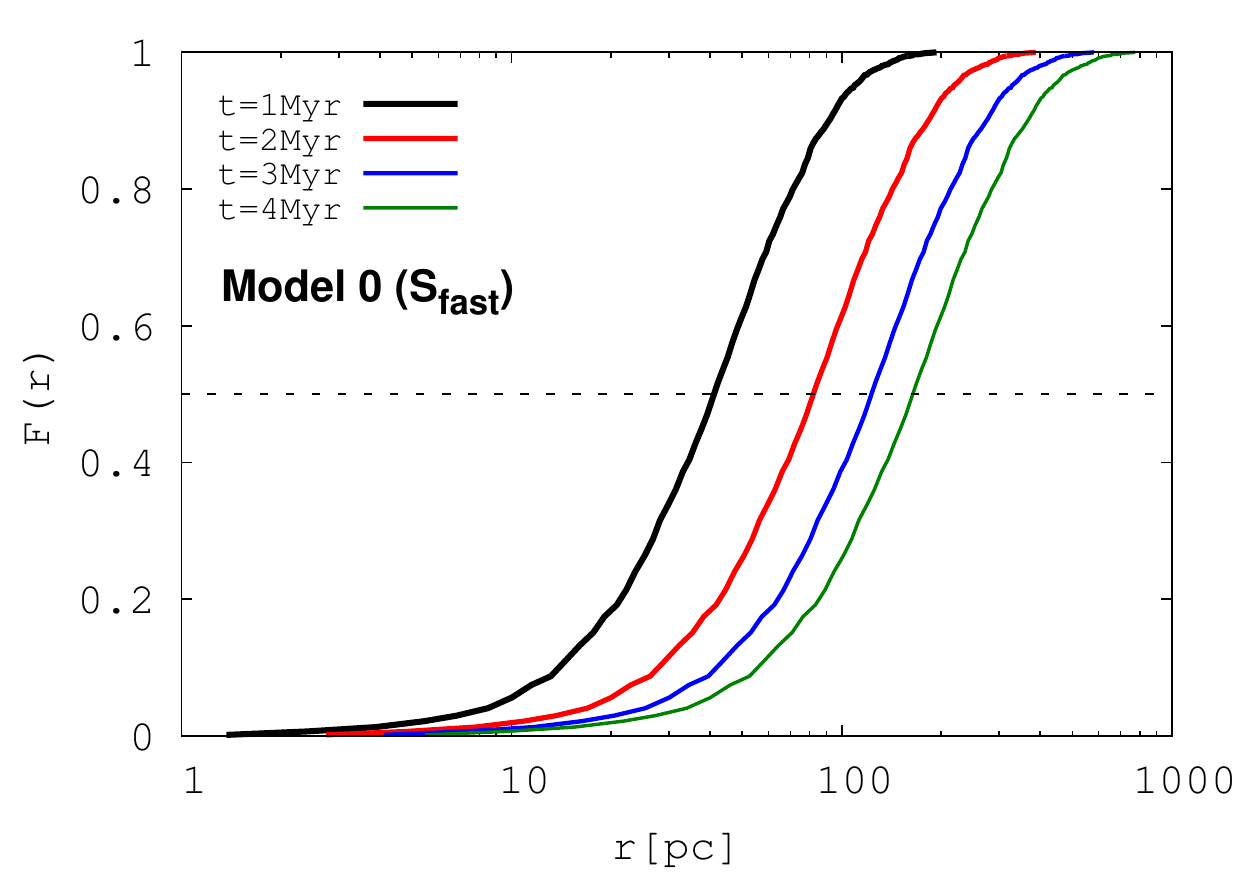}					\includegraphics[width=7.8cm]{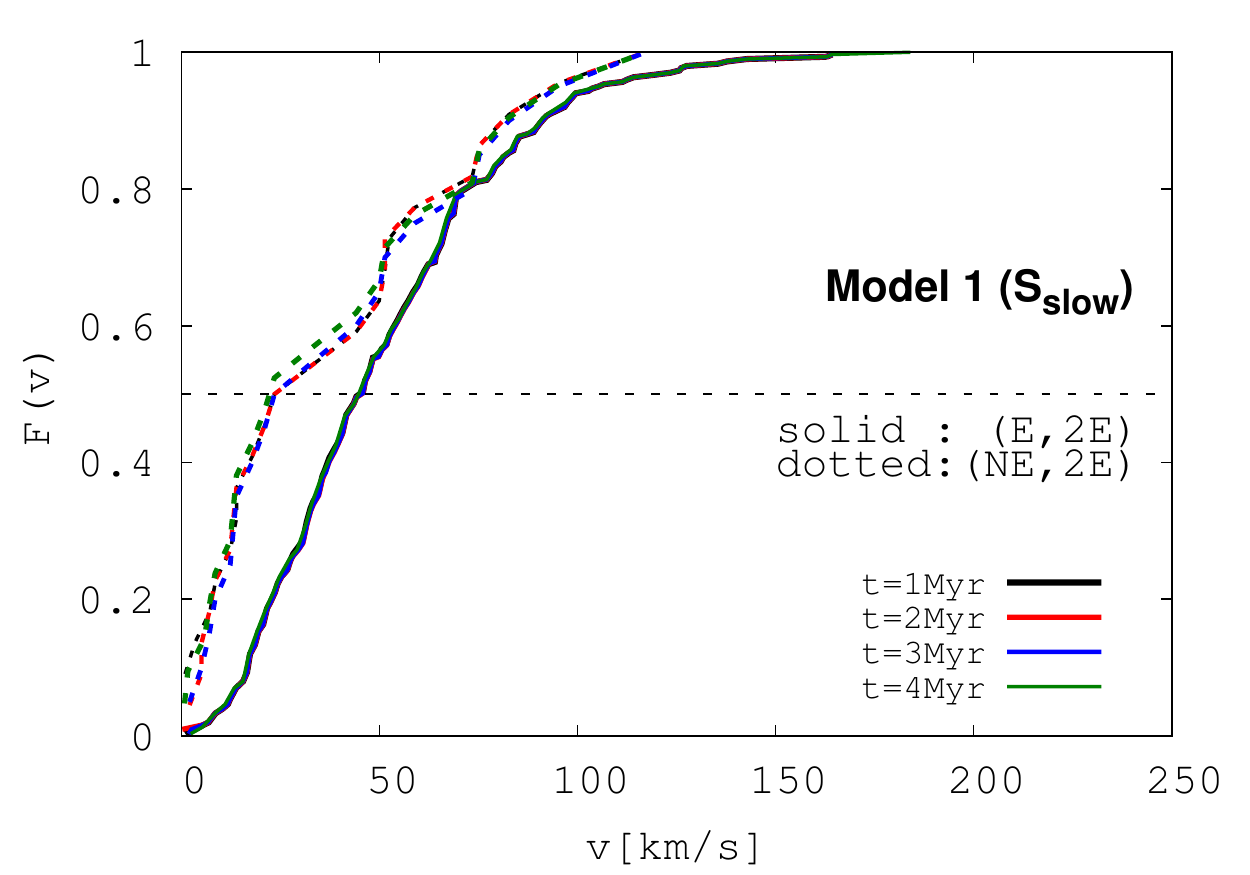}			
	\includegraphics[width=7.8cm]{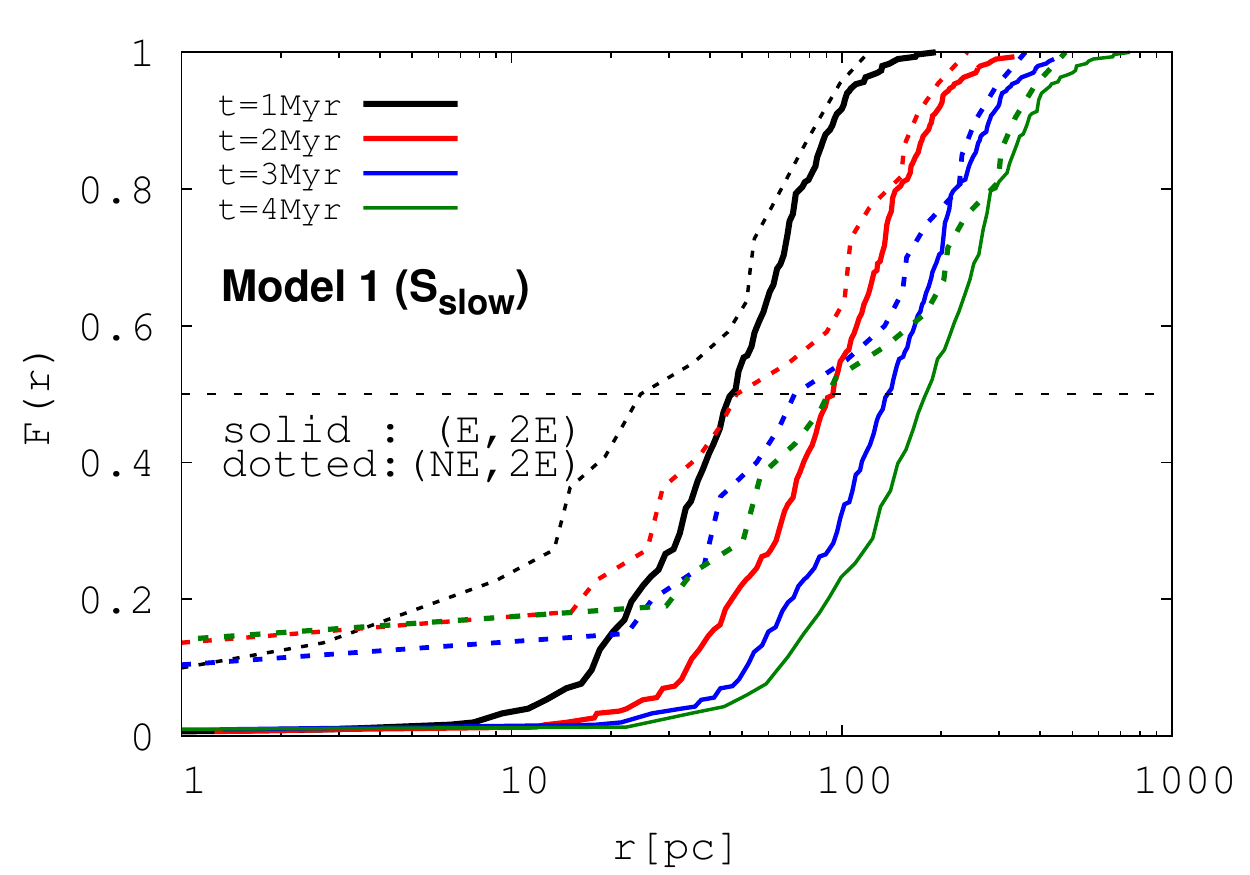}					\includegraphics[width=7.8cm]{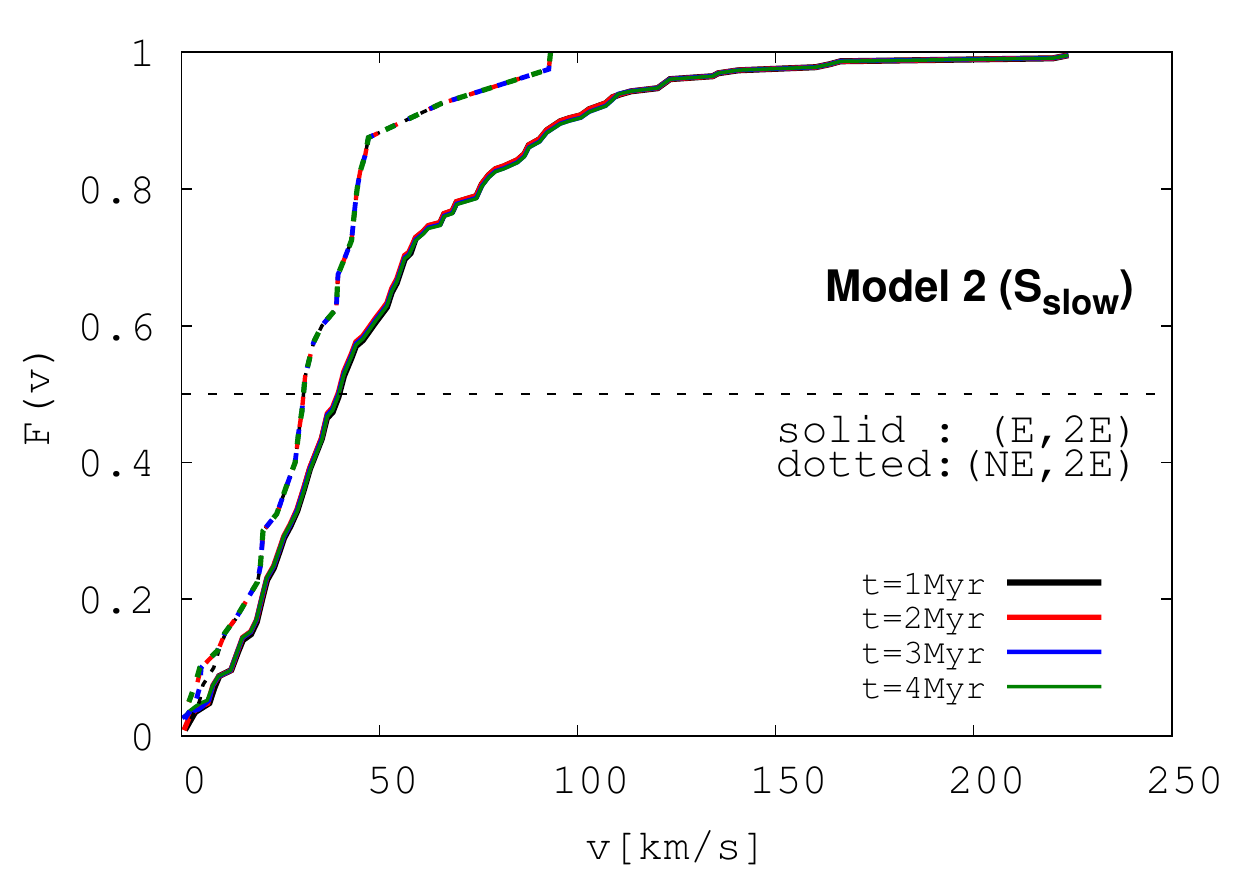}			
	\includegraphics[width=7.8cm]{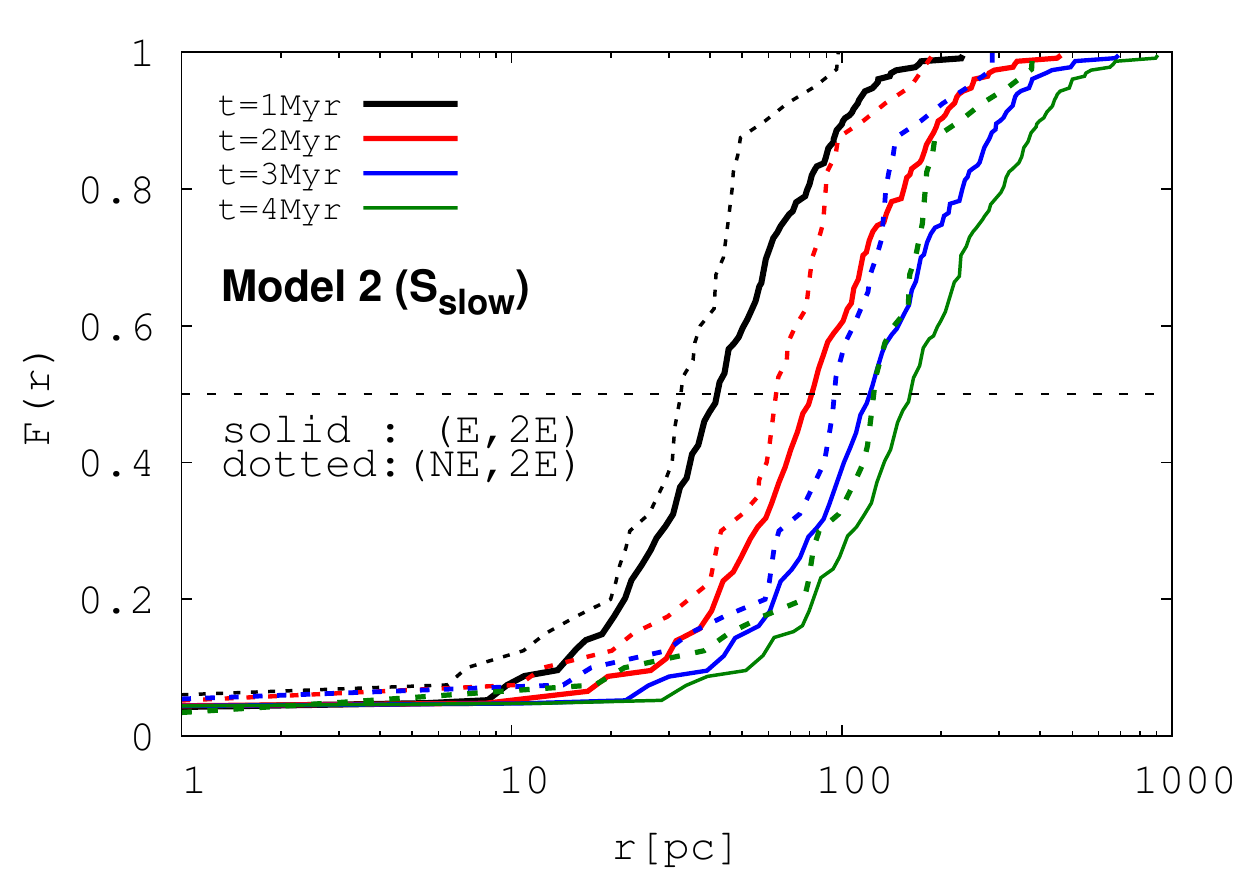}					\includegraphics[width=7.8cm]{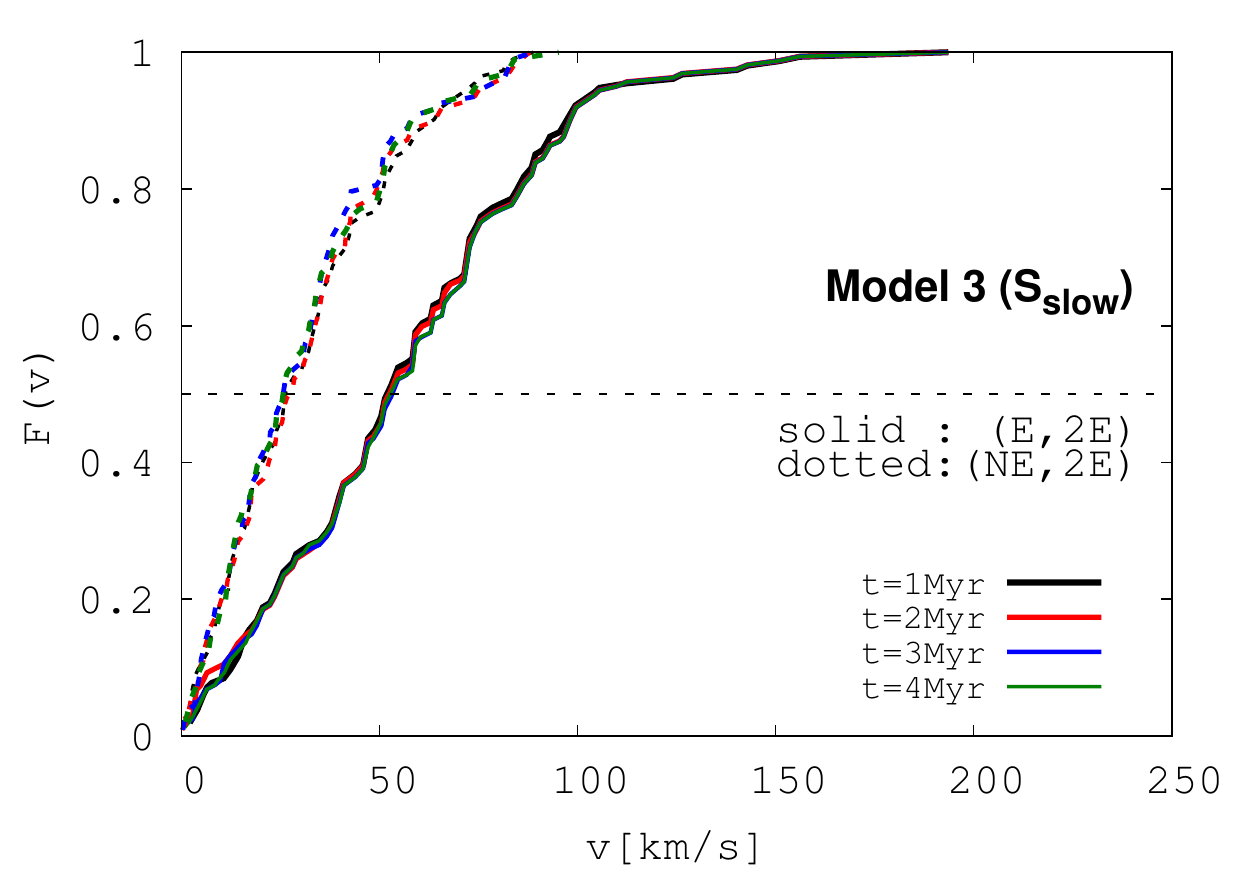}			
	\includegraphics[width=7.8cm]{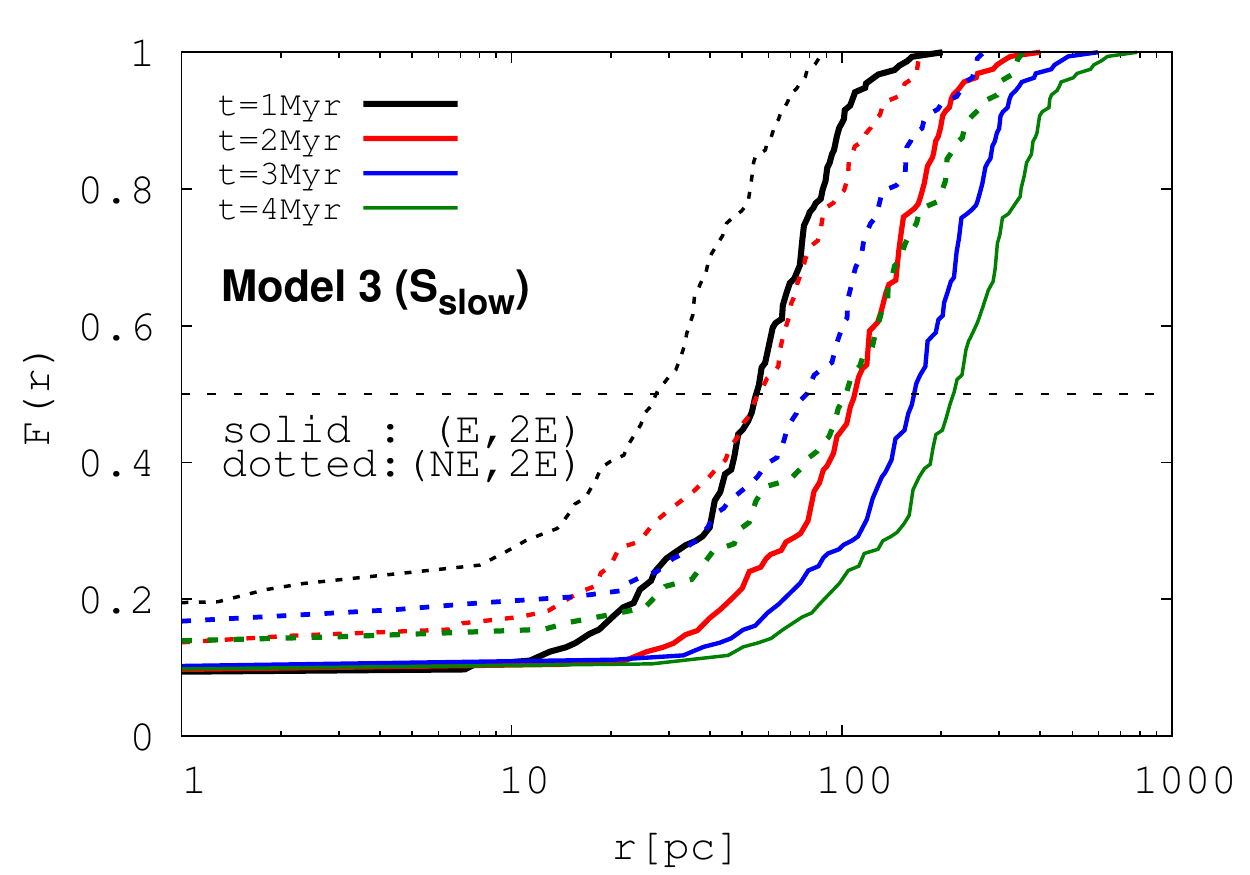}				
	\caption{\textit{slower single star} :The cumulative distributions of the final velocities (\textit{left} column) and the radial distances from the system CM (\textit{right} column) for the slower single stars ($S_{\rm slow}$) for all the models at $t=1\Myr$ (black), $2\Myr$ (red), $3\Myr$ (blue) and $4\Myr$ (green) from the drop-in time. We use the same formats used as in Figure \ref{fig:cumulative_S_fast}.}
	\label{fig:cumulative_S_slow}
\end{figure*}

We provide the cumulative distributions of the spatial distances from the system CM and the speeds for the escaped binaries (Figure \ref{fig:cumulative_binary}), $S_{\rm fast}$ (Figure \ref{fig:cumulative_S_fast}) and $S_{\rm slow}$ (Figure \ref{fig:cumulative_S_slow}) for all models for $t=1,~2,~3$ and $4\Myr$.

\bsp	
\label{lastpage}
\end{document}